\documentclass[aps, pra, twocolumn, 10pt, balancelastpage, amsmath, amssymb,superscriptaddress]{revtex4-2}

\usepackage{graphicx}%
\usepackage{adjustbox} %
\usepackage{ragged2e}%
\usepackage{dcolumn}%
\usepackage{bm}%
\usepackage{svg}
\usepackage{dsfont} %
\usepackage{subcaption} %
\usepackage[colorlinks=true]{hyperref}%
\hypersetup{
    allcolors = {blue}
}
\usepackage{amsmath}
\usepackage{cleveref}%
\usepackage{amsfonts}
\usepackage{braket}
\usepackage{enumitem}
\usepackage{bm} %
\usepackage{bbold} %
\usepackage{comment} %

\usepackage{orcidlink}

\newcommand{\hopintra}{J_{o}} %
\newcommand{\hopinter}{J_{e}} %

\usepackage[normalem]{ulem}
\newcommand{\add}[1]{{#1}}
\newcommand{\del}[1]{}
\newcommand{\rep}[2]{\del{#1}\add{#2}}

\begin{document}

\title{Dynamics of  edge modes in monitored Su-Schrieffer-Heeger Models}

\author{Giulia Salatino~\orcidlink{0009-0005-6913-6920}}
 \affiliation{Scuola Superiore Meridionale, Via Mezzocannone, 4 80138, Napoli, Italy.}
 \email{giulia.salatino@unina.it} 

\author{Gianluca Passarelli~\orcidlink{0000-0002-3292-0034}}%
\affiliation{Dipartimento di Fisica E. Pancini, Universit\`a di Napoli "Federico II",
Complesso di Monte S. Angelo, via Cinthia, I-80126 Napoli, Italy}

\author{Angelo Russomanno~\orcidlink{0009-0000-1923-370X}}
\affiliation{Dipartimento di Fisica E. Pancini, Universit\`a di Napoli "Federico II",
Complesso di Monte S. Angelo, via Cinthia, I-80126 Napoli, Italy}

\author{Giuseppe E. Santoro~\orcidlink{0000-0001-6854-4512}}
\affiliation{SISSA, Via Bonomea 265, I-34136 Trieste, Italy.}
\affiliation{ The Abdus Salam International Centre for Theoretical Physics, I-34014 Trieste, Italy}

\author{Procolo Lucignano~\orcidlink{0000-0003-2784-8485}}
\affiliation{Dipartimento di Fisica E. Pancini, Universit\`a di Napoli "Federico II",
Complesso di Monte S. Angelo, via Cinthia, I-80126 Napoli, Italy}

\author{Rosario Fazio~\orcidlink{0000-0002-7793-179X}}
\affiliation{ The Abdus Salam International Centre for Theoretical Physics, I-34014 Trieste, Italy}
\affiliation{Dipartimento di Fisica E. Pancini, Universit\`a di Napoli "Federico II",
Complesso di Monte S. Angelo, via Cinthia, I-80126 Napoli, Italy}

\date{\today}%

\begin{abstract}
We investigate the effect of dissipation on the dynamics of edge modes in the monitored Su-Schrieffer-Heeger (SSH) model. Our study considers both a linear observable and a nonlinear entanglement measure, namely the two-point correlation function and the Disconnected Entanglement Entropy (DEE), as diagnostic tools. While dissipation inevitably alters the entanglement properties observed in the closed system, statistical analysis of quantum trajectories reveals that by protecting the chain’s edges from dissipation, it is possible to recover characteristic features analogous to those found in the unitary limit. This highlights the fundamental role of spatial dissipation patterns in shaping the dynamics of edge modes in monitored systems.
\end{abstract}

\maketitle

\section{Introduction}
\label{sec:introduction}

Condensed matter physics has seen significant advancements from the discovery and investigation of systems characterized by non-trivial topological 
phases~\cite{bernevig2013topological}.  While  the main focus has been centered in the study to the ground state  of both free and interacting 
many-body systems, recent attention has been devoted to understand topological properties of  non-Hermitian~\cite{Ashida_2020, okuma2023nonhermitian} 
and open quantum systems governed by Lindblad dynamics~\cite{Diehl_2011, bardyn2013topologybydissipation, cooper, Altland_2021},
where the interplay between topology and non-unitary dynamics  gives rise to novel phases and dynamics not observed in closed systems. Examples are 
the breakdown of bulk-boundary correspondence and the emergence of the non-Hermitian skin effect~\cite{Ashida_2020}.
Symmetries play a crucial role in determining the topological properties of these systems and the tenfold way classification, based on 
Altland-Zirnbauer classes~\cite{altland} has been extended to the steady-state of systems governed by non-Hermitian~\cite{Kawabata_2019,Bernard_2002} 
and Lindblad~\cite{Prosen_2012, Buča_2012, Liang_2014, Rabl_2020, Altland_2021, Kawasaki_2022, Mao_2024,cooper} dynamics. Still, defining 
a suitable topological marker for open topological insulators remains an open challenge. Several quantities have been proposed, including non-Hermitian 
topological invariants~\cite{Kunst_2018, Gong_2018, Shen_2018}, the Uhlmann phase~\cite{Delgado_2014, Viyuela_2018, carollo2017uhlmanncurvaturedissipativephase}, 
the ensemble geometric phase~\cite{Bardyn_2018, Unayan_2020}, the  mixed-state topological order parameter of~\cite{huang2024mixedstatetopologicalorder}, the long-time entanglement negativity~\cite{chen2024universalentanglementrevivaltopological}, as well as the imbalance and current properties~\cite{nava2023, campagnano2024}.

Inspired by these studies on the interplay of decoherence/dissipation and topology, aim of the present work is to investigate the topological properties of monitored  
quantum systems, i.e. a stochastic dynamics given by a smooth evolution interrupted by quantum jumps~\cite{carmichael2013statistical,Daley_2014, 
Plenio_1998,Molmer:93, breuer}, and whose average quantum state (the density matrix) is governed by the Lindblad equation. More specifically, we will analyze 
the dynamics of topological edge states under the combined effect of a quantum quench {\em and} the random fluctuations induced by quantum jumps.  

Very recently,  monitored dynamics of many-body systems has been actively started to be investigated in quantum circuits and open systems. Two independent works~\cite{li2018quantumzenoeffect,skinner2019measurementinducedphase}  showing the existence of a measurement-induced phase transition 
visible only at the level of single trajectories triggered an  intense activity scrutinizing many different facets of this phenomenology (the interested reader can find 
additional references on this activities in the review articles~\cite{vasseur2022entanglementhybrid} and~\cite{Fazio2024}). In essence, by observing the system at the level of single trajectories
(i.e. monitoring the systems) it is possible to have access to nonlinear functions of the quantum states, showing a plethora of new phenomena,  that cannot be 
extracted by looking at the density matrix. This is the case of topological markers.

Indeed, averaging over trajectories is {\em not} equivalent to the study of the topology embedded in the density matrix because the topological markers are
usually nonlinear functions of the quantum states. This means that looking at monitored systems we are able to probe the interplay of topology and dissipation from a
new angle. The protocol we follow, on the other hand,  is rather standard, that of a quantum quench. After having prepared the system in a given quantum 
states (topological or trivial) we will follow the evolution of the signatures of topology along a given trajectory, where the smooth (non-Hermitian) evolution is 
interrupted at random times by quantum jumps.  We do not confine ourselves to the steady-state, on the contrary we are interested in understanding 
how these properties evolve in time, along a given quantum trajectory, and subject both to Hamiltonian evolution and dissipation/decoherence.  

A key aspect to bear in mind is the fact that along a single trajectory the state is always pure and therefore we can use topological markers that have been 
devised for the unitary case. For this reason, after a preliminary analysis of the two-point correlator between the edge sites of the chain, we move on and study a quantity that is nonlinear in the density matrix of the system, namely the Disconnected Entanglement Entropy (DEE), first introduced in Ref.~\cite{zeng2018quantum}. DEE is a 
robust measure  of symmetry-protected topological phases, while traditional entanglement measures, such as the von Neumann entropy, fall 
short~\cite{Chen_2015, Isakov_2011, Jiang_2012, Kitaev_2006, Levin_2006}. Unlike the winding number, the DEE is not a bulk topological invariant. Instead, 
it quantifies the entanglement between topological edge states by partitioning the system into disjoint regions and measuring the entropy of the reduced density 
matrix for these regions. DEE has been shown to detect long-range entanglement in certain phases that are not symmetry-protected topological 
 phases~\cite{torre2023longrangeentanglementtopologicalexcitations}, as well as Majorana zero modes in semiconductor-superconductor heterostructures~\cite{arora2023conclusivedetectionmajoranazero}. This order parameter, closely related to entanglement entropy, is also experimentally 
 accessible, as discussed in Refs.~\cite{Zoller_PRX_2016, Zoller_PRL_2018, Zoller_Science_2019}.
Of particular relevance for the present work, are Refs.~\onlinecite{Dalmonte_PhysRevB.101.085136, Micallo_2020, Mondal_2022} where the dynamics of the 
DEE  is considered, here  extended to the stochastic dynamics of quantum jumps.

We investigate a paradigmatic case, the Su-Schrieffer-Heeger (SSH) model, well-known for its simplicity and the rich variety of phenomena it presents. Initially 
proposed to describe electron behavior in polyacetylene~\cite{ssh_1979, ssh_1980}, it serves as a one-dimensional example of a system with topological edge 
states protected by chiral symmetry~\cite{Hasan_2010, Qi_2011}. Specifically, we examine the DEE of an SSH chain under different types of dissipation, 
classified according to Ref.~\cite{cooper}.

We first examine the time evolution of the average value of the DEE over multiple trajectories. Then, we perform a statistical analysis of individual trajectories, focusing on the variation of the DEE caused by each quantum jump in the initial phase of the dynamics. We consider both uniform and non-uniform types of dissipation, acting differently on different segments of the chain.  

Our results show that while bulk dissipation has a limited
impact on the system, quantum jumps at the boundaries, as expected,  have a disruptive effect on the edge modes. These results emerge both in the average properties and in the fluctuations of the DEE. 

On one hand, through the study of the average value of the DEE, we extend the results
of~\cite{Micallo_2020}, where it was shown that an initial doublet of entangled topological modes persists for a time linear in the system size when evolving under
a local Hamiltonian. Indeed, here we extend this result to quadratic dissipators, demonstrating that the entanglement of the topological modes remains for a time linear 
in system size as long as the dissipator does not affect the boundary. Notably, in the transient regime of the dynamics, 
the spatial localization of the dissipation plays a more crucial role than its symmetry class in determining the stability of the edge modes. Furthermore, when dissipation affects the boundary of the chain, the dynamics induced by the coherent part of the Lindbladian also appears to be crucial. 

On the other hand, the statistical analysis reveals that, when the quantum jumps are localized, there is a peak in the probability distribution of the DEE variation that signals the destruction of the topological entanglement. This peak is primarily attributable to the effect of the first local jump occurring on one of the two edge sites. 

The paper is organized as follows: In~\autoref{sec:model}, we review the SSH model and its topological properties, introduce the types of dissipative dynamics 
we will examine, describe the quantum jump unraveling of the Lindblad equation, and outline our methodology for calculating the DEE. In~\autoref{sec:dee}, 
we present the DEE and the method to calculate its time evolution. In~\autoref{sec:results}, we discuss our main results, including the effects of different dissipative 
dynamics on the edge modes characterizing the system's initial topological phase. We also explore the time evolution of the DEE and the scaling behavior of a 
DEE-related quantity: the time at which the DEE deviates from its quantized initial value due to dissipation. Lastly, we conduct a statistical analysis of the DEE's 
variation after a quantum jump, considering both the timing and location of the jump for a comprehensive time- and index-resolved analysis. In~\autoref{sec:conclusion}, 
we discuss the implications of our findings for the understanding of dissipation in topological phases and propose potential directions for future research.

\section{The model and its monitored dynamics}
\label{sec:model}

As mentioned in the Introduction, we will focus our attention on a one-dimensional free-fermion model governed by the SSH Hamiltonian. 
The Hamiltonian of the SSH model $\hat{H}_{\mathrm{\scriptscriptstyle SSH}}$ describes a 1D atomic chain with two atoms per unit cell, on which 
electrons hop with staggered hopping amplitude~\cite{Asb_th_2016}.  The sites of the lattice $i \equiv (j,A/B)$ are identified by their coordinates
$j$ in one of the two sub-lattices $A/B$.  In the following, we will study the dynamics of the monitored system along a quantum 
trajectory~\cite{carmichael2013statistical,Daley_2014, Plenio_1998,Molmer:93, breuer}. In this case, the dynamics, under the action of the
environment, is given by a smooth evolution, governed by non-Hermitian Hamiltonian, interrupted by quantum jumps described by the corresponding
Lindblad operators $\hat{L}_{i}$.

\subsection{Stochastic dynamics and Quantum jumps} 

 The stochastic Schr{\"o}dinger equation, governing the wavefunction $\ket{\psi(t)}$, has the form ($\hbar=1$) 
\begin{equation}  
	\begin{split} &\text{d}\ket{\psi(t)}=
	\Bigg\{\text{d}t \left(-i\hat{H}_{\mathrm{\scriptstyle eff}}+\textstyle{\frac{1}{2}}\sum_i\braket{\hat{L}^{\dagger}_{i}\hat{L}_{i}}\right)+
	\\
	&\hspace{15mm}+
	\sum_i \text{d}N_i(t)
	\Bigg(
	\frac{\hat{L}_{i}}{\sqrt{\braket{\hat{L}^{\dagger}_{i}\hat{L}_{i}}_t}}-\hat{\mathds{1}}
	\Bigg)
	\Bigg\}
	\ket{\psi(t)} \;,
	\end{split}
    \label{eq:stocastic_schrod}
\end{equation}
where we define $\langle \hat{L}^{\dagger}_{i}\hat{L}_{i}\rangle_t = \langle \psi(t) | \hat{L}^{\dagger}_{i}\hat{L}_{i} |\psi(t)\rangle$. In Eq.~\eqref{eq:stocastic_schrod}, the
term $-i \hat{H}_{\mathrm{\scriptstyle eff}} \ket{\psi(t)} \text{d}t$ gives a deterministic contribution to the evolution of the state with a non-Hermitian Hamiltonian of the form
\begin{equation}
    	\hat{H}_{\mathrm{\scriptstyle eff}}=\hat{H}_{\mathrm{\scriptscriptstyle SSH}}-\frac{i}{2}\sum_i \hat{L}^{\dagger}_{i}\hat{L}_{i} \;,
\end{equation}
the c-number term in the first bracket is needed in order to keep the state normalized.  The second term on the r.h.s. --- the fluctuation term --- is the one 
that makes the differential equation stochastic. Indeed, $\text{d}N_i(t)=N_i(t+\text{d}t)-N_i(t)$ represents the number of \textit{jumps}  of type $i$ the state 
goes across as a result of a \textit{measurement} made by the environment in the time interval $\text{d}t$.  This is a stochastic variable with expectation value
\begin{equation}
    	\text{E}[\text{d}N_i(t)]=\braket{\hat{L}^{\dagger}_{i}\hat{L}_{i}}_t\text{d}t \;.
\end{equation}
Since we consider $\text{d}t\rightarrow0$, we can assume $\text{d}N_i(t)$ will follow a Poissonian distribution~\cite{gal2024entanglement}, so that
\begin{equation}
    	\text{d}N_i(t)=\left\{ \begin{array}{ll}
    	1 \hspace{2mm}\mbox{\small{with probability}}\hspace{2mm} \braket{\hat{L}^{\dagger}_{i}\hat{L}_{i}}_t\text{d}t \vspace{3mm}\\
    	0 \hspace{2mm}\mbox{\small{with probability}}\hspace{2mm} 1-\braket{\hat{L}^{\dagger}_{i}\hat{L}_{i}}_t\text{d}t
    	\end{array}
    	\right. \;.
\end{equation}
The stochastic noise acting on each trajectory can be seen as a result of action (generalized measurement) of the environment on the systems. 

Each realization of this stochastic process is a quantum trajectory, the ensemble (average state) evolution being provided by the Lindblad equation~\cite{lindblad, breuer} 
\begin{align}
    	\frac{\mathrm{d}}{\mathrm{d} t} \hat{\rho}(t) &= -i\big[\hat{H}_{\mathrm{\scriptscriptstyle SSH}},\hat{\rho}(t)\big]+
    	\nonumber \\
    	&\phantom{=} 
    	+ \sum_i \Big( \hat{L}_{i}\hat{\rho}(t) \hat{L}^{\dagger}_{i}
    	-\textstyle{\frac{1}{2}}\big\{\hat{L}^{\dagger}_{i}\hat{L}_{i},\hat{\rho}(t)\big\} \Big) \;.
    	\label{eq:lindblad}
\end{align}
As long as one is interested in the evolution of physical observables, the averaging over the trajectories or the study of the Lindblad dynamics is equivalent
(the advantage of the unraveling method is that it requires evolving a pure state instead of a density matrix, resulting in a substantial reduction in computational complexity).
The situation changes drastically when one is interested in signatures that are embodied in quantities that are nonlinear function of the quantum states. Topological
markers are an example of this sort.  In this case following the dynamics of topological properties of a monitored quantum system may differ considerably from the
topological properties of its average dynamics.  This is what we are going to study in the next Sections.

\subsection{The model}

\begin{figure}
    \centering
    \includegraphics[width=\columnwidth]{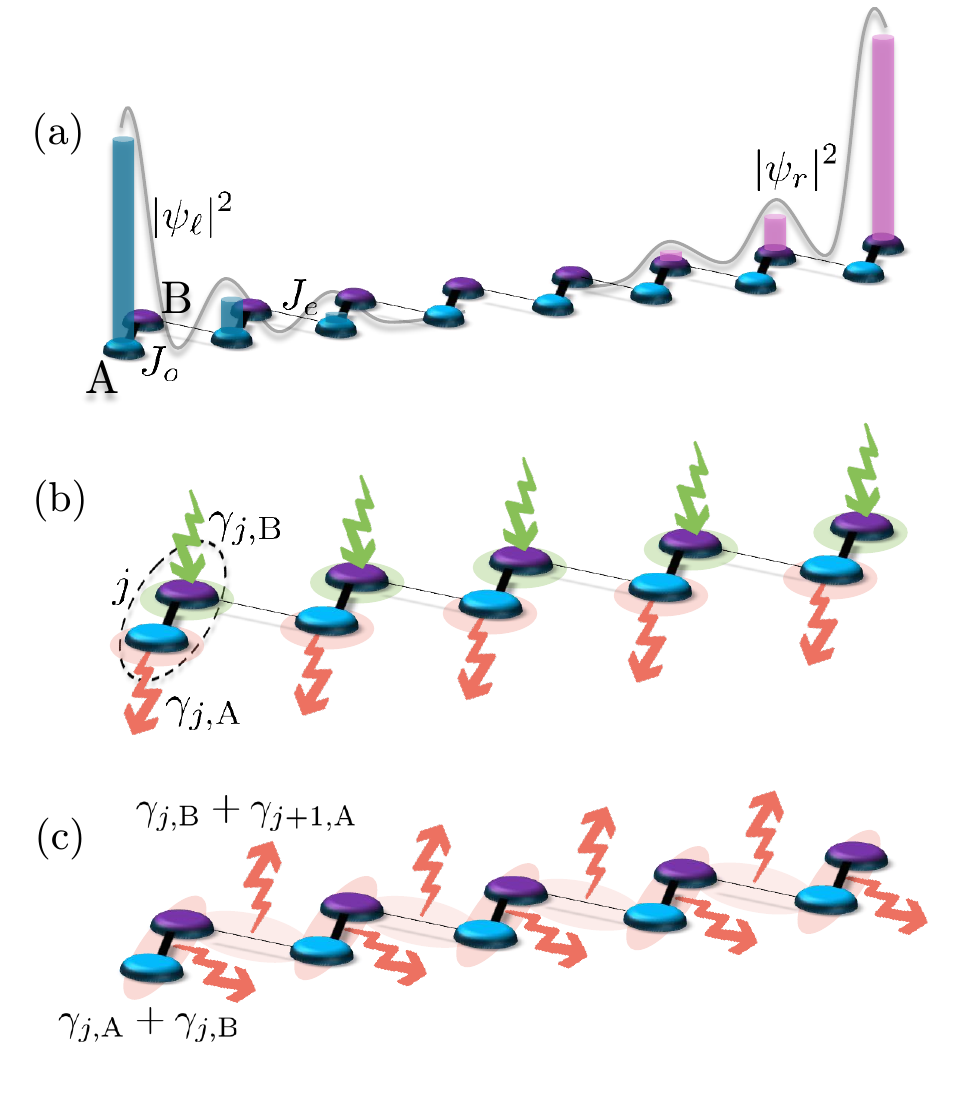}
    \caption{\justifying (a) Sketch of the probability amplitude of the zero-mode states $\ket{\psi_\ell}$ and $\ket{\psi_r}$ that are exponentially 
    localized at the edges of the SSH chain. (b) Sketch of the symmetry-preserving environment. The dotted oval represents the $j$-th unit cell. 
    The green and orange shaded shapes represent the sites involved by the single Lindblad operators. (c) Sketch of the symmetry-breaking 
    environment.}\label{fig:sketch_dissipation}
\end{figure}

The Hamiltonian of the systems, defined on a one-dimensional lattice with open boundary conditions (OBC), is 
\begin{equation}
	\hat{H}_{\mathrm{\scriptscriptstyle SSH}} = 
	-\hopintra \sum_{j=1}^N 
	{\hat{c}^{\dagger}}_{j,\mathrm{\scriptstyle A}} \hat{c}^{\phantom \dagger}_{j,\mathrm{\scriptstyle B}}
	-\hopinter \sum_{j=1}^{N-1} 
	 {\hat{c}^{\dagger}}_{j+1,\mathrm{\scriptstyle A}} \hat{c}^{\phantom \dagger}_{j,\mathrm{\scriptstyle B}}+\text{h.c.},
	\label{eq:ssh_hamiltonian}
\end{equation}
where $\hopintra, \hopinter >0$, A and B indicate the type of atom within the unit cell (the sublattice index), $N$ is the number of unit cells and 
$L=2N$ is the number of sites in the chain.  At half filling, the ground state of $\hat{H}_{\mathrm{\scriptscriptstyle SSH}}$ displays two distinct
topological phases. When $ \hopintra/\hopinter> 1$ the phase is topologically trivial, while it is topological for $ \hopintra/\hopinter< 1$.
In the topological phase, the density profile of each edge shows one localized fermion on the left end and one on the right end of the chain -- 
see sketch in Fig.~\ref{fig:sketch_dissipation}(a), and \Cref{app:dee} and Refs.~\cite{Asb_th_2016,Simon_2019} for further details. This is related to the fact
that, in this topological, fully-dimerized limit, one fermion is localized at each inter-cell link, and two unpaired fermions are left on the extreme sites.

In the thermodynamic limit these topological boundary excitations are degenerate and are at zero energy. For any finite size they hybridize, and an
exponentially small gap open between the even and the odd superposition of them. Therefore at finite size, half filling and zero temperature -- 
when all the lower band is occupied -- only the even superposition $\frac{1}{\sqrt{2}}\left({\hat{c}^{\dagger}}_{N,\mathrm{\scriptstyle B}}+
{\hat{c}^{\dagger}}_{1,\mathrm{\scriptstyle A}}\right)$ of the topological boundary modes is occupied, while the odd is empty.
This even superposition of edge modes can be written as a Bell state, as it is explained in detail in~\Cref{app:dee} and is such that the two distant 
edges are entangled. This is the only long-range entanglement of the whole chain that can be detected when the system is in a topological
phase~\cite{Micallo_2020}. 

Beyond it, there is only short-range entanglement across each intercell link. The Hamiltonian of the SSH 
model~\eqref{eq:ssh_hamiltonian} falls in the BDI symmetry class of the AZ tenfold way~\cite{altland}, which leads to a $\mathbb{Z}$ type of 
topological invariant in 1D. This means that the edge states are protected by time-reversal, chiral and particle-hole symmetries and that there is 
an infinite countable number of distinct topological phases with the symmetries preserved (see also~\Cref{app:symmetry_classifications}). 
In the presence of periodic boundary conditions (PBC) translation symmetry allows to detect the topological phase using non-local topological 
invariants like the winding number or Zak phase in $k$ space~\cite{M.V.Berry03081984, Zak1989}.

The dynamics introduced in Eq.~\eqref{eq:stocastic_schrod} requires the definition of the jump operators (Lindblad operators). 
In this manuscript we consider two different cases.  
The first type of jumps we analyze is the one that lead to the global Symmetry-Preserving Dissipation (SPD) dynamics~\cite{cooper}. 
The jump operators are defined as:
\begin{equation} \label{eq:spd_jump}
	\hat{L}_{i} = \left\{
	\begin{array}{l}
	\hat{L}_{j,\mathrm{\scriptstyle A}}=\sqrt{\gamma_j} \, \hat{c}^{\phantom \dagger}_{j,\mathrm{\scriptstyle A}} \,,\vspace{5mm} \\
	\hat{L}_{j,\mathrm{\scriptstyle B}}=\sqrt{\gamma_j} \, {\hat{c}^{\dagger}}_{j,\mathrm{\scriptstyle B}}\,,
	\end{array}
	\right.
\end{equation}
In Eq.~\eqref{eq:spd_jump} we assumed that the associated dissipation strength $\gamma_j$ can be site-dependent since we will also study how 
jumps impact topological edge states and for this it is convenient to consider non-homogenous coupling to the external environment.  
The non-Hermitian Hamiltonian associated to this dynamics (no-jump trajectory)  has been first theoretically proposed in Ref.~\cite{Schomerus:13} 
and then experimentally implemented~\cite{Weimann2017TopologicallyPB} in a photonic lattice with engineered gain and loss, and robust topological 
edge states have been observed. We sketch this dissipation in Fig.~\ref{fig:sketch_dissipation}(b) emphasizing that in each cell there is one site 
where fermions are injected by the Lindbladian, and the other where fermions leak out, with the same rate.

The second type of jump operators that we will analyze corresponds to the Lindbladian with  global Symmetry-Breaking Dissipation (SBD)~\cite{cooper},
\begin{equation} \label{eq:sbd_jump}
	\hat{L}_{i} = \left\{
	\begin{array}{l}
    	\hat{L}_{j,\mathrm{\scriptstyle A}}=\sqrt{\gamma_j} \left(\hat{c}^{\phantom \dagger}_{j,\mathrm{\scriptstyle A}}+
	\hat{c}^{\phantom \dagger}_{j,\mathrm{\scriptstyle B}}\right)\;,
    	\vspace{5mm} \\
    	\hat{L}_{j,\mathrm{\scriptstyle B}}=\sqrt{\gamma_j} \left(\hat{c}^{\phantom \dagger}_{j,\mathrm{\scriptstyle B}}+
	\hat{c}^{\phantom \dagger}_{j+1,\mathrm{\scriptstyle A}}\right)\,.
	\end{array}
	\right.
\end{equation}
Fig.~\ref{fig:sketch_dissipation}(c) shows that each Lindblad operator acts on pairs of sites and has the effect of losses, that is, fermions leak out
of the system.
Jump operators in Eq.~\eqref{eq:spd_jump} and Eq.~\eqref{eq:sbd_jump} lead to a Lindblad dynamics that respectively satisfies/breaks the generalized
symmetry relations that characterize the BDI class in presence of dissipation, as proposed in Ref.~\cite{cooper}. Further details about this topological 
classification of dissipators (the so-called dissipative tenfold way) are provided in~\Cref{app:symmetry_classifications}. 

As briefly mentioned above, in order to assess the impact of dissipation on the edge modes,  we will consider the possibility that the jump
operators involve only a portion of $\lfloor\alpha L\rfloor$ ($\lfloor\cdot\rfloor$ is the floor function) central sites, with $\alpha\in[0,1]$ such that, once we fix the parameters of the system and the
range of lengths we will deal with, the edges of the chain are left untouched. This can be achieved by adjusting the site-dependent values of
$\gamma_{j,\mathrm{\scriptstyle A}/\mathrm{\scriptstyle B}}$ so that they are different from zero on a limited portion of central sites. We have 
numerically verified that choosing $\alpha=0.8$ is sufficient to separate the bulk dissipation from the boundary. 
Fig.~\ref{fig:sketch_dissipation}(a) is a sketch representing how the zero-energy modes are exponentially localized at the edges of the chain in
such a way that a non-homogeneous SPD or SBD environment with a suitable choice of $\alpha$ might not touch them. 

The Lindbladian that we take into account is quadratic, since the Hamiltonian of the SSH model is quadratic and the jump operators are
linear in the fermionic operators. Given a Gaussian state as initial condition of a quadratic dissipative dynamics, its Gaussian character is preserved
in time, so that Wick's theorem holds for its whole evolution in time (see~\Cref{app:free_fermions} for more details on Gaussian states).

\section{Disconnected Entanglement Entropy along quantum trajectories}
\label{sec:dee}
Having defined the model and its dynamics, we now discuss the quantity that we will use to study the dynamics of edge modes, the DEE,
employed for quantum quenches in Refs.~\cite{Dalmonte_PhysRevB.101.085136, Micallo_2020, Mondal_2022}. 

\begin{figure}[b]
    \centering
    \includegraphics[width=\columnwidth]{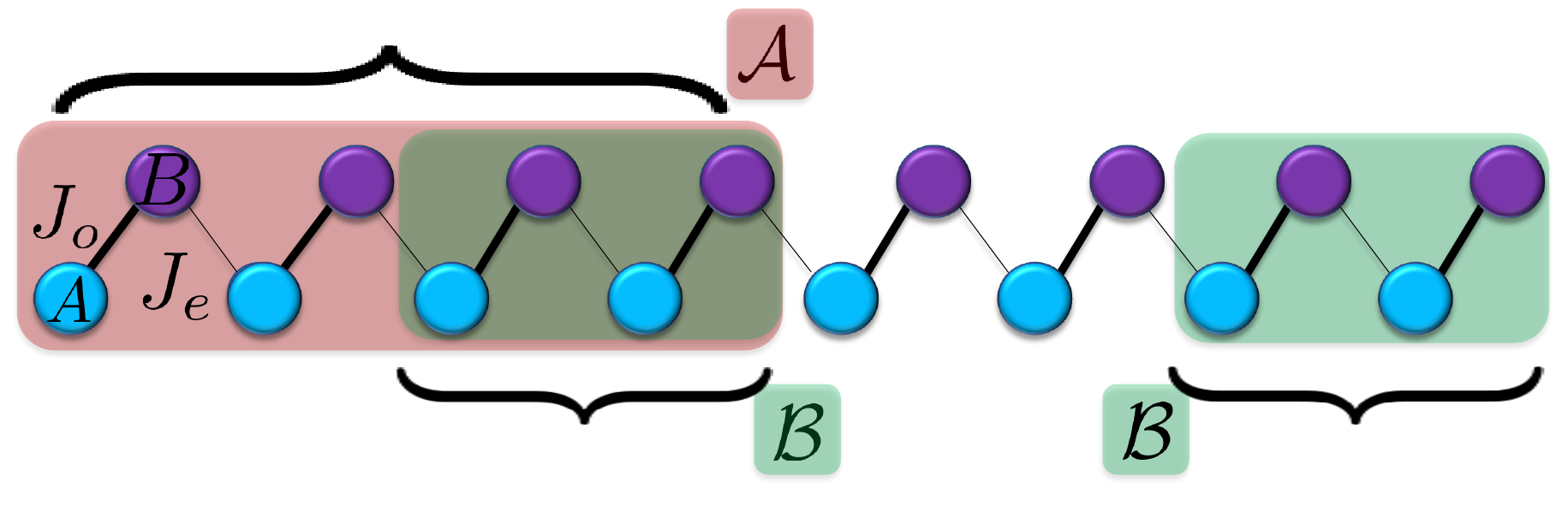}
    \caption{\justifying 
    Disconnected partition of the SSH chain. The red connected subset is named $\mathcal{A}$, while the green disconnected one is named $\mathcal{B}$. 
    The blue (odd) sites represent the sublattice indices $A$, while the purple (even) sites represent the sublattice indices $B$. $\hopintra$ and 
    $\hopinter$ are the intra-cell and inter-cell hopping amplitudes, respectively.}
    \label{fig:partition}
\end{figure}

Let us  consider a connected bipartition of a system into two subsets $X$ and $\Bar{X}$. The von Neumann bipartite entanglement entropy of the 
subsystem in $X$ is defined as 
\begin{equation}
    	S_X=-\mathrm{Tr}_{\scriptscriptstyle{\mathrm X}} \left(\hat{\rho}_{\scriptscriptstyle{\mathrm X}} \log_2 \hat{\rho}_{\scriptscriptstyle{\mathrm X}} \right) ,
    	\label{eq:dee_spectrum}
\end{equation}
where $\hat{\rho}_{\scriptscriptstyle{\mathrm X}}=\mathrm{Tr}_{\Bar{\scriptscriptstyle{\mathrm X}}} \hat{\rho}$ is the reduced density matrix of the system 
in $X$ and $\{\lambda\}$ is the set of its eigenvalues and $\hat{\rho}$ is a pure state. 
By choosing the different bi-partitions as shown in Fig.~\ref{fig:partition}, the DEE is defined as \footnote{Other choices could also be considered, but 
with less experimental relevance and a more complicated interpretation in terms of mutual information~\cite{Dalmonte_PhysRevB.101.085136}.}
\begin{equation}
    	S^D=S_{\mathcal{A}}+S_{\mathcal{B}}-S_{\mathcal{A}\cup \mathcal{B}}-S_{\mathcal{A}\cap \mathcal{B}}.
    	\label{eq:dee}
\end{equation}
The DEE is able to detect the presence of edge modes, thus being equal to $2$ in the topological phase or to $0$ in the trivial phase for the SSH chain.  
A disconnected partition is necessary for the definition of a marker of topological phases since the entanglement spectrum of single connected partitions 
is not able to distinguish the topological character of wave functions~\cite{Dalmonte_PhysRevB.101.085136}. 
A more detailed explanation of the properties of DEE is provided in \Cref{app:dee}.

For Gaussian states, computing the entanglement entropy for a subset $X$ of a partition is equivalent to computing the spectrum $\{\zeta\}$ of the 
reduced covariance matrix $\textbf{G}^{(\mathrm{\scriptstyle traj})}$ of the same subsystem. The elements of this covariance matrix are
\begin{equation}
	\mathrm{G}^{(\mathrm{\scriptstyle traj})}_{i,i'}=\braket{\psi(t)|{\hat{c}^{\dagger}}_{i'}\hat{c}^{\phantom \dagger}_{i}|\psi(t)}=
	\mathrm{Tr}(\hat{\rho}_{\scriptscriptstyle{\mathrm X}}{\hat{c}^{\dagger}}_{i'}\hat{c}^{\phantom \dagger}_{i}),
    	\label{eq:cov_mat}
\end{equation}
where $i,i'\in X$~\cite{IngoPeschel_2003} and $\ket{\psi(t)}$ is the state along the considered quantum trajectory. 

Since the dissipative dynamics is quadratic, hence preserving the Gaussian character of the initial state, we directly evolve the covariance matrix of the reduced system \eqref{eq:cov_mat} on each single trajectory. Details on the calculations are provided in \Cref{app:numerical_degeneracy} and in \Cref{app:trajectories}. Resorting to the 
properties of free fermions (\Cref{app:free_fermions}), we obtain  $S^{(D,\mathrm{\scriptstyle traj})}$, and we can analyze its statistical properties because it is a stochastic 
variable itself. 
In particular,  we can compute its expected value averaging over trajectories 
$$
	S^D=\overline{S^{(D,\mathrm{\scriptstyle traj})}}.
$$ 
as well as the probability distribution function of its variations under quantum jumps $P(\Delta S^D)$, evaluated over time and trajectories (see Sec.~\ref{Del:sec}).

\section{\label{sec:results}Results}
In this section, we investigate the stability of the topological phase under dissipation by 
evaluating the DEE along quantum trajectories.

In order to get a first insight on the impact of Lindblad dynamics on the edge states, we first discuss the spreading of the edge modes due to dissipation. 
Next, we investigate the time-evolution of the expectation value of the DEE under the two types of dissipation --- SPD and SBD --- outlined in the previous sections.

\subsection{Spreading of the edge modes}
The existence of edge states is a well-known  characteristics of topological insulators under open boundary conditions. The localization of the edge 
modes depends on the choice of the ratio $\hopintra/\hopinter$ and on the length of the chain~\cite{Asb_th_2016}. In the limit of perfectly dimerized chain, 
as shown in~\Cref{app:dee}, the edge modes are exactly localized at the two edge sites. On the contrary, when $\hopintra/\hopinter>0$, the modes have 
an exponential decay within the bulk of the chain.  As a consequence, owing to the long-range entanglement, the existence of a non-zero correlation among 
the edge sites of the chain can also characterize a topological state. For instance, the topological order would be disrupted if the boundary modes vanish 
under the action of a dissipation, and this would be probed by the correlations going to zero as well. 

For this reason, we look at the time evolution of the covariance matrix 
$$
	\textbf{G} = \overline{\textbf{G}^{(\mathrm{\scriptstyle traj})}} , 
$$
averaged over trajectories (the covariance matrix is a linear function of the state and its average over trajectories coincides with its evaluation using the density matrix).

\begin{figure}
    \centering
    \includegraphics[width=\columnwidth]{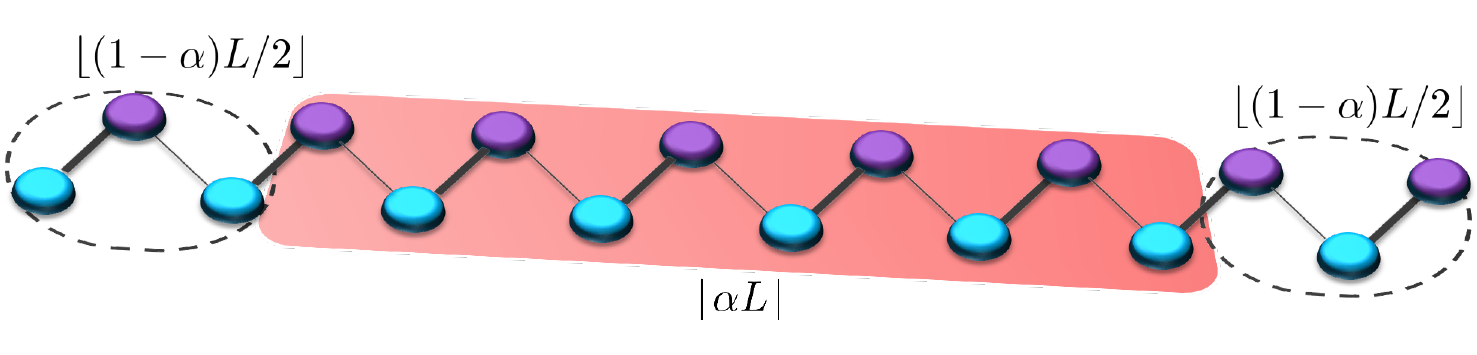}
    \caption{\justifying Range of the non-homogeneous SPD dissipation. When $\lfloor\alpha L\rfloor$ central sites are affected by dissipation, there are $n=\lfloor(1-\alpha)L/2\rfloor$ sites 
    left of each edge on the chain.}\label{fig:sketch_alpha}
\end{figure}

Not surprisingly, not all the quantum-jump protocols have the same effect on the edge modes. To show this, we 
consider a site-dependent decay rate. This  non-homogeneous dissipation leaves $n=\lfloor\frac{L(1-\alpha)}{2}\rfloor$ sites untouched near each edge, 
as shown in Fig.~\ref{fig:sketch_alpha}. Defining $\xi$ as the characteristic localization length of the edge modes 
(see~\Cref{app:dee} for details), we study the evolution of the correlator $G_{1,L}(t)$ for different values of the ratio $\nu =n/\xi$. 

\begin{figure}[t]
    \centering
    \includegraphics[width=0.95\columnwidth]{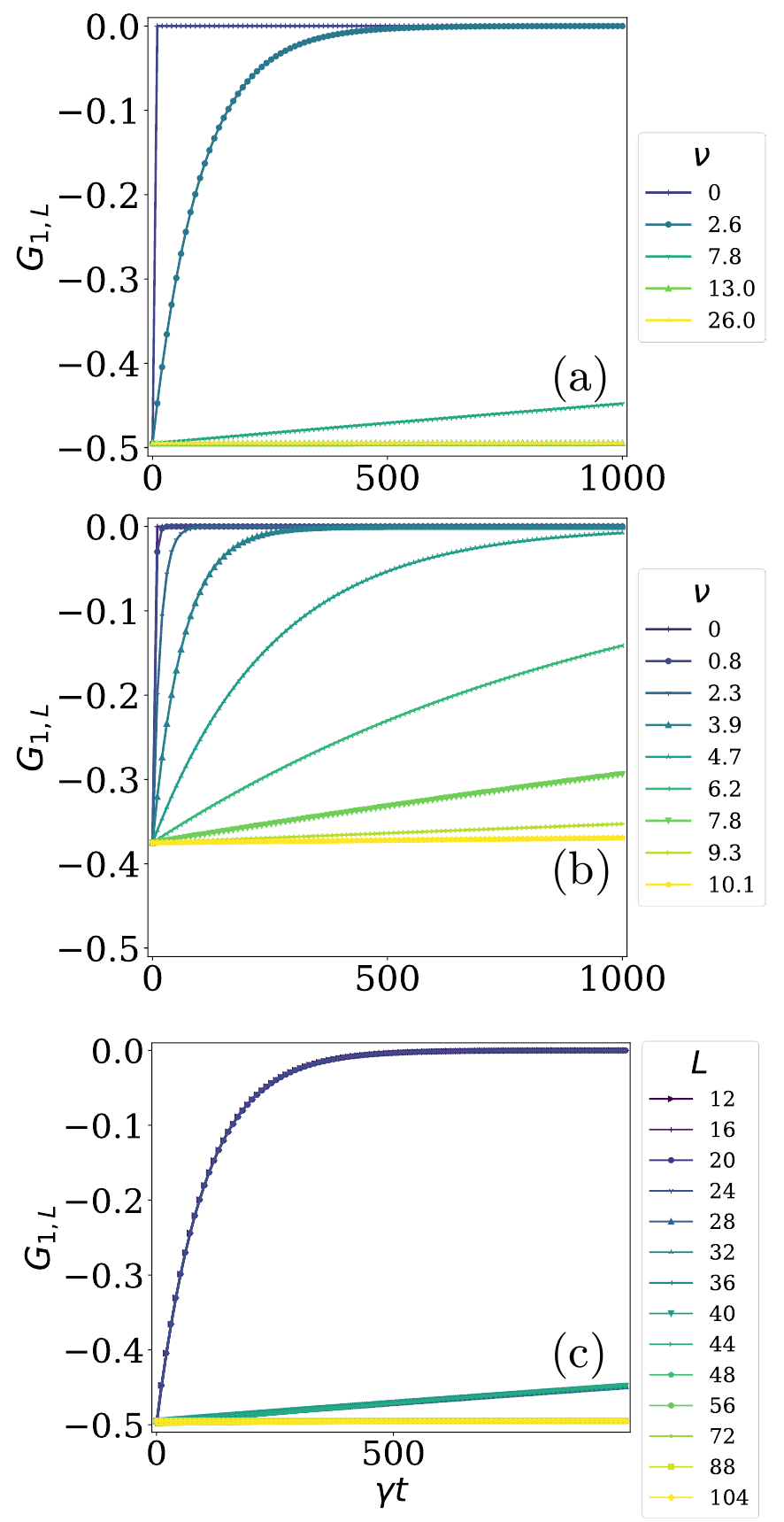}
    \caption{\justifying 
    Two-point correlator for the $\alpha$-SPD dynamics for $L=112$ and different values of $\alpha$ decreases. For decreasing $\alpha$ a growing fraction $\nu = \frac{(1-\alpha)L}{2\xi}$ of sites near each edge is left untouched. The system is initialized in the topological half-filled ground state of the Hamiltonian of the isolated SSH chain with parameters $\hopintra/\hopinter=0.1$ --- for which $\xi = 0.43$ --- in Panel (a) and $\hopintra/\hopinter=0.5$ --- for which $\xi = 1.44$ --- in Panel (b), with $N=56$ unit cells ($L=112$ sites). In Panel (c) $\alpha$ is fixed at $0.8$ and $L$ is varied. The system is initialized with $\hopintra/\hopinter=0.1$. }
    \label{fig:two_point_correlator}
\end{figure}

In Fig.~\ref{fig:two_point_correlator} we show the time evolution of $G_{1,L}$ for different values of $\alpha$ and  $\nu$ at fixed $\hopintra/\hopinter$ 
and  size $L$ of the chain.  The initial correlation between the two edge modes is lost with time. The time scale over which this happens depends 
strongly on $\nu$, the larger is the number of boundary sites not touched by dissipation, the slower the suppression of the correlation occurs. 

For $\alpha$ small enough so that $n\gg \xi$, the Lindblad operators act sufficiently far from the range where the edge modes at $t=0$ are  
localized, in this case the correlation is very weakly dependent on time and correlations are preserved [see Figs.~\ref{fig:two_point_correlator}(a,b)]. 
This behavior does not depend on the particular type of dissipation considered.

In this case, we can safely conclude that for values of $\alpha \approx 0.8$ the boundaries are unaffected by dissipation.

In Fig.~\ref{fig:two_point_correlator}(c), we present the time evolution of the two-point correlator for different system sizes, having fixed $\alpha = 0.8$. The plot shows that by selecting an appropriate value of $\alpha$, a sufficient number of sites remain unaffected by dissipation, allowing us to observe a decay in the two-point correlator over time, yet with significantly large decay times. For large enough system size, the correlator can be considered effectively constant within the timescales of interest. This result highlights a key distinction: while the introduction of a uniform dissipation drastically alters the correlation properties of the original topological system, carefully confining dissipation to the central region of the chain allows us to recover the same behavior observed in the unitary case, where correlations remain stable in time in the thermodynamic limit.

This preliminary analysis gives the flavor of the dynamics of the edge modes following a quench and evolving under the effect of dissipation. How does the 
dynamics of correlations between the edge modes reflect in the properties of the disconnected entanglement entropy? Being a nonlinear function of the 
state, it will contain additional information compared to those encoded in the density matrix.

\subsection{Time-evolution of the disconnected entanglement entropy}

In Ref.~\cite{Micallo_2020}, it has been shown that, in the case of a local, unitary ($\gamma =0$)  and symmetry-preserving quench, the time at which 
the DEE deviates from $2$ (the initial state being in the topological phase) scales linearly with the system size, proving that the DEE is a good non-local 
order parameter for isolated topological systems in the thermodynamic limit. 

\begin{figure}
\centering
\includegraphics[width=\columnwidth]{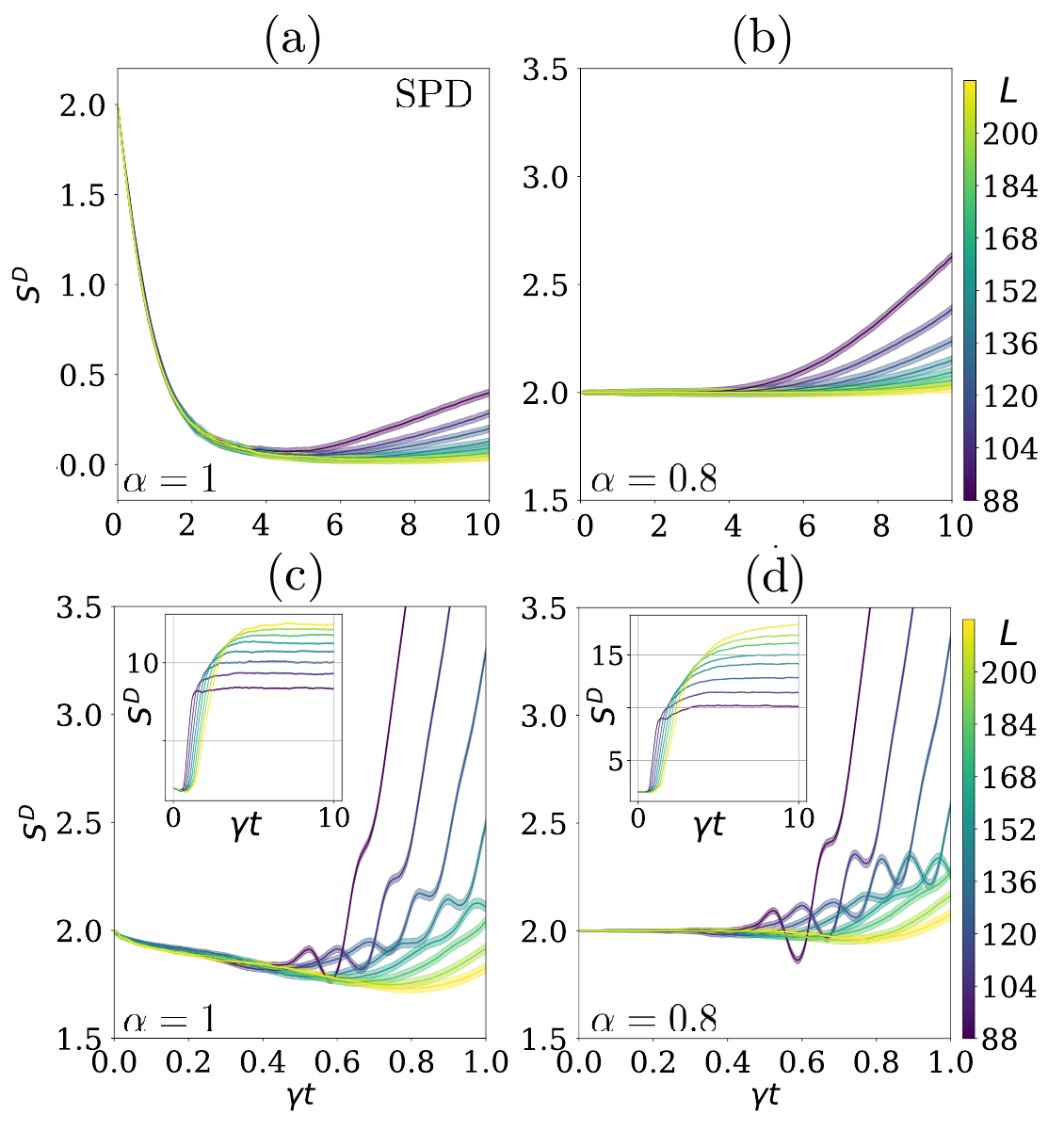} 
\caption{\justifying Dynamics of the DEE under the SPD dynamics with $\gamma=1$, $\hopinter/\gamma=20$ until $\gamma t =10.0$. (a) SPD with $\alpha=1$ and \rep{topological}{unquenched} Hamiltonian ($\hopintra/\hopinter=0.1$),  (b) SPD with $\alpha=0.8$ and \rep{topological}{unquenched} Hamiltonian ($\hopintra/\hopinter=0.1$). (c) SPD with $\alpha=1$ and \rep{trivial}{quenched-to-trivial} Hamiltonian ($\hopintra/\hopinter=1.5$) until $\gamma t = 1.0$. (d) SPD with $\alpha = 0.8$ and \rep{trivial}{quenched-to-trivial} Hamiltonian ($\hopintra/\hopinter=1.5$) until $\gamma t=1.0$. (c,d) Inset plots: whole dynamics until $\gamma t=10.0$. 
}
\label{fig:dee_spd}
\end{figure}

\begin{figure}
\centering
\includegraphics[width=\columnwidth]{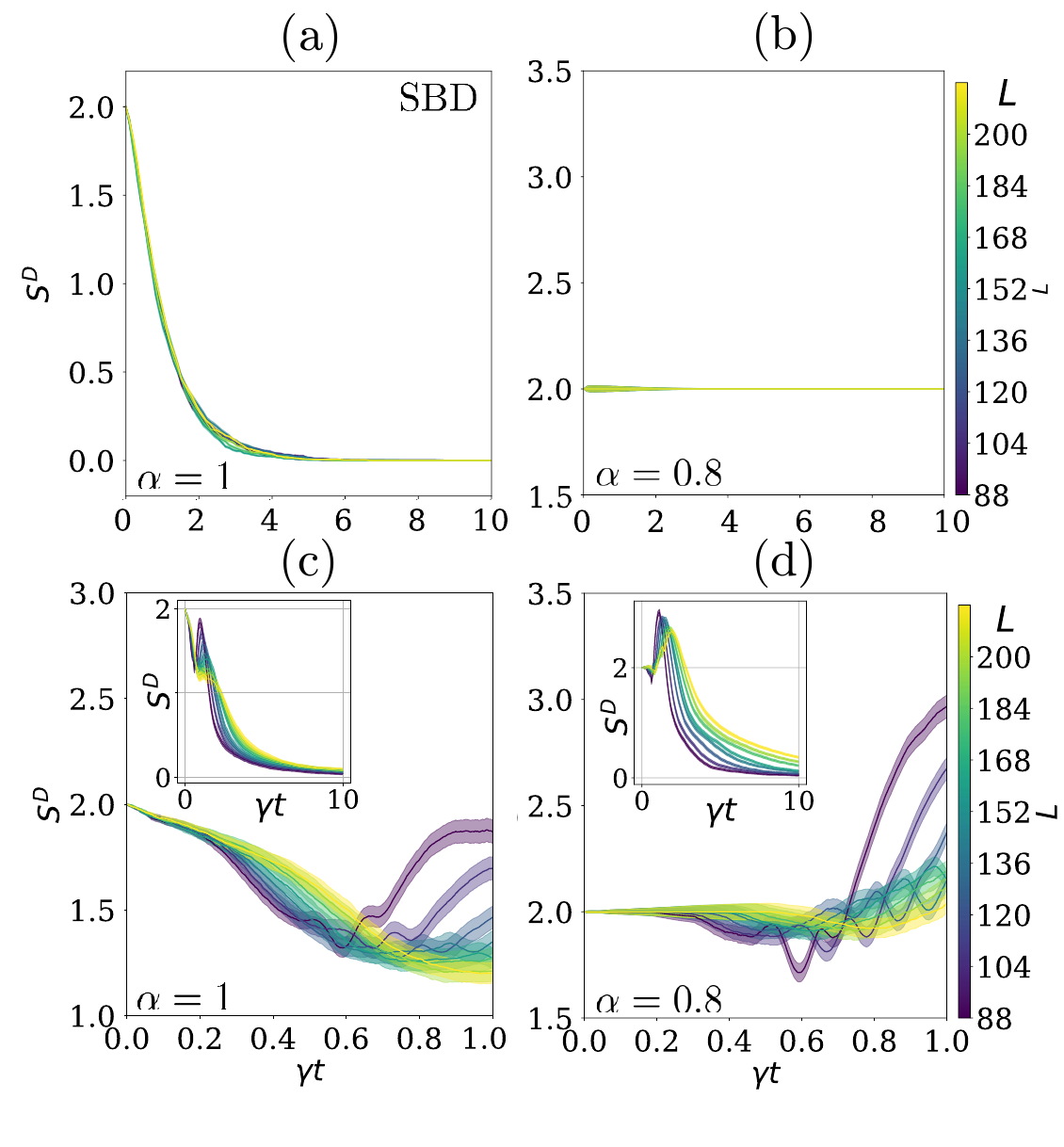} 
\caption{\justifying Dynamics of the DEE under the SBD dynamics with $\gamma=1$, $\hopinter/\gamma=20$ until $\gamma t =10.0$. (a) 
SBD with $\alpha=1$ and \rep{topological}{unquenched} Hamiltonian ($\hopintra/\hopinter=0.1$),  (b) SBD with $\alpha=0.8$ and \rep{topological}{unquenched} Hamiltonian 
($\hopintra/\hopinter=0.1$). (c) SBD with $\alpha=1$ and \rep{trivial}{quenched-to-trivial} Hamiltonian ($\hopintra/\hopinter=1.5$) until $\gamma t = 1.0$. (d) SBD 
with $\alpha = 0.8$ and \rep{trivial}{quenched-to-trivial} Hamiltonian ($\hopintra/\hopinter=1.5$) until $\gamma t=1.0$. (c,d) Inset plots: whole dynamics until $\gamma t=10.0$.}
\label{fig:dee_sbd}
\end{figure}

Our aim is to extend this study to the dissipative case. Adopting the techniques described in \Cref{app:free_fermions}, we calculate the time evolution 
of the reduced covariance matrix for the subsystems $\mathcal{A}$, $\mathcal{B}$, $\mathcal{A}\cup \mathcal{B}$, and $\mathcal{A}\cap \mathcal{B}$ for a 
single trajectory. From the diagonalization of the reduced covariance matrix, we calculate the different contributions needed to evaluate the DEE. We average 
the result over multiple trajectories ($N_{\mathrm{\scriptstyle traj}}=960$), calculating the error as the standard error.
In the following, in Fig.~\ref{fig:dee_spd}  and Fig.~\ref{fig:dee_sbd}, we show the time evolution of the DEE  for $\alpha=1$ (uniform) an  $\alpha=0.8$. 
In both cases, we consider growing system sizes. 

 \add{Our goal is to explore how the dynamics of the DEE changes when its evolution is governed not only by a Hamiltonian inducing a unitary quench, but by a full Lindbladian. We set the initial state of the dynamics as the ground state of the SSH chain in the topological phase, with an initial DEE value of $2$. With this initial state we consider the dynamics under different Lindbladians. We take SPD and SBD Lindbladians, considering both the cases with bulk and boundary dissipation. We take also different Hamiltonian parts of the Lindbladian. We consider two cases. In one case we evolve with the same topological Hamiltonian as the one that has the initial state as ground state. We call this case ``unquenched topological'', because the initial state is ground state of the evolution Hamiltonian and there is no quench in the sense of~\cite{Polkovnikov_RMP11}. In the other case we evolve with a trivial Hamiltonian, so the initial state is not the ground state of the evolution Hamiltonian and we are therefore performing a quench. We call this case ``quenched-to-trivial''. Applying a quench, the system gets excited and entangled (see for instance~\cite{Caleb_SciPostPhysLectNotes20}), and this effect can counteract the measurements of the environment that occur through the quantum jumps and tend to destroy entanglement.}

Figs.~\ref{fig:dee_spd} and~\ref{fig:dee_sbd} illustrate the impact of dissipation on the stability of the entangled topological edge modes for $\alpha=1$ [panels (a, c)] and $\alpha=0.8$ [panels (b, d)]. Fig.~\ref{fig:dee_spd} shows the results for the first type of dynamics (SPD), while Fig.~\ref{fig:dee_sbd} presents the same analysis for the second type (SBD). Among all pairs of panels with the same letter in the two figures, we observe clear similarities, leading us to conclude that there are no substantial differences between the two types of dissipation in terms of their impact on the properties of the average DEE.

In particular, in panels (a) of both figures, where the Hamiltonian governing the dynamics is topological, \add{there is no quench}, and $\alpha = 1$, meaning that the dissipation is global, the DEE initially starts at $2$ but quickly deviates from this value as time progresses. This variation is a direct consequence of dissipation --- whether strictly local (SPD) or localized between two nearest-neighbor sites (SBD) --- which tends to destroy entanglement both at short range (within the bulk) and, more significantly, at long range (between the two topological modes). Since the DEE detects only the latter, the observed variations in the plots are entirely due to the destruction of long-range topological entanglement. This behavior can be intuitively understood in the fully-dimerized limit of the chain and will be further analyzed in subsequent sections and in~\Cref{app:effect_of_jumps}.

In these cases, since the initial state is the half-filled ground state of the topological Hamiltonian, and the coherent part of the Lindbladian \add{is the very same} topological Hamiltonian, \add{no quench contribution is present to excite and entangle the system counteracting the measurement effects of the environment manifested through quantum jumps.} As a result, quantum jumps dominate the dynamics, causing the DEE to steadily decrease to zero. However, while in the SBD case --- where dissipation eventually depletes the entire chain --- this value remains asymptotic, in the SPD case, the curves exhibit a recovery over a timescale that increases with system size. Since our focus is not on asymptotic entanglement properties, we do not investigate this scaling further.

Conversely, in panels (b), where $\alpha = 0.8$, effectively isolating the edges from dissipation, the behavior of the DEE changes drastically. Here, the DEE remains at $2$, either persistently in the SBD case or for a time that scales linearly with system size in the SPD case, ensuring stability in the thermodynamic limit. The key difference between the $\alpha = 1$ and $\alpha = 0.8$ cases is that, in the latter, the edge modes remain unaffected by dissipation. Given that dissipation is spatially localized in both cases, this is sufficient to preserve the topological entanglement associated with the edge modes.

\add{Panels (c) introduce a crucial difference: here, the coherent part of the Lindbladian is a trivial Hamiltonian. The initial state is not the ground state of this Hamiltonian, so there is a quench that leads to a coherent contribution to the generation of entanglement that competes with the dissipative effects dominating panels (a) and (b). This contribution plays a central role in enhancing the robustness of the DEE: The coherent quench dynamics counteracts the entanglement-degrading effects of quantum jumps, leading to a more sustained entanglement signal. As a result, the DEE exhibits a behavior reminiscent of a plateau during the early stages of evolution.}

\add{Strictly speaking, we use the term \textit{plateau} in analogy with the unitary case, where it denotes an initial segment with zero slope. In our dissipative setting, the plateau is tilted — meaning the DEE still decays, but more slowly — and reflects a transient regime where the coherent evolution helps delay the onset of decoherence-induced decay. This dynamical competition illustrates that, even though the Hamiltonian part of the Lindbladian is trivial, this does not imply that the system is dynamically in a trivial phase. On the contrary, the DEE remains more robust precisely because of the quench, confirming that the nature of the Hamiltonian does not directly determine the dynamical behavior of the DEE under Lindbladian evolution, in analogy with what happens in the unitary case~\cite{Micallo_2020}.}

\add{In summary, panels (c) highlight a key physical insight: the introduction of a quench through the coherent Hamiltonian term — even if it is \emph{trivial} in the ground-state sense — can enhance entanglement stability through purely dynamical mechanisms. This is reflected in the delayed decay of the DEE and establishes a meaningful generalization of the behavior observed in the unitary case.}

Finally, panels (d) reveal another intriguing result: even when \add{there is a quench} and the coherent part of the Lindbladian corresponds to a trivial Hamiltonian, the DEE remains robust as long as the edges of the chain are \add{not affected by dissipation}. This confirms that, since long-range entanglement is encoded in the occupation number basis of the fermionic edge modes, \add{a dissipation acting on the boundary sites} inevitably destroys it. The only way to preserve persistence of the topological edge modes is to shield them from dissipation. Once this protection is in place, the DEE remains topological for a time increasing with the system size, regardless of the coherent part of the Lindbladian dynamics, mirroring the unitary case where stability under symmetry-preserving quenches was demonstrated~\cite{Micallo_2020}.

On the other hand, as discussed for panels (c), when dissipation affects the edges, the coherent part of the Lindbladian becomes crucial in determining the DEE's robustness. Within a single trajectory, the dynamic contributions from quantum jumps and those from the non-Hermitian Hamiltonian component compete with each other~\cite{gal2024entanglement}. As a result, introducing a quench to a trivial Hamiltonian enhances stability of the topological value of the DEE, as the destructive effect of quantum jumps is mitigated \add{by the coherent generation of excitations and entanglement} in the initial stages of the evolution.

\begin{figure}[t]
\centering
\includegraphics[width=\columnwidth]{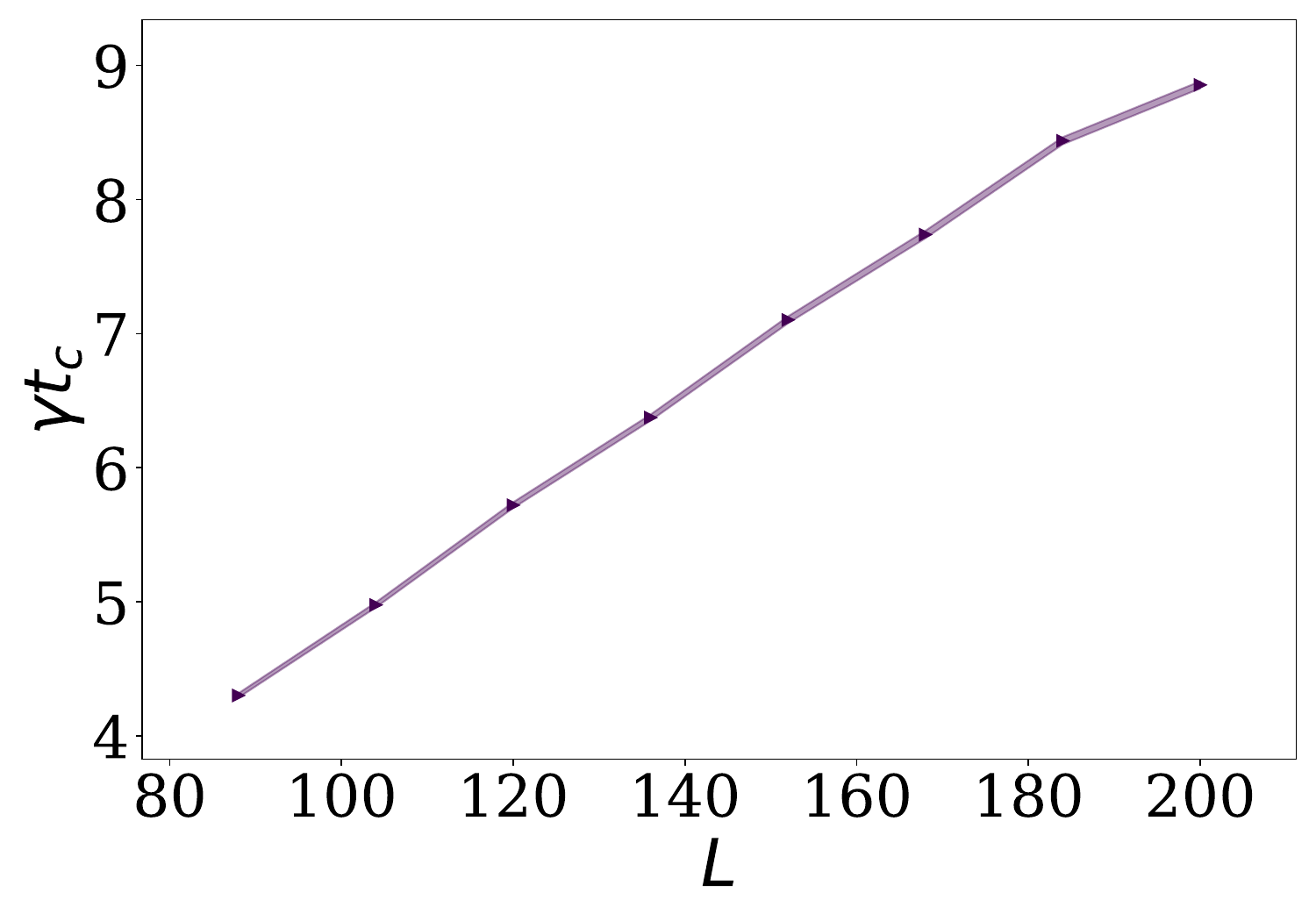} 
\caption{\justifying Linear scaling of $\gamma t_c$ as function of $L$ corresponding to the curves of Fig.~\ref{fig:dee_spd}(b). Non-homogeneous SPD dissipation and  \rep{topological}{unquenched} Hamiltonian driving the evolution with $\hopintra/\hopinter=0.1$, $\alpha=0.8$, $\gamma=1$, and $\hopinter/\gamma=20$.
}
\label{fig:time_scaling_center_no_quench}
\end{figure}

 As already mentioned, in the time traces of Fig.~\ref{fig:dee_spd} and~\ref{fig:dee_sbd} the DEE departs from the initial (topological) value at a 
 given  time scale. In the numerics, we estimate it in the following way: for each trajectory, we evaluate $\gamma t_c^{(\mathrm{\scriptstyle traj})}$ 
 as the first time at which the threshold condition $\Big|S^{(D,{\mathrm{\scriptstyle traj}})}(t)|-S^{(D,{\mathrm{\scriptstyle traj}})}(0)\Big|<(2\log_2 2)/100$ 
 is met. We then average it over trajectories and get the characteristic time $\gamma t_c = \overline{\gamma t_c^{(\mathrm{\scriptstyle traj})}}$. We have checked that different 
 threshold levels provide qualitatively similar results.   Fig.~\ref{fig:time_scaling_center_no_quench} shows $\gamma t_c$ versus $L$ for the case of 
  $\alpha =0.8$ SPD dissipation and \rep{topological Hamiltonian driving the evolution}{unquenched} [see Fig.~\ref{fig:dee_spd}(b)]. We see that $t_c$ linearly 
 increases with $L$ and this shows that when the quantum jumps do not involve the edges of the chain, the DEE remains quantized in the 
 thermodynamic limit.

\begin{figure}
\centering   
\includegraphics[width=\columnwidth]{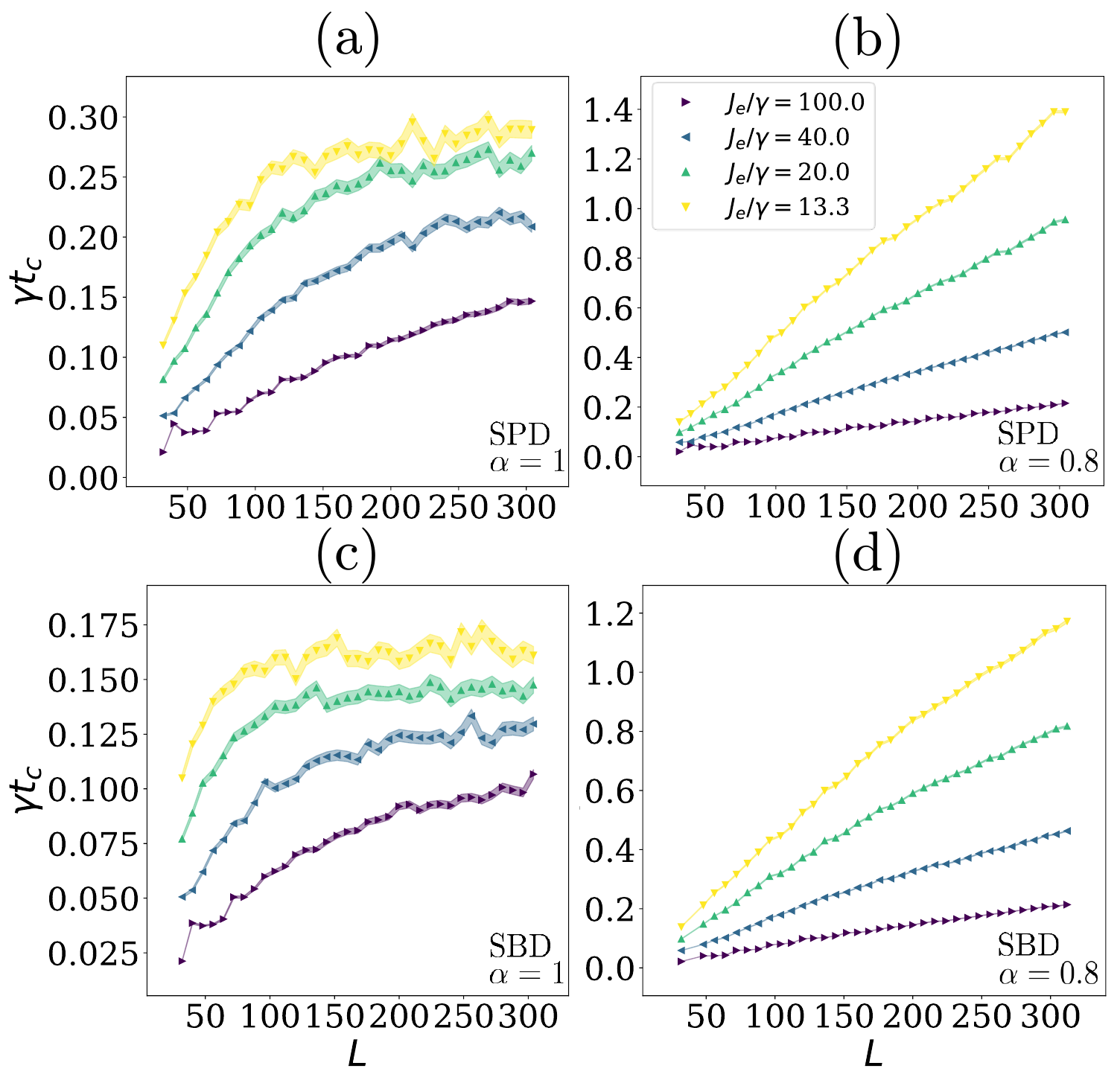}
\caption{\justifying $\gamma t_c$ versus $L$ with various choices of $\hopinter/\gamma$. (a) SPD dynamics with $\alpha=1$. (b) SPD 
dynamics with $\alpha=0.8$. (c) SBD dynamics with $\alpha=1$. (d) SBD dynamics with $\alpha=0.8$. \rep{Trivial}{Quenched-to-trivial} case 
($\hopintra/\hopinter=1.5$), $\gamma=1$, and $\hopinter/\gamma=20$.} 
\label{fig:tc_scaling}
\end{figure}

Fig.~\ref{fig:tc_scaling} shows how $\gamma t_c$ scales with the system size for the  \add{dynamics with different dissipation profiles} and a \rep{trivial}{quenched to the trivial} Hamiltonian, namely corresponding to the curves of Figs.~\ref{fig:dee_spd}(c),~\ref{fig:dee_spd}(d),~\ref{fig:dee_sbd}(c) and ~\ref{fig:dee_sbd}(d). 
The analysis of $\gamma t_c$, shown in Fig~\ref{fig:tc_scaling}, reveals different scaling behaviors depending on the dissipation profile. When $\alpha =0.8$ [Fig.~\ref{fig:tc_scaling}(b,d)], the dissipation is confined to the bulk and $\gamma t_c$ increases linearly with system size, indicating that the entangled edge states remain stable for increasingly long times. This scaling matches what is observed in the unitary case, where topological edge modes persist for a time that increases \add{linearly with the system size when a quench is applied.}

\add{At variance with that}, when $\alpha=1$ [Fig.~\ref{fig:tc_scaling}(a,c)], the dissipation acts on the boundaries and $\gamma t_c$ saturates to a finite value, signaling that the entanglement between the edge modes is eventually lost after a finite time, regardless of system size. This marks a crucial difference from the unitary scenario, where the DEE deviation time scales linearly with size. Here, instead, the presence of dissipation at the edges imposes a strict limit on the survival of the topological correlations, leading to their complete suppression within a finite timescale. However, again, imposing $\alpha<1$ as in Fig.~\ref{fig:tc_scaling}(b,d), provides results similar to those of the unitary case.

In summary, when examining the time evolution of the DEE starting from a topological state, the action of the dissipation on the boundary appears to be the most relevant aspect to consider. On the contrary, belonging to one class rather than another within the 
tenfold way framework appears to have no significant relevance in the study of the time evolution of the DEE. In the next section we 
inquire more deeply this fact by looking at the changes in DEE due to the effect of quantum jumps.

\subsection{Time-resolved statistics of $\Delta S^D$}\label{Del:sec}
\begin{figure}
    \centering
    \includegraphics[width=\columnwidth]{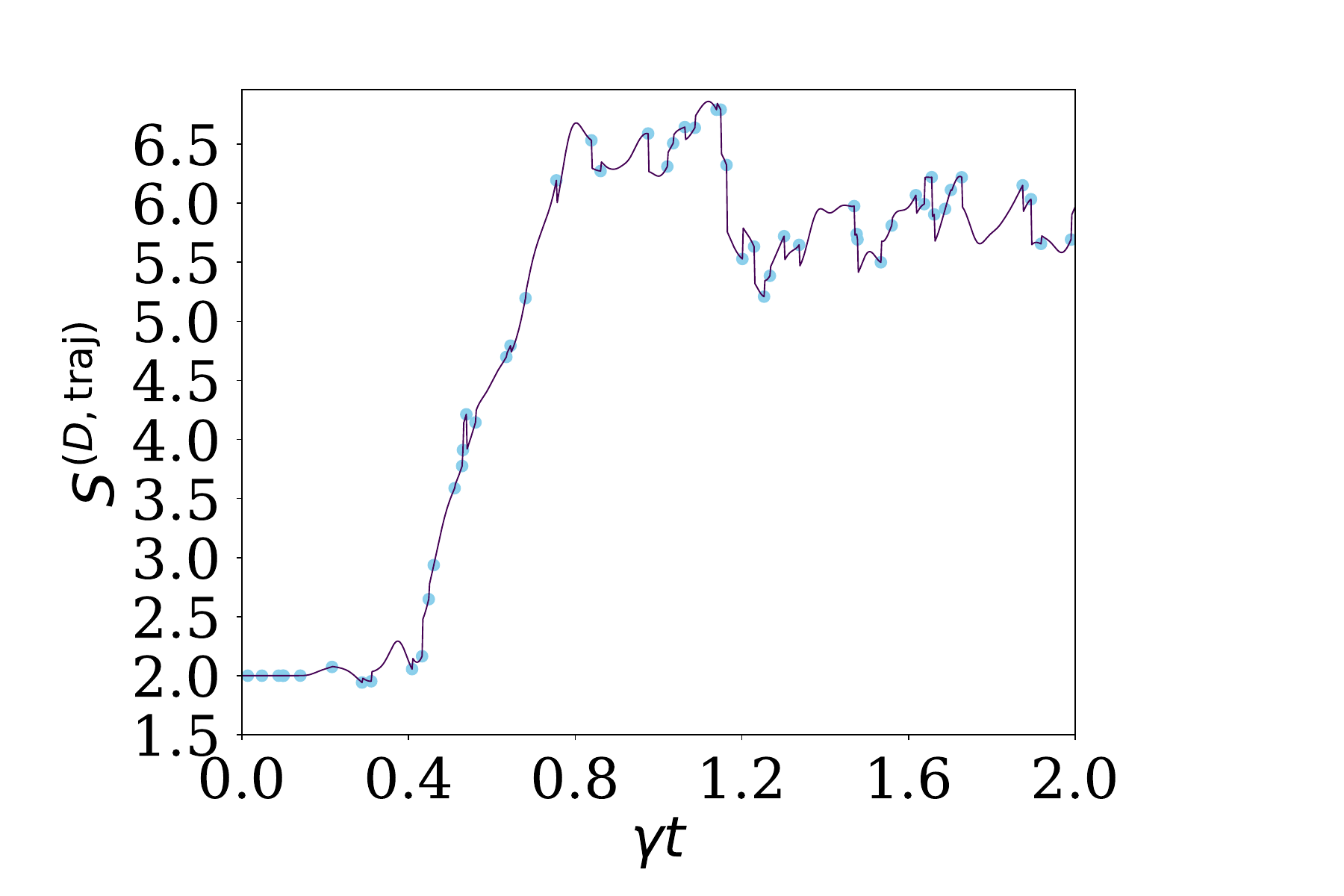}
    \caption{\justifying Time evolution of the DEE over a single trajectory, namely $S^{(D,\mathrm{\scriptstyle traj})}$. The blue dots correspond to 
    the occurrence of a quantum jump on any site apart form the two edges, where the occurrence of the jump is signaled by a blue dot.}
    \label{fig:single_traj}
\end{figure}

To gain a deeper understanding of how the DEE evolves along individual quantum trajectories, we analyze the statistical distribution of its variations  due to quantum jumps. The typical pattern consists of an evolution driven by the non-Hermitian Hamiltonian, abruptly interrupted by a discontinuous change in entanglement entropy at the occurrence of a jump. These changes, which can either increase ($\Delta S^D>0$) or decrease ($\Delta S^D<0$) the DEE, are illustrated in Fig.~\ref{fig:single_traj}~\cite{gal2024entanglement}.
To systematically characterize this behavior, we sample the changes in entanglement entropy at each jump event along a trajectory and repeat this process over multiple realizations to construct the histogram $P(\Delta S^D)$. The aim is to identify whether the variations follow a structured pattern and to determine whether specific jump events are responsible for significant entanglement loss. 

\begin{figure}
\centering
\includegraphics[width=0.95\columnwidth]{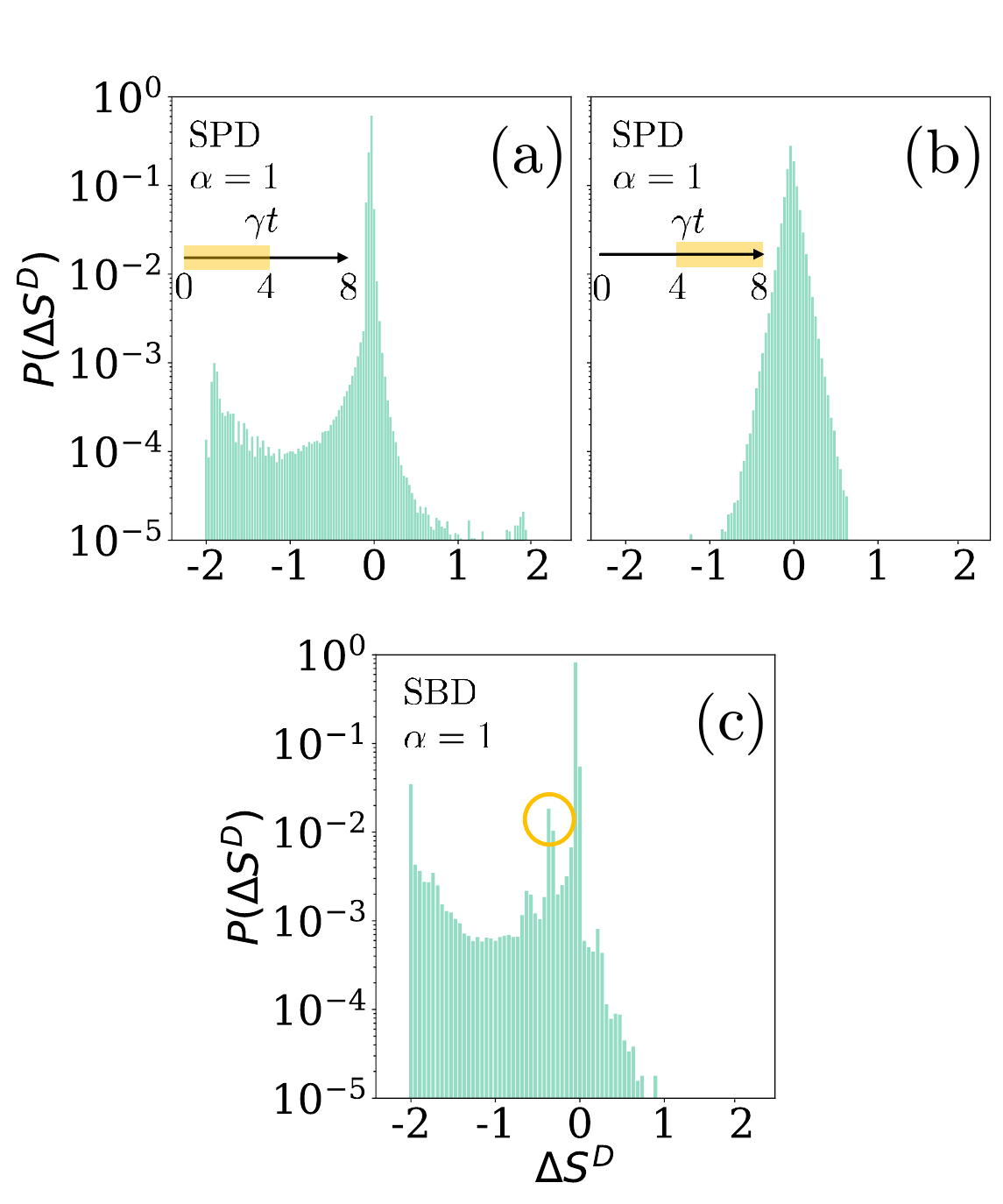} 
\caption{\justifying $P(\Delta S^D)$ for the two different time windows of the two homogeneous dynamics. (a) SPD with $\alpha=1$, $t_0=0.0$, $t_f=4.0$ (yellow highlighted time interval). (b) SPD with $\alpha=1$, $t_0=4.0$, $t_f=8.0$ (yellow highlighted time interval). (c) SBD with $\alpha=1$, $t_0=0.0$, $t_f=4.0$. We do not show the statistical analysis related to the second time window for the global SBD dynamics since the chain is emptied by dissipation and no more jumps occur. The initial state is the ground state of $\hat{H}_{\mathrm{\scriptscriptstyle SSH}}$ with $L=56$ sites, $\hopintra/\hopinter=0.1$ and the parameters of the dissipative evolution are $\gamma=1$, $\hopintra/\hopinter=0.1$, $\hopinter/\gamma=20$ and $N_{\mathrm{\scriptstyle traj}}=28800$ trajectories. The yellow circle signals the peak at $\Delta S^D=\Delta S^D=2-2(-\frac{3}{4}\log_2(3)+2)\approx-0.38$. 
}
\label{fig:time_stat_global}
\end{figure}

Figure~\ref{fig:time_stat_global} presents the statistical analysis for the two homogeneous dissipative dynamics, namely with $\alpha =1$. We consider a system of $L=56$ sites with the \rep{topological}{unquenched} Hamiltonian as the coherent contribution. The average dynamics generated by Eq.~\eqref{eq:stocastic_schrod} under these conditions allows us to distinguish two different time windows --- an initial transient regime, and a later stage where the DEE approaches its stationary value. We perform a statistical analysis of these windows separately. More precisely, in the SBD case, we show only the statistical analysis related to the first time window, because later the chain is emptied 
by dissipation and no more jumps occur.

As shown in Fig.~\ref{fig:time_stat_global}, the distribution of $\Delta S^D$ reveals distinct signatures depending on the type of dissipation. In the SPD case [Fig.~\ref{fig:time_stat_global}(a)], the early stage shows a bimodal distribution with a pronounced peak at $\Delta S^D=-2$, signaling the destruction of topological entanglement. In the later stage [Fig.~\ref{fig:time_stat_global}(b)], this peak disappears and the distribution becomes unimodal, indicating that the destruction of entanglement has already occurred. In contrast, the SBD case [Fig.~\ref{fig:time_stat_global}(c)] exhibits a similar peak at $\Delta S^D=-2$, but an additional peak at $\Delta S^D=\Delta S^D=2-2(-\frac{3}{4}\log_2(3)+2)\approx-0.38$ (signaled by a yellow circle) emerges, which can be attributed to the extended nature of the jump operators in this dissipation model. We will further analyze it in \Cref{app:effect_of_jumps}.

\begin{figure}
\centering
\includegraphics[width=\columnwidth]{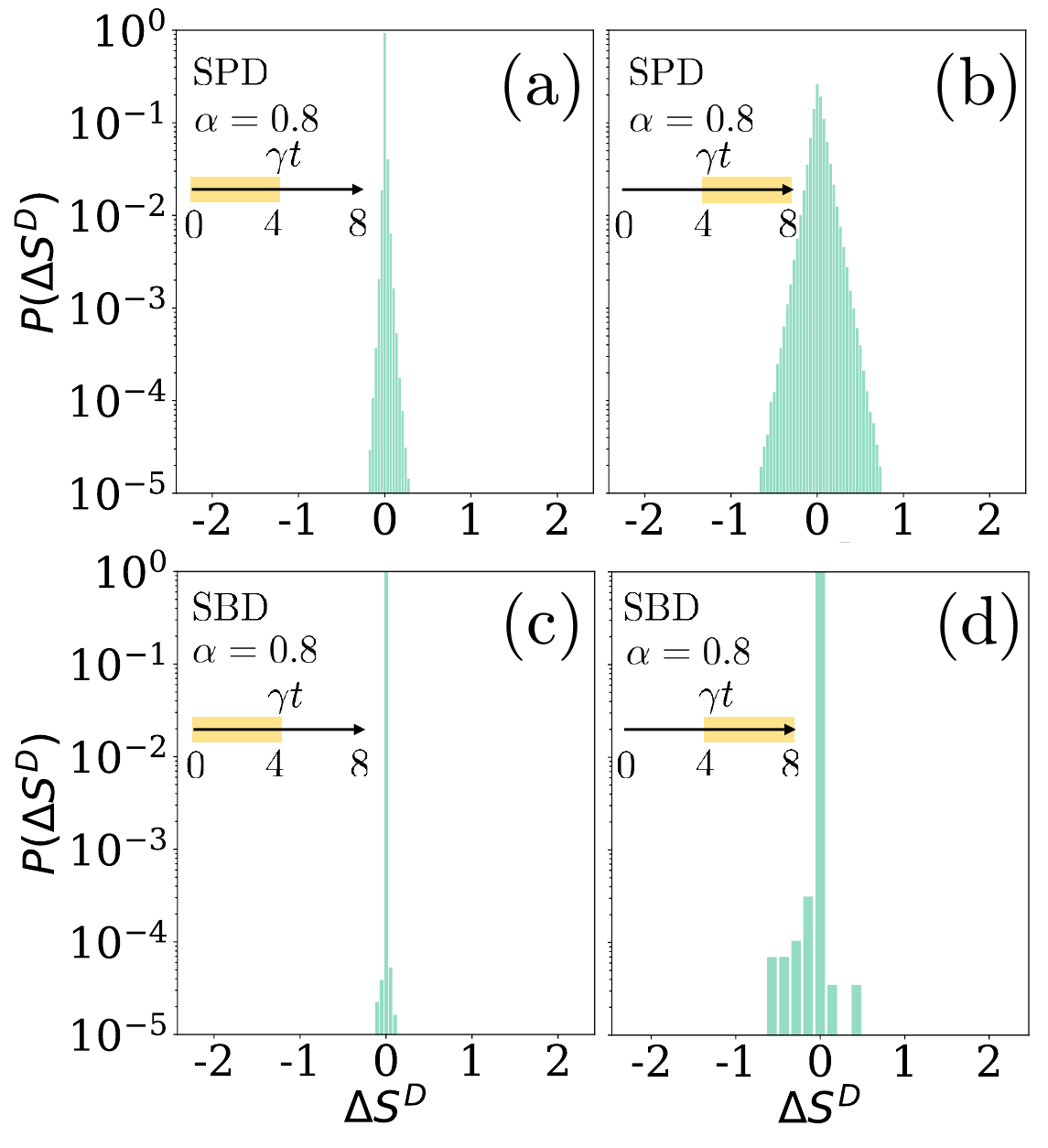} 
\caption{\justifying $P(\Delta S^D)$ for the two different time windows of the SPD dynamics with $\alpha =0.8$. (a) SPD, $t_0=0.0$, $t_f=4.0$. (b) SPD, $t_0=4.0$, $t_f=8.0$. (c) SBD, $t_0=0.0$, $t_f=4.0$. (d) SBD, $t_0=4.0$, $t_f=8.0$.
}
\label{fig:time_stat_center}
\end{figure}

To further investigate the role of spatially localized dissipation, we analyze the SPD case that does not affect the edges, i.e., with $\alpha = 0.8$, shown in Fig.~\ref{fig:time_stat_center}.
In this case, where dissipation is confined to the central region of the chain, the distributions remain unimodal at all times. The absence of the $\Delta S^D=-2$ peak confirms that the edge states are unaffected by the quantum jumps. This result strongly suggests that the entanglement between boundary modes is disrupted only when the dissipation acts directly on the edges.

A key observation is that the pronounced peaks observed in the homogeneous dissipation cases indicate that the effect of quantum jumps on the DEE manifests in a discrete manner for the SPD dynamics. This suggests that when a quantum jump is lozalized on one of the two edge sites, the entanglement entropy undergoes an abrupt and quantized reduction, particularly in the early transient regime. To further confirm this interpretation, we perform a site-resolved analysis of $\Delta S^D$ to explicitly verify whether jumps occurring at the boundary sites are responsible for these discrete entanglement changes.

\subsection{Site-resolved statistics of $\Delta S^D$}
The analysis of the time-resolved statistics of $\Delta S^D$ in the previous section revealed the presence of distinct peaks, particularly at $\Delta S^D=-2$ for the homogeneous ($\alpha=1$) SPD dynamics, suggesting that specific quantum jumps play a dominant role in destroying topological entanglement. To verify whether these critical jumps occur at specific sites in the chain, we now investigate the site-resolved distribution of $\Delta S^D$.

Since the characteristic peaks in $P(\Delta S^D)$ appear primarily in the first time window of the SPD and SBD homogeneous dissipations ($\alpha = 1$), we focus on this interval. For each trajectory, we record the site index $j$ where the jump leading to a $\Delta S^D$ occurs and analyze the conditional probability distribution $P_j(\Delta S^D)$ for different sites. 
\begin{figure}
\centering
\includegraphics[width=\columnwidth]{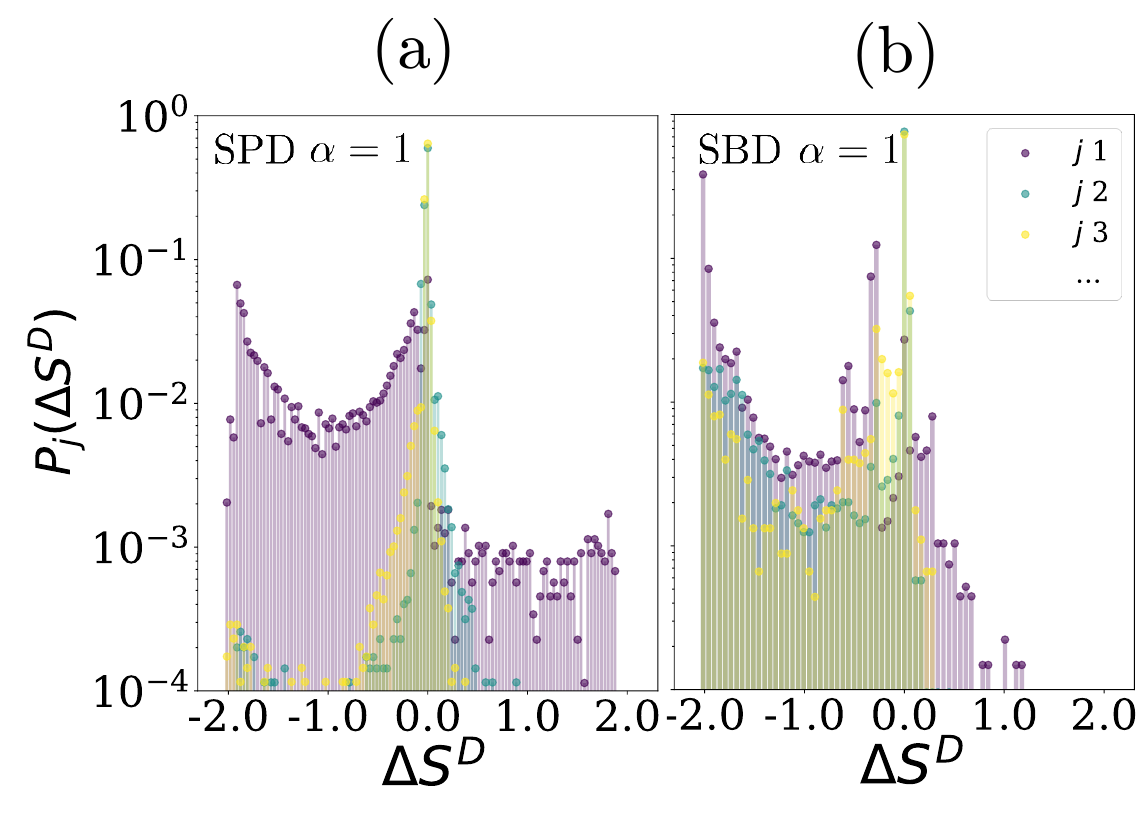}
\caption{\justifying   $P(\Delta S^D)$ conditioned on the site where the jump has occurred. (a) SPD with $\alpha=1$. (b) SBD with $\alpha=1$. 
We focus the statistical analysis in the first time window going from $\gamma t_0 = 0.0$ to $\gamma t_f = 4.0$. 
The statistical analysis is done considering a sample of $N_{\mathrm{\scriptstyle traj}}=28800$ trajectories.}
\label{fig:index_stat}
\end{figure}
Figure~\ref{fig:index_stat}(a) shows the site-resolved histogram for the SPD dynamics with $\alpha = 1$. We observe that the peak at $\Delta S^D$ is exclusively associated with jumps occurring at the boundary sites ($j=1$). This result strongly supports our previous hypothesis --- the destruction of topological entanglement occurs due to quantum jumps acting directly on the edge modes, where their Bell-pair entanglement is lost. Furthermore, the data suggest that the first jump occurring at the boundary site in each trajectory contributes predominantly to this peak, as further discussed in \Cref{app:effect_of_jumps}.

In contrast, for the SBD case with $\alpha=1$ shown in Fig.~\ref{fig:index_stat}(b), while a peak at $\Delta S^D=-2$ is still observed for boundary jumps, an additional peak emerges at $\Delta S^D\approx-0.38$. This secondary peak originates from the non-local nature of the SBD jump operators, which affect entanglement more gradually rather than in a single discrete step. As demonstrated in \Cref{app:effect_of_jumps}, the first jump does not immediately destroy the boundary entanglement but instead leads to a partial entanglement reduction, reflected in the peak at $\Delta S^D\approx-0.38$. The later jumps at the boundaries eventually lead to the complete destruction of the edge modes, giving rise to the $\Delta S^D=-2$ peak.

Overall, this site-resolved analysis confirms that, in the SPD case, the loss of topological entanglement occurs in a discrete manner --- a single jump at the boundary is sufficient to destroy the edge modes, producing a sharp and quantized change in the DEE. This result highlights a key difference from the SBD case, where the effect of quantum jumps is more gradual due to the extended nature of the dissipation process.

\section{\label{sec:conclusion}Discussion and Conclusions}

In this manuscript, we explored the topological properties of the monitored Su-Schrieffer-Heeger model, our focus being on how monitored dynamics 
can provide insights into topology in open quantum systems. By employing the quantum jump approach, we  accessed to nonlinear functions of 
the quantum state, revealing aspects of topological behavior that may be hidden in a density matrix formulation.

The central quantity we analyzed is the Disconnected Entanglement Entropy, averaged over  trajectories, and studied the  evolution of this average under 
different types of couplings to the environment (i.e. different types of quantum jumps). Furthermore, we also examined the discrete changes in the DEE 
induced by individual quantum jumps (in particular,  we also considered time-resolved statistics, distinguishing between the early and late stages of the 
evolution).

Summarizing:  
\begin{itemize}
\item By analyzing the evolution of the average DEE, we showed that the effects of dissipation are primarily governed by its spatial arrangement 
rather than its symmetry classification within the tenfold way framework.  When dissipation extends homogeneously throughout the chain, the coherent contribution to the dynamics becomes crucial in determining the robustness of the DEE, that anyway deviates from its topological value after a finite time. Conversely, when dissipation is restricted to the central region of the chain, the entangled 
edge states remain stable for a time that scales linearly with system size, regardless of the coherent dynamical contribution. This finding aligns with previous works on the unitary dynamics of topological 
systems and demonstrates that edge states can be protected from decoherence as long as boundary dissipation is avoided.
\item Through the statistical analysis of quantum jumps, we revealed that boundary jumps play a dominant role in the destruction of topological order.
Sometimes --- when they are local --- they lead to discrete changes in the DEE associated with the sudden loss of the topological long-range entangled Bell pair.
 \end{itemize}

Furthermore, our results reinforce the conclusions of previous works~\cite{cooper}, suggesting that the tenfold way classification for quadratic Lindbladians 
is insufficient for predicting the dynamical properties of open systems. While this classification provides insight into the single-particle spectrum of the 
Lindbladian, it does not capture the evolution of entanglement properties or the stability of edge modes under dissipation. A similar result has been observed in the unitary case, where \rep{topological}{unquenched} and \rep{non topological}{quenched-to-trivial} Hamiltonians provide a similar evolution for the DEE~\cite{Micallo_2020}.

\begin{acknowledgments}
We would like to thank Dario Bercioux, Marcello Dalmonte, Sebastian Diehl, Gabriele Campagnano and Vittorio Vitale for very helpful discussions. 
G.\,P. acknowledges computational resources from the CINECA award under the ISCRA initiative, and from MUR, PON Ricerca e Innovazione 
2014-2020, under Grant No.~PIR01\_00011 - (I.Bi.S.Co.). This work was supported by PNRR MUR project~PE0000023 - NQSTI, by the European 
Union's Horizon 2020 research and innovation programme under Grant Agreement No~101017733, by the MUR project~CN\_00000013-ICSC (P.\,L.),
by the MUR project PRIN 2022H77XB7 (G.\,E.\,S.), 
and by the  QuantERA II Programme STAQS project that has received funding from the European Union's Horizon 2020 research and innovation 
programme under Grant Agreement No~101017733 (P.\,L. and G.\,E.\,S.). This work is co-funded by the European Union (ERC, RAVE,~101053159) (R.\,F.). Views 
and opinions expressed are however those of the authors only and do not necessarily reflect those of the European Union or the European Research 
Council. Neither the European  Union nor the granting authority can be held responsible for them.
\end{acknowledgments}

\appendix

\section{\label{app:dee} Details on the fully dimerized limit of the SSH chain and DEE}

\subsection{Edge modes}\label{app:modes}
\begin{figure}
    \centering
    \includegraphics[width=0.9\linewidth]{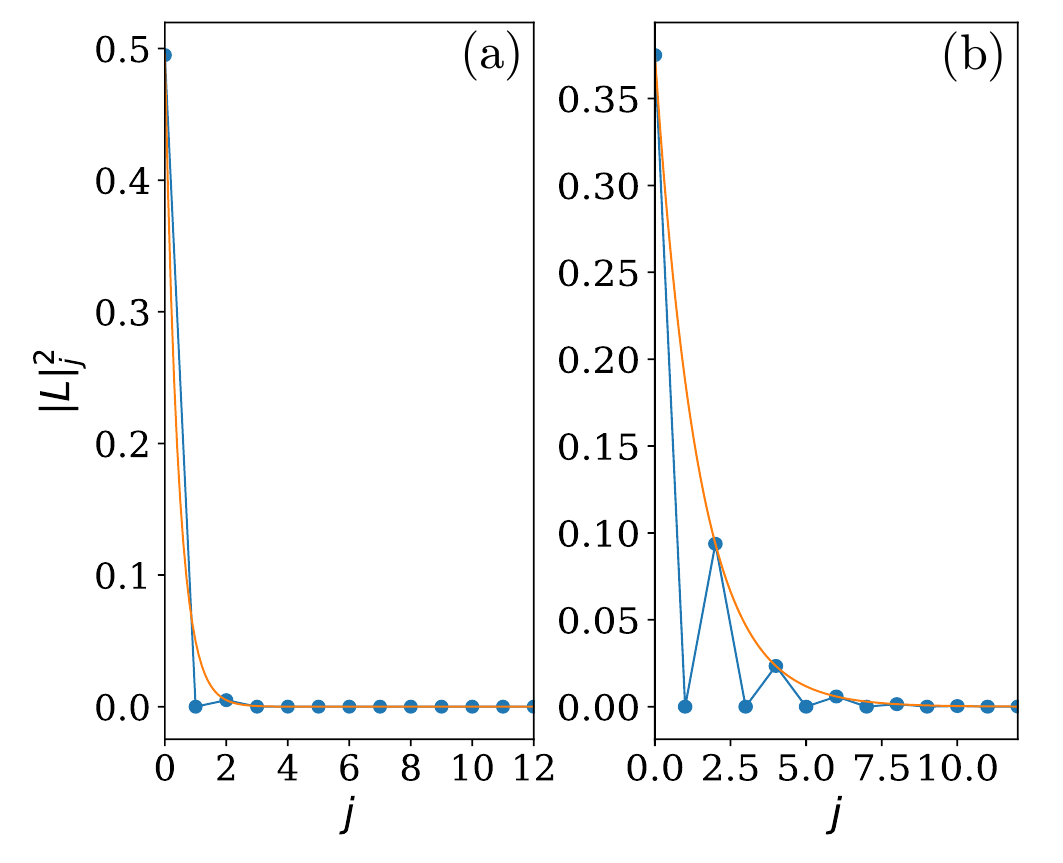}
    \caption{\justifying Edge modes distribution for a \rep{topological}{unquenched} Hamiltonian with (a) $\hopintra/\hopinter $= 0.1 (b) $\hopintra/\hopinter$ = 0.5. Blue curve: probability amplitude. Orange curve: exponential fit $|L|^2_j = e^{-j/\xi}$.}
    \label{fig:edge_mode_distribution}
\end{figure}

Considering the Hamiltonian of Eq.~\eqref{eq:ssh_hamiltonian} with space-dependent couplings~\cite{Asb_th_2016},

it is straightforward to see that, in the thermodynamic limit, two zero-energy eigenmodes of the Hamiltonian can be found in the form 
\begin{subequations}
    \begin{equation}
    \ket{L}=\sum_{i=1}^N a_i{\hat{c}^{\dagger}}_{i,\scriptscriptstyle{\mathrm A}}\ket{0}=\sum_{i=1}^N a_i\ket{i,A},
    \end{equation}
    \begin{equation}
    \ket{R}=\sum_{i=1}^N b_i{\hat{c}^{\dagger}}_{i,\scriptscriptstyle{\mathrm B}}\ket{0}=\sum_{i=1}^N b_i\ket{i,B},
    \end{equation}
    \label{eq:edgemodes}
\end{subequations}

which are states exponentially localized at either the first site $A$ or the last site $B$ of the chain. In \eqref{eq:edgemodes}, the coefficients are given by
\begin{subequations}
    \begin{equation}
        a_i=-\prod_{j=1}^{i-1}\frac{{\hopintra}_j}{{\hopinter}_j}a_1 \hspace{5mm}i=2,...,N
    \end{equation}
    \begin{equation}
    b_i=b_N\frac{-{\hopintra}_N}{{\hopinter}_i}\prod_{j=i+1}^{N-1}\frac{-{\hopintra}_j}{{\hopinter}_j} \hspace{5mm}i=1,...,N-1
    \end{equation}
    \begin{equation}
    b_1=a_N=0.
    \label{eq:impossible_condition}
    \end{equation}
\end{subequations}
The condition Eq. \eqref{eq:impossible_condition} is instead incompatible with the existence of zero-energy modes and one must consider the small lift $\Delta$ --- which is exponentially decaying in the system's size --- in the degeneracy between the two edge states. The best approximations of the two edge states are thus the two orthogonal real equal-weighted superpositions of the two. This superposition generates an additional saturated contribution to the entanglement entropy of a partition that includes one edge without the second, like $\mathcal{A}$ and $\mathcal{B}/\mathcal{A}$ in Fig. \ref{fig:partition}. 
More precisely, in the fully dimerized topological limit, we can write
\begin{subequations}
\begin{equation}
    \ket{L}={\hat{c}^{\dagger}}_{1,\scriptscriptstyle{\mathrm A}}\ket{0}=\ket{1_{1\scriptscriptstyle{\mathrm A}} 0_{N\scriptscriptstyle{\mathrm B}}}
\end{equation} 
\begin{equation}
    \ket{R}={\hat{c}^{\dagger}}_{N,\scriptscriptstyle{\mathrm B}}\ket{0}=\ket{0_{1\scriptscriptstyle{\mathrm A}} 1_{N\scriptscriptstyle{\mathrm B}}}
\end{equation}
\end{subequations}
we can write the density matrix of the superposition (singlet) state as
\begin{equation}
\begin{split}
    \hat{\rho}_{\scriptscriptstyle{\mathrm A}\scriptscriptstyle{\mathrm B}}&=\frac{1}{2}\left[\ket{1_{1\scriptscriptstyle{\mathrm A}} 0_{N\scriptscriptstyle{\mathrm B}}}\bra{0_{N\scriptscriptstyle{\mathrm B}} 1_{0\scriptscriptstyle{\mathrm A}}}+\ket{0_{1\scriptscriptstyle{\mathrm A}} 1_{N\scriptscriptstyle{\mathrm B}}}\bra{1_{N\scriptscriptstyle{\mathrm B}} 0_{1\scriptscriptstyle{\mathrm A}}}+\right.\\
    &+\left.\ket{1_{1\scriptscriptstyle{\mathrm A}} 0_{N\scriptscriptstyle{\mathrm B}}}\bra{1_{N\scriptscriptstyle{\mathrm B}} 0_{1\scriptscriptstyle{\mathrm A}}}+\ket{0_{1\scriptscriptstyle{\mathrm A}} 1_{N\scriptscriptstyle{\mathrm B}}}\bra{0_{N\scriptscriptstyle{\mathrm B}} 1_{1\scriptscriptstyle{\mathrm A}}}\right].
\end{split}
\end{equation}
When the chain is not in the perfectly dimerized limit, the edge modes (specifically, their probability distribution $|L|^2_j$ or $|R|^2_j$) decay exponentially into the bulk as function of the site index $j$, as shown in Fig.~\ref{fig:edge_mode_distribution} for the left edge mode, without loss of generality. By fitting an exponential function, we estimate the characteristic localization length $\xi$ of the decay $|L|^2_j = e^{-j/\xi}$, where $j$ is the site index.
\subsection{Disconnected entanglement entropy}
It is thus possible to analytically understand the role of the DEE in the fully-dimerized and topological limit of the SSH chain and show that the ground state always contains the maximally entangled superposition of the two edge states in the topological phase. The reduced system is thus
\begin{equation}
    \hat{\rho}_{\scriptscriptstyle{\mathrm B}} = \frac{1}{2}\left[\ket{0_{N\scriptscriptstyle{\mathrm B}}}\bra{0_{N\scriptscriptstyle{\mathrm B}}}+\ket{1_{N\scriptscriptstyle{\mathrm B}}}\bra{1_{N\scriptscriptstyle{\mathrm B}}}\right],
\end{equation}
so that $S_{{\mathcal{B}}_{\mathrm{\scriptstyle{edge}}}} = S_{{\mathcal{A}}_{\mathrm{\scriptstyle{edge}}}} = \log_2 2$. This extra contribution is added to the other bulk contributions which depend on the cuts of the chosen partitions. 

In the thermodynamic limit it can be easily shown~\cite{Micallo_2020} that the latter give a zero contribution, so that only the edge contribution survives and $\lim_{L\rightarrow\infty}S^D=2\log_22$. More generally, in the thermodynamic limit, the value of $S^D$ is fixed by the number of edge states $\mathcal{D}$, which is in turn fixed by the bulk-boundary correspondence~\cite{PhysRevB.91.115118}, i.e.,
$\lim_{L\to \infty}S^D=\log_2\mathcal{D}$. For this reason, the DEE is said to correspond to the thermodynamic entropy at zero temperature of one edge~\cite{Micallo_2020}. 

\section{\label{app:symmetry_classifications}Details on symmetry classifications for open systems}
 
\subsection{Nambu Formalism}
Let $\hat{c}^{\phantom \dagger}_{j}$ be the destruction operator for a system of spinless fermions labelled by $j=1,\cdots,L$.
We define a Nambu column vector $\hat{{\mathbf C}}^{\phantom \dagger}_{}\!\!$ and its Hermitian 
conjugate row vector $\hat{{\mathbf C}}^{\dagger}_{}$, each of length $2L$, by~\cite{Mbeng_2024} 
\begin{equation}
\hat{{\mathbf C}}^{\phantom \dagger}_{} = \left(
\begin{array}{c}
 \hat{c}^{\phantom \dagger}_{1}  \\ \vdots \\ \hat{c}^{\phantom \dagger}_{L} \\ {\hat{c}^{\dagger}}_{1} \\ \vdots \\ {\hat{c}^{\dagger}}_{L} 
\end{array}
\right) 
\;,
\label{eq:Nambu_op}
\end{equation}
and
\begin{equation}
\hat{{\mathbf C}}^{\dagger}_{} = \left( {\hat{c}^{\dagger}}_{1} \,,\, \cdots \,,\, {\hat{c}^{\dagger}}_{L} \,,\, \hat{c}^{\phantom \dagger}_{1} \,,\, \cdots \,,\,\hat{c}^{\phantom \dagger}_{L} \right) \;. 
\label{eq:Nambu_op_dag}
\end{equation}

Majorana fermions are Hermitian combinations of ordinary complex fermions:
\begin{equation}
\check{\mathbf{c}}^{\phantom \dagger}_{} = \left( 
\begin{array}{c} \check{\mathbf{c}}^{\phantom \dagger}_{1} \\ \check{\mathbf{c}}^{\phantom \dagger}_{2} \end{array} \right) \;,
\end{equation}
where $\check{\mathbf{c}}^{\phantom \dagger}_{1}$ and $\check{\mathbf{c}}^{\phantom \dagger}_{2}$ are $L$-dimensional (column) vectors whose elements are:
\begin{equation} \label{eqn:Majorana_def}
\check{c}_{\scriptstyle{1},j}=({\hat{c}^{\dagger}}_{j} + \hat{c}^{\phantom \dagger}_{j})   \hspace{5mm} \mbox{and} \hspace{5mm} \check{c}_{\scriptstyle{2},j}= i ({\hat{c}^{\dagger}}_{j} - \hat{c}^{\phantom \dagger}_{j}) \;.
\end{equation}
These operators are manifestly Hermitian. They allow us to express the original fermions as:
\begin{equation}
\hat{c}^{\phantom \dagger}_{j} = \textstyle{\frac{1}{2}}(\check{c}_{\scriptstyle{1},j} + i \check{c}_{\scriptstyle{2},j}) \hspace{3mm} \mbox{and} \hspace{3mm} {\hat{c}^{\dagger}}_{j} = \textstyle{\frac{1}{2}}(\check{c}_{\scriptstyle{1},j} - i \check{c}_{\scriptstyle{2},j}) \;, 
\label{eq:majoranas}
\end{equation}
and satisfy the anti-commutation relations:
\begin{equation}
\{ \check{c}_{\scriptstyle{\alpha},j}, \check{c}_{\scriptstyle{\alpha}',j'} \} = 2 \delta_{\scriptstyle{\alpha},\scriptstyle{\alpha}'} \delta_{j,j'}  \;.
\end{equation}

To be consistent with the Nambu notation for the ordinary fermions, we better define the Majorana column vector
\footnote{The standard definition by Kitaev, which in row-vector form would read:
\begin{equation}
\begin{split}
&\check{\mathbf{c}}^{\phantom \dagger}_{} = (\check{c}_{1}, \check{c}_{2}, \check{c}_{3}, \check{c}_{4}, \cdots, \check{c}_{2N-1}, \check{c}_{2L}  )^{\scriptscriptstyle{\mathrm{T}}} \equiv \\
&\equiv (  \check{c}_{1,1},  \check{c}_{2,1},  \check{c}_{1,2}, \check{c}_{2,2}, \cdots,  \check{c}_{1,N}, \check{c}_{2,L})^{\scriptscriptstyle{\mathrm{T}}} \;,
\end{split}
\end{equation}
mixes the different blocks of the Nambu fermions in a way that makes the algebra more complicated. 
}:
\begin{equation}
\check{\mathbf{c}}^{\phantom \dagger}_{} = {\mathbf{W}} \, \hat{{\mathbf C}}^{\phantom \dagger}_{} \;,
\end{equation}
where we defined the $2L \times 2L$ block matrix:
\begin{equation}
\mathbf{W} = \left( \begin{array}{r | r}  \textbf{1} & \textbf{1} \\ \hline -i\textbf{1} & i\textbf{1} \end{array} \right) 
\hspace{3mm} \mbox{such that} \hspace{3mm}
\mathbf{W} \mathbf{W}^{\dagger} = \mathbf{W}^{\dagger} \mathbf{W} = 2 \, \textbf{1} \;.
\end{equation}

\subsection{Symmetry classes for open systems:\\review of the tenfold way classification of quadratic Lindbladians}
We review the symmetry classification for quadratic open Markovian systems proposed in Ref.~\cite{cooper}. 
We start by writing the quadratic Hamiltonian in 
Eq.~\eqref{eq:ssh_hamiltonian} and the linear jump operators in terms of the $2L$ Majorana operators $\{\check{c}_{j}\}$ defined in Eq. \eqref{eq:majoranas} such that
\begin{subequations}
\begin{equation}
    \hat{H}_{\mathrm{\scriptscriptstyle SSH}}=\sum_{j,j'=1}^{2L} \check{c}_{j} {\mathrm H}_{j,j'}^{\mathrm{\scriptscriptstyle M}} \check{c}_{j'} \;,
    \label{quadraticHamiltonian}
\end{equation}
\begin{equation}
    \hat{L}_{\mu}=\sum_{j=1}^{2L} \ell_{\mu j} \check{c}_{j} \;,
    \label{qlind}
\end{equation}
\end{subequations}
where $\mathrm{H}_{j,j'}^{\mathrm{\scriptscriptstyle M}}$ are the matrix elements of a 
$2L\times 2L$ matrix $\mathbf{H}^{\mathrm{\scriptscriptstyle M}}$ which is purely imaginary and anti-symmetric, ${\mathbf{H}^{\mathrm{\scriptscriptstyle M}}}^{\scriptscriptstyle{\mathrm{T}}}=-\mathbf{H}^{\mathrm{\scriptscriptstyle M}}$, while $\ell_{\mu j} \in \mathbb{C}$. The superscript in $\mathbf{H}^{\mathrm{\scriptscriptstyle M}}$ underlines the fact that the Hamiltonian is written in the Majorana operators basis.
The dissipation is encoded into a $2L\times 2L$ complex 
(semi)-positive definite Hermitian matrix $\mathbf{M}$ whose elements are 
${\mathrm M}_{j,j'}=\sum_\mu \ell_{\mu j}\ell^*_{\mu j'}$. 
Since $\mathbf{M}^*=\mathbf{M}^{\scriptscriptstyle{\mathrm{T}}}$, the real part of $\mathbf{M}$, $\mathbf{M}_{\scriptscriptstyle{\mathrm R}}=\frac{1}{2}(\mathbf{M} + \mathbf{M}^*)$, is a (semi)-positive definite symmetric matrix, 
while the imaginary part, $\mathbf{M}_{\scriptscriptstyle{\mathrm I}}=\frac{1}{2i}(\mathbf{M} - \mathbf{M}^*)$, is anti-symmetric.  
(One can show that $\mathbf{M}_{\scriptscriptstyle{\mathrm R}}$ and $\mathbf{M}_{\scriptscriptstyle{\mathrm I}}$  are associated, respectively, to dissipation and driving~\cite{Prosen_2010}.)

We now rely on the third-quantization formalism~\cite{prosen, Prosen_2010} to vectorize the Lindblad equation as 
\begin{equation}
\frac{\mathrm{d}}{\mathrm{d} t}|\hat{\rho}\rangle = \hat{\mathcal{L}} |\hat{\rho}\rangle \;.
\end{equation}
According to this formalism, one expands $\ket{\hat{\rho}}$ into a basis of vectors (Hermitian operators) $\left\{\ket{P_{\underline{\alpha}}}\right\}$ of a $2^{2L}=4^L$-dimensional space:
\begin{equation}
\ket{P_{\underline{\alpha}}} \stackrel{\mathrm{def}}{=} 2^{-L/2} \, 
\check{c}_{1}^{\, \alpha_1} \cdots \check{c}_{2L}^{\, \alpha_{2L}} 
\;, \hspace{5mm} \alpha_j={0,1} \;.
\end{equation}
We then define~\cite{prosen, Prosen_2010} the action of fermionic superoperators on this space as follows:
\begin{equation} \label{eq:fermionic_superoperators}
\left\{
\begin{array}{l}
    \hat{a}^{\phantom \dagger}_{j}\ket{P_{\underline{\alpha}}}=\delta_{\alpha_j,1}\ket{\check{c}_{j} P_{\underline{\alpha}}}
    \vspace{5mm} \\
    \hat{a}^{\dagger}_{j}\ket{P_{\underline{\alpha}}}=\delta_{\alpha_j,0}\ket{\check{c}_{j} P_{\underline{\alpha}}}.
\end{array}
\right.
\end{equation}
In terms of these, the Lindbladian can be represented as a quadratic superoperator of the form
\begin{equation}
\hat{\mathcal{L}} = 
\left( \hat{{\mathbf a}}^\dagger, \hat{{\mathbf a}}^{\scriptscriptstyle{\mathrm{T}}} \right) 
\left( 
\begin{array}{cc} 
-\mathbf{X}^{\scriptscriptstyle{\mathrm{T}}} & i\mathbf{Y} \\ 
\mathbf{0} & \mathbf{X} 
\end{array} 
\right)
\left( \begin{array}{c}  \hat{{\mathbf a}} \\ 
\hat{{\mathbf a}}^{\dagger \scriptscriptstyle{\mathrm{T}}} \end{array} \right)
-\mathrm{Tr} \mathbf{X} 
\end{equation}
Here, $\hat{{\mathbf a}}=(\hat{a}^{\phantom \dagger}_{1},\cdots,\hat{a}^{\phantom \dagger}_{2L})^{\scriptscriptstyle{\mathrm{T}}}$ is a $2L$-dimensional column vector formed with the $\hat{a}^{\phantom \dagger}_{j}$ operators, 
$\hat{{\mathbf a}^\dagger}=(\hat{a}^{\dagger}_{1},\cdots,\hat{a}^{\dagger}_{2L})$ the corresponding row vector, the $2L\times 2L$ real matrix $\mathbf{X}$ is given by
\begin{equation} \label{eq:lindblad_cooper}
\mathbf{X} = -2i \mathbf{H}^{\mathrm{\scriptscriptstyle M}} + 2 \mathbf{M}_{\scriptscriptstyle{\mathrm R}} 
\;.
\end{equation}
and the $2L\times 2L$ real anti-symmetric matrix
$\mathbf{Y}=4\mathbf{M}_{\scriptscriptstyle{\mathrm I}}$.
Quite importantly, $\mathbf{X}+\mathbf{X}^{\scriptscriptstyle{\mathrm{T}}}=4\mathbf{M}_{\scriptscriptstyle{\mathrm R}}$ is 
(semi)-positive definite.

Due to the upper triangular nature of the Lindbladian, its spectral properties are completely determined by the spectrum of $\mathbf{X}$, i.e., by the set of \textit{rapidities} $\{\beta_j\}$~\cite{prosen, Prosen_2010}, with $\mathrm{Re} \beta_j\ge 0$. Hence, it is possible to state symmetry relations for $\mathbf{X}$ that generalize the tenfold way classification of topological insulators~\cite{Kitaev_2009, chiu} when in interaction with an environment, as long as the dissipative dynamics can be described by quadratic Lindbladians. Like the Hamiltonian for closed systems, $\mathbf{X}$ becomes now the landmark to check whether the symmetries are preserved or broken and, consequently, if the topological features of the open systems can be preserved or not. 
Indeed, for closed systems, according to the presence or absence of the following three time-reversal, particle-hole and chiral symmetries (TRS, PHS, Chiral)
\begin{align}
\nonumber
&\mathbf{H}^{\mathrm{\scriptscriptstyle M}}=\mathbf{U}_{\scriptscriptstyle{\mathrm T}} {\mathbf{H}^{\mathrm{\scriptscriptstyle M}}}^*\mathbf{U}_{\scriptscriptstyle{\mathrm T}}^\dagger &\mathbf{U}_{\scriptscriptstyle{\mathrm T}} \mathbf{U}_{\scriptscriptstyle{\mathrm T}}^*=\pm\textbf{1} & &\text{(TRS)}\\
&\mathbf{H}^{\mathrm{\scriptscriptstyle M}}=-\mathbf{U}_{\scriptscriptstyle{\mathrm C}} {\mathbf{H}^{\mathrm{\scriptscriptstyle M}}}^*\mathbf{U}_{\scriptscriptstyle{\mathrm C}}^\dagger &\mathbf{U}_{\scriptscriptstyle{\mathrm C}} \mathbf{U}_{\scriptscriptstyle{\mathrm C}}^*=\pm\textbf{1} & &\text{(PHS)}\\
\nonumber
&\mathbf{H}^{\mathrm{\scriptscriptstyle M}}=-\mathbf{U}_{\scriptscriptstyle{\mathrm S}} \mathbf{H}^{\mathrm{\scriptscriptstyle M}} \mathbf{U}_{\scriptscriptstyle{\mathrm S}}^\dagger &\mathbf{U}_{\scriptscriptstyle{\mathrm S}}^2=\textbf{1} & &\text{(Chiral)}
\end{align}
where $\mathbf{U}_{\scriptscriptstyle{\mathrm T},\scriptscriptstyle{\mathrm C},\scriptscriptstyle{\mathrm S}}$ are all unitary operators, and $\mathbf{U}_{\scriptscriptstyle{\mathrm S}}=\mathbf{U}_{\scriptscriptstyle{\mathrm C}} \mathbf{U}_{\scriptscriptstyle{\mathrm T}}$, the systems fall in one of the ten symmetry classes related to specific types ($\mathbb{Z}$, $\mathbb{Z}_2$, $2\mathbb{Z}$) of topological states, according to the dimension of the system~\cite{Kitaev_2009}. For open systems, the generalized equations become
\begin{align}
\nonumber
&\mathbf{X}=\mathbf{U}_{\scriptscriptstyle{\mathrm T}} \mathbf{X}^{\scriptscriptstyle{\mathrm{T}}}\mathbf{U}_{\scriptscriptstyle{\mathrm T}}^\dagger &\mathbf{U}_{\scriptscriptstyle{\mathrm T}} \mathbf{U}_{\scriptscriptstyle{\mathrm T}}^*=\pm\textbf{1} & &\text{(TRS)}\\
&\mathbf{X}=\mathbf{U}_{\scriptscriptstyle{\mathrm C}} \mathbf{X}^*\mathbf{U}_{\scriptscriptstyle{\mathrm C}}^\dagger &\mathbf{U}_{\scriptscriptstyle{\mathrm C}} \mathbf{U}_{\scriptscriptstyle{\mathrm C}}^*=\pm\textbf{1} & &\text{(PHS)}\label{eq:cooper_symmetries}\\
\nonumber
&\mathbf{X}=\mathbf{U}_{\scriptscriptstyle{\mathrm S}} \mathbf{X}^\dagger \mathbf{U}_{\scriptscriptstyle{\mathrm S}}^\dagger &\mathbf{U}_{\scriptscriptstyle{\mathrm S}}^2=\textbf{1} & &\text{(PAH)}
\end{align}
where the unitaries have to be the same as those of the closed classification and the Pseudo-Anti-Hermiticity (PAH) symmetry replaces Chiral symmetry. These equations are obtained imposing some physical constraints on the spectrum of the Lindbladian. 
According to this classification, it is possible to see that the jump operators in Eqs.~\eqref{eq:spd_jump} are symmetry-preserving, while the ones in Eq.~\eqref{eq:sbd_jump} are symmetry-breaking.

\subsection{Hamiltonian}
Let us start considering the SSH Hamiltonian of Eq.~\eqref{eq:ssh_hamiltonian}. In Nambu formalism it reads as:
\begin{equation}
\hat{H}_{\mathrm{\scriptscriptstyle SSH}} = \hat{{\mathbf C}}^{\dagger}_{} \mathbf{H}_{\mathrm{\scriptscriptstyle SSH}} \, \hat{{\mathbf C}}^{\phantom \dagger}_{}
\;,
\end{equation}
with
\begin{equation}
\mathbf{H}_{\mathrm{\scriptscriptstyle SSH}} = 
\left( 
\begin{array}{cc} \mathbf{A}_{\scriptscriptstyle{\mathrm H}} & \mathbf{0} \\
\mathbf{0} & -\mathbf{A}_{\scriptscriptstyle{\mathrm H}}
\end{array}
\right) \;,
\end{equation}
where the $L\times L$ real symmetric matrix $\mathbf{A}_{\scriptscriptstyle{\mathrm H}}$ reads:
\begin{equation}
\mathbf{A}_{\scriptscriptstyle{\mathrm H}} = \frac{1}{2} 
\left(
\begin{array}{ccccccc}
0 & -\hopintra& 0 & \cdots & \cdots & \cdots & 0 \\
-\hopintra& 0 & -\hopinter & 0 & \cdots & \cdots & 0 \\
0 & -\hopinter & 0 & -\hopintra & 0 & \cdots & 0 \\
0 & 0 & -\hopintra & 0 & -\hopinter & \cdots & 0 \\
\vdots & \vdots & \cdots & \ddots & \ddots & \ddots & 0 \\
\vdots & \vdots & \vdots & 0 & -\hopinter & 0 & -\hopintra \\
0 & \cdots & \cdots & \cdots & 0 & -\hopintra & 0 
\end{array}
\right) \;.
\end{equation}
Equivalently, in terms of Majorana operators:
\begin{equation}
\hat{H}_{\mathrm{\scriptscriptstyle SSH}} = \hat{{\mathbf C}}^{\dagger}_{} \, {\mathbf{H}_{\mathrm{\scriptscriptstyle SSH}}} \, \hat{{\mathbf C}}^{\phantom \dagger}_{} 
= \check{\mathbf{c}}^{\scriptscriptstyle{\mathrm{T}}} \, 
{\mathbf{H}}_{\mathrm{\scriptscriptstyle SSH}}^{\mathrm{\scriptscriptstyle M}} \, \check{\mathbf{c}} \;.
\end{equation}
with
\begin{equation}
{\mathbf{H}}_{\mathrm{\scriptscriptstyle SSH}}^{\mathrm{\scriptscriptstyle M}} = 
\frac{i}{2}\left( 
\begin{array}{cc} \mathbf{0} & \mathbf{A}_{\scriptscriptstyle{\mathrm H}} \\
-\mathbf{A}_{\scriptscriptstyle{\mathrm H}} & \mathbf{0}  
\end{array}
\right) .
\end{equation}
The symmetries that hold for the SSH chain, written in terms of Majorana, are
\begin{align}
\nonumber
\mathbf{H}_{\mathrm{\scriptscriptstyle SSH}}^{\mathrm{\scriptscriptstyle M}}&=-\mathbf{H}_{\mathrm{\scriptscriptstyle SSH}}^{\mathrm{\scriptscriptstyle M}*}=\mathbf{U}_{\scriptscriptstyle{\mathrm T}} \mathbf{H}_{\mathrm{\scriptscriptstyle SSH}}^{\mathrm{\scriptscriptstyle M}*}\mathbf{U}_{\scriptscriptstyle{\mathrm T}}^\dagger & &\text{(TRS)}\\
\nonumber
\mathbf{H}_{\mathrm{\scriptscriptstyle SSH}}^{\mathrm{\scriptscriptstyle M}}&=\mathbf{\Sigma}_z \mathbf{H}_{\mathrm{\scriptscriptstyle SSH}}^{\mathrm{\scriptscriptstyle M}*}\mathbf{\Sigma}_z=\\
&=-\mathbf{U}_{\scriptscriptstyle{\mathrm T}} \mathbf{\Sigma}_z\mathbf{H}_{\mathrm{\scriptscriptstyle SSH}}^{\mathrm{\scriptscriptstyle M}*}( \mathbf{U}_{\scriptscriptstyle{\mathrm T}}\mathbf{\Sigma}_z)^\dagger & &\text{(PHS)}\label{eq:unitaries_ssh}\\
\nonumber
\mathbf{H}_{\mathrm{\scriptscriptstyle SSH}}^{\mathrm{\scriptscriptstyle M}}&=-\mathbf{\Sigma}_z \mathbf{H}_{\mathrm{\scriptscriptstyle SSH}}^{\mathrm{\scriptscriptstyle M}}\mathbf{\Sigma}_z^\dagger & &\text{(Chiral)}
\end{align}
where $\mathbf{U}_{\scriptscriptstyle{\mathrm T}}=\mathbf{\Sigma}_z$ and 
\begin{equation}
    \mathbf{\Sigma}_z=\hat{\sigma}^z \otimes \textbf{1}
\end{equation}
so that $\mathbf{U}_{\scriptscriptstyle{\mathrm C}}=\mathbf{U}_{\scriptscriptstyle{\mathrm T}} \mathbf{\Sigma}_z=\textbf{1}$ is the operator related to the particle-hole symmetry (PHS) and 
$\mathbf{U}_{\scriptscriptstyle{\mathrm S}}=\mathbf{\Sigma}_z$ is the one related to the chiral symmetry (CS). 
These operators are properly the same we will use to check the symmetries in the dissipative case. The fact that the three symmetries are satisfied, together with the fact that
\begin{flalign}
    \nonumber
    \mathbf{U}_{\scriptscriptstyle{\mathrm T}}^2&=\mathbf{U}_{\scriptscriptstyle{\mathrm T}} \mathbf{U}_{\scriptscriptstyle{\mathrm T}}^*=\textbf{1}\\
    \mathbf{U}_{\scriptscriptstyle{\mathrm C}}^2&=\mathbf{U}_{\scriptscriptstyle{\mathrm C}} \mathbf{U}_{\scriptscriptstyle{\mathrm C}}^*=\textbf{1}\\
    \nonumber
    \mathbf{U}_{\scriptscriptstyle{\mathrm S}}^2&=\textbf{1}
\end{flalign}
makes the system fall into the topological class BDI which provides for a $\mathbb{Z}$-type of topological invariants, according to the periodic table of topological insulators~\cite{chiu}. 

\subsection{SPD dynamics}
In the SPD dynamics, the dissipation matrix,
in terms of Majorana operators, is:
\begin{equation}
    \mathbf{M} = \frac{\gamma}{4}
\left( 
\begin{array}{cc} \textbf{1} &  -i(\mathbf{1}_{\scriptscriptstyle o}-\mathbf{1}_{\scriptscriptstyle e}) \\
 i(\mathbf{1}_{\scriptscriptstyle o}-\mathbf{1}_{\scriptscriptstyle e}) &  \textbf{1}
\end{array}
\right)   \;.
\end{equation}
where $\mathbf{1}_{\scriptscriptstyle o}$ is an identity only on the odd (A) sites, 
and similarly $\mathbf{1}_{\scriptscriptstyle e}$ for the even (B) sites, hence
$\textbf{1}=\mathbf{1}_{\scriptscriptstyle o}+\mathbf{1}_{\scriptscriptstyle e}$.
The real matrix $\mathbf{X}$ appearing in Eq. \eqref{eq:lindblad_cooper} is therefore given by: 
\begin{equation}
\mathbf{X} \equiv -2i{\mathbf{H}}_{\mathrm{\scriptscriptstyle SSH}}^{\mathrm{\scriptscriptstyle M}} + 2\mathbf{M}_{\scriptscriptstyle{\mathrm R}} = 
\left( 
\begin{array}{cc} \frac{\gamma}{2} \mathbf{1} &  \mathbf{A}_{\scriptscriptstyle{\mathrm H}} \\
-\mathbf{A}_{\scriptscriptstyle{\mathrm H}} &  \frac{\gamma}{2} \mathbf{1}
\end{array}
\right) \;.
\end{equation}
The latter satisfies equations~\eqref{eq:cooper_symmetries} with the same unitaries of equations~\eqref{eq:unitaries_ssh}.

\subsection{SBD dynamics}
In the SBD dynamics the dissipation matrix is:
\begin{equation}
\mathbf{M} = \frac{\gamma}{4}
\left( 
\begin{array}{cc} \textbf{M}_{\scriptscriptstyle{\mathrm A}} &  -i \textbf{M}_{\scriptscriptstyle{\mathrm A}} \\
 i \textbf{M}_{\scriptscriptstyle{\mathrm A}} &  \textbf{M}_{\scriptscriptstyle{\mathrm A}}
\end{array}
\right) \;,
\end{equation}
with the complex $L\times L$ matrix $\textbf{M}_{\scriptscriptstyle{\mathrm A}}$ given by:
\begin{equation}
\textbf{M}_{\scriptscriptstyle{\mathrm A}} = 
\left(
\begin{array}{ccccccc}
1 &1 & 0 & \cdots & \cdots & \cdots & 0 \\
1 & 2 & 1 & 0 & \cdots & \cdots & 0 \\
0 & 1 & 2 & 1 & 0 & \cdots & 0 \\
0 & 0 & 1 & 2 & 1 & \cdots & 0 \\
\vdots & \vdots & \cdots & \ddots & \ddots & \ddots & 0 \\
\vdots & \vdots & \vdots & 0 & 1 & 2 & 1 \\
0 & \cdots & \cdots & \cdots & 0 & 1 & 1 
\end{array}
\right) \;.
\end{equation}
which does not allow to see the symmetries preserved for $\mathbf{X}$.

\section{\label{app:free_fermions}Free-fermions techniques}
\subsection{Quadratic Hamiltonians}
Let us consider the most general form for a quadratic Hamiltonian
\begin{equation}
\begin{split}
    &\hat{\Theta}=\sum_{i,j}\left[A_{i,j}{\hat{c}^{\dagger}}_{i} \hat{c}^{\phantom \dagger}_{j}-A^*_{i,j} \hat{c}^{\phantom \dagger}_{i} {\hat{c}^{\dagger}}_{j}+\right.\\
    &\left.+B_{i,j}\hat{c}^{\phantom \dagger}_{i} \hat{c}^{\phantom \dagger}_{j} -B^*_{i,j}{\hat{c}^{\dagger}}_{i} {\hat{c}^{\dagger}}_{j}\right],
\end{split}
\label{eq:quadratic_hamiltonian}
\end{equation}
where $\textbf{A}$ is a Hermitian matrix and $\textbf{B}$ is a skew-symmetric matrix. This general quadratic Hamiltonian can be also rewritten as~\cite{Mbeng_2024}
\begin{equation}
    \hat{\Theta}=\hat{{\mathbf C}}^{\dagger}_{} \mathbb{\Theta}\hat{{\mathbf C}}^{\phantom \dagger}_{}, 
\end{equation}
where 
\begin{equation}
    \mathbb{\Theta}=\begin{pmatrix}
        \textbf{A} &\textbf{B}\\
        -\textbf{B}^* &\textbf{A}^*
    \end{pmatrix}
\end{equation}
and $\hat{{\mathbf C}}^{\phantom \dagger}_{}$ is the Nambu operator defined in Eq.~\eqref{eq:Nambu_op}.

\subsection{\label{sec:Gaussian_states}Gaussian states}
A Gaussian state is any state whose density matrix can be written as
\begin{equation}
    \hat{\rho}(t) = \frac{1}{\mathcal{Z}(t)} \mathrm{e}^{- \hat{\Theta}(t) }
    \label{eq:gaussian_state}
\end{equation}
where $\mathcal{Z}(t)=\mathrm{Tr} \, \mathrm{e}^{- \hat{\Theta}(t) }$ enforces the normalization.
For Gibbs states, $\hat{\Theta}$ is the real quadratic Hamiltonian of the equilibrium state, while, in general, it plays the role of an effective Hamiltonian, which we will refer to as \textit{entanglement Hamiltonian}. 

A simpler expression for the Gaussian state can be derived considering a number-preserving $\hat{\Theta}$, i.e.,
\begin{equation}
    \hat{H}_{\mathrm{\scriptstyle np}}=\sum_{i,j}H_{i,j}{\hat{c}^{\dagger}}_{i}\hat{c}^{\phantom \dagger}_{j}.
    \label{eq:quadratic_hamiltonian_particle}
\end{equation}
If $U_{k,\alpha}$ is the i-th component of the k-th eigenstate of $\textbf{H}$ with eigenvalues $\epsilon_k$, the unitary transformation
\begin{equation}
    \hat{c}^{\phantom \dagger}_{i}=\sum_k U_{k,i}\hat{a}^{\phantom \dagger}_{k},
    \label{eq:unitary_transformation}
\end{equation} 
allows to diagonalise $\hat{H}_{\mathrm{\scriptstyle np}}$ and write the Gaussian state as~\cite{IngoPeschel_2003, Surace_2022}
\begin{equation}
\begin{split}
    &\hat{\rho}=\frac{1}{\mathcal{Z}}e^{-\sum_k\frac{\epsilon_k}{2}\left(\hat{a}^{\dagger}_{k} \hat{a}^{\phantom \dagger}_{k}-\hat{a}^{\phantom \dagger}_{k} \hat{a}^{\dagger}_{k}\right)}=\\
    &=\bigotimes_{k=1}^N\frac{e^{-\frac{\epsilon_k}{2}\left(\hat{a}^{\dagger}_{k} \hat{a}^{\phantom \dagger}_{k}-\hat{a}^{\phantom \dagger}_{k} \hat{a}^{\dagger}_{k}\right)}}{\mathcal{Z}_k}=\bigotimes_{k=1}^N \frac{\hat{\rho}_k}{\mathcal{Z}_k},
    \end{split}
\end{equation}

where

\begin{equation}
\begin{split}    &\mathcal{Z}_k=\mathrm{Tr}\left[e^{\frac{\epsilon_k}{2}(\hat{a}^{\dagger}_{k} \hat{a}^{\phantom \dagger}_{k}-\hat{a}^{\phantom \dagger}_{k} \hat{a}^{\dagger}_{k})}\right]=\\ &=2\cosh{\epsilon_k/2}=e^{\frac{\epsilon_k}{2}}+e^{-\frac{\epsilon_k}{2}}.
\end{split}
\end{equation}
It is thus trivial to derive the connection between the spectrum of the entanglement Hamiltonian and that of the density matrix.

\subsection{Covariance matrix}
As previously done, let us restrict our study to the number-preserving quadratic Hamiltonian of Eq.~ \eqref{eq:quadratic_hamiltonian_particle}. Let us consider the covariance matrix of Eq.~ \eqref{eq:cov_mat} whose average is computed over the Gaussian state with effective Hamiltonian \eqref{eq:quadratic_hamiltonian_particle}.

Since $\hat{\rho}$ is a Gaussian state, Wick's theorem holds, and all the higher correlation functions can be expressed in terms of the Hermitian matrix $\textbf{G}$. This means that the two-point covariance matrix $\textbf{G}$ encodes all the necessary information of the Gaussian state~\cite{sieberer2023universalitydrivenopenquantum, Altland_2021}. As a limiting case, $\hat{\rho}$ can also be a pure state related to a Slater determinant, i.e., an eigenstate of an effective Hamiltonian.
Diagonalizing $\hat{\Theta}$ with the unitary transformation of Eq.~ \eqref{eq:unitary_transformation} means having
\begin{equation}
    \sum_{i,j} U_{k,i}^* H_{i,j}U_{k',j}=\epsilon_k \delta_{k,k'}.
\end{equation}
From the latter equation, we can derive
\begin{equation}
    H_{i,j}=\sum_k U_{k,i}^*U_{k,j}\epsilon_k.
    \label{eq:hrel}
\end{equation}
Considering Eq.~\eqref{eq:hrel} and the Gaussian state of Eq.~\eqref{eq:gaussian_state}, together with the Wick's theorem, we can obtain
\begin{equation}
    G_{i,j}=\sum_k U_{k,i}^*U_{k,j}\frac{1}{1+e^{\epsilon_k}}.
    \label{eq:correlation}
\end{equation}
Comparing Eq.~\eqref{eq:hrel} and Eq.~\eqref{eq:correlation}, we can deduce that the eigenvalues $\left\{\epsilon_k\right\}$ of $\textbf{H}$ and those $\left\{\zeta_k\right\}$ of $\textbf{G}$ are related by
\begin{equation}
    \zeta_k=\frac{1}{1+e^{\epsilon_k}},
\end{equation}
i.e.,
\begin{equation}
    \textbf{H}=\ln\left[\frac{\textbf{1}-\textbf{G}}{\textbf{G}}\right].
\end{equation}
Eventually, what we have to do is diagonalizing the Hamiltonian. From the unitary diagonalisation matrix  $\textbf{U}$, we can find the expression of the effective creation and annihilation operators, so that

\begin{equation}
    G_{i,j}=\braket{{\hat{c}^{\dagger}}_{j} \hat{c}^{\phantom \dagger}_{i}}=\sum_{k,k'}\braket{\hat{a}^{\dagger}_{k'}U_{k',j}U_{i,k}^*\hat{a}^{\phantom \dagger}_{k}},
\end{equation}
which means that

\begin{equation}
    G_{i,j}=\sum_{k=1}^{N}U_{i,k}^*U_{k,j},
\end{equation}
where $N$ is the number of particles we consider in the system, which, in our half-filling case at zero temperature, coincides with the number of unit cells of the chain.

This approach is faster than direct diagonalisation and allows to compute efficiently,  $-$ i.e., by the $N$x$N$ matrix $\textbf{G}$ $-$ the reduced density matrix and its entanglement spectra and $S^D$.

The representation of the reduced density matrix can be thus written as
\begin{equation}
\begin{split}
    &\bm{\rho}=\bigotimes_{k=1}^N\frac{1}{\mathcal{Z}_k}\begin{pmatrix}
        \rho_{k_{11}} &\rho_{k_{10}}\\
        \rho_{k_{01}} &\rho_{k_{00}}
    \end{pmatrix}=\\
    &=\bigotimes_{k=1}^N\begin{pmatrix}
        \frac{e^{-\frac{\epsilon_k}{2}}}{e^{\frac{\epsilon_k}{2}}+e^{-\frac{\epsilon_k}{2}}} &0\\
        0 &\frac{e^{\frac{\epsilon_k}{2}}}{e^{\frac{\epsilon_k}{2}}+e^{-\frac{\epsilon_k}{2}}}
    \end{pmatrix}=\\
        &=\bigotimes_{k=1}^N \begin{pmatrix}
            \zeta_k &0\\
            0 &(1-\zeta_k)
        \end{pmatrix},
\end{split}
\label{eq:rho}
\end{equation}
which leaves us with the seeked connection between the spectra $\{\lambda_k\}$ of $\bm{\rho}$ and $\{\zeta_k\}$ of $\textbf{G}$.

\subsection{Reduced system}
When we consider a subsystem of $\{\alpha,\beta\}\in X$ sites, the reduced density matrix $\hat{\rho}_{\scriptscriptstyle{\mathrm X}}$ allows to reproduce all expectation values in the subsystem and, as long as it remains Gaussian, so does the \textit{reduced two-point covariance matrix}
\begin{equation}
    G_{{\scriptscriptstyle{\mathrm X}}\alpha,\beta}=\mathrm{Tr}(\hat{\rho}_{\scriptscriptstyle{\mathrm X}} {\hat{c}^{\dagger}}_{\beta}\hat{c}^{\phantom \dagger}_{\alpha}).
    \label{eq:reduced_cov_mat}
\end{equation}
In order for $\hat{\rho}_{\scriptscriptstyle{\mathrm X}}$ to be Gaussian, it is still required to be the exponential of a quadratic effective Hamiltonian, i.e.,
\begin{equation}
    \hat{\rho}_{\scriptscriptstyle{\mathrm X}}=\mathcal{K}e^{-\hat{H}_{\scriptscriptstyle{\mathrm X}}},
    \label{eq:reduced_entanglement_hamiltonian}
\end{equation}
with
\begin{equation}
    \hat{H}_{\scriptscriptstyle{\mathrm X}}=\sum_{\alpha,\beta}H_{{\scriptscriptstyle{\mathrm X}}\alpha,\beta}{\hat{c}^{\dagger}}_{\alpha}\hat{c}^{\phantom \dagger}_{\beta}.
\end{equation}
Computing the spectrum of the reduced correlation matrix \eqref{eq:reduced_cov_mat} is equivalent to exactly diagonalizing the entanglement Hamiltonian of the reduced density matrix \eqref{eq:reduced_entanglement_hamiltonian}, yet faster.

\section{\label{app:numerical_degeneracy}Treating the degeneracy of the edge modes}

In the preparation of the initial state for the dynamics, we encounter a numerical issue due to the edge mode degeneracy. Specifically, when diagonalizing the Hamiltonian, we find two states with energies close to zero, separated by a gap that decreases exponentially with the system size. Due to this degeneracy, diagonalization routines arbitrarily combine the two eigenstates associated with the nearly degenerate eigenvalue. This leads to problems in constructing the initial correlator, as the zero-energy eigenstates are arbitrarily ordered and combined. As a consequence, the DEE in the initial state is not equal to $2$. To overcome this issue, we decide to exploit the parity symmetry of the Hamiltonian. 

Indeed, we know that $[\hat{H}_{\mathrm{\scriptscriptstyle SSH}},\hat{P}]=0$ where $\hat{P}$ is the parity operator whose matrix representation in one dimension acts as
\begin{equation}
    \textbf{P}=\begin{pmatrix}
        0 &0 &\cdots &0 &1\\
        0 &0 &\cdots &1 &0\\
         & &\ddots & &\\
        0 &1 &\cdots &0 &0\\
        1 &0 &\cdots &0 &0
    \end{pmatrix},
\end{equation}
 so to invert the first with the last site, the second with the second-last site and so on. Due to this symmetry relation of the SSH Hamiltonian, since both $\hat{H}_{\mathrm{\scriptscriptstyle SSH}}$ and $\hat{P}$ are Hermitian, we know it is possible to find a common basis of eigenvectors $\textbf{V}_\textrm{P}$ such that $\textbf{P}$ becomes diagonal and $\textbf{H}_{\mathrm{\scriptscriptstyle SSH}}$ is reduced into two diagonal blocks of different parity, i.e., $\textbf{H}_\textrm{P}=\textbf{V}_\textrm{P}^T \textbf{H}_{\mathrm{\scriptscriptstyle SSH}} \textbf{V}_\textrm{P}$ such that
 \begin{equation}
     \textbf{H}_\mathrm{P}=\begin{pmatrix}
         \textbf{H}_{\scriptscriptstyle\textrm{odd}} &\\
          &\textbf{H}_{\scriptscriptstyle\textrm{even}}
     \end{pmatrix}.
 \end{equation}
Reducing the Hamiltonian into two blocks allows us to split the two degenerate zero modes --- one will go in the even block and the other in odd block. Hence, we can diagonalize the two blocks and find the eigenvectors, $\textbf{U}_{\scriptscriptstyle\textrm{odd}}$ and $\textbf{U}_{\scriptscriptstyle\textrm{even}}$, separately, so to avoid arbitrary numerical superpositions of the two quasi-zero-energy modes. The total matrix of eigenvectors in the rotated basis is
\begin{equation}
    \textbf{U}_\mathrm{P} = \begin{pmatrix}
        \textbf{U}_{\scriptscriptstyle\textrm{odd}} &\\
         &\textbf{U}_{\scriptscriptstyle\textrm{even}}
    \end{pmatrix}.
\end{equation}
When ordering the eigenvalues in ascending order, we treat the two near-zero-energy modes as degenerate since they differ only beyond the threshold of machine precision ($\sim10^{-12}$). We place the zero mode of the even block first, followed by the odd-block one. This is because, when constructing the ground state correlator at half-filling, we will sum over all negative energy modes from both the even and odd blocks, up to the sole even zero mode. We choose to populate only the even mode to achieve a spatial configuration of the Bell-like pair between the two edge modes that corresponds to a triplet state ${\scriptscriptstyle \frac{1}{\sqrt{2}}} (\ket{0_1,1_L}+\ket{1_1,0_L})$. Once the eigenvalues and eigenstates are ordered in the rotated basis, we return to the original basis by rotating the eigenstates with $\textbf{V}_\textrm{P}$. With these eigenstates, we can finally construct the ground state correlator at half-filling.

\section{\label{app:trajectories}Details on the quantum-jump unraveling for quadratic Lindblad equations}

\subsection{Quantum-jump unraveling for Gaussian states}
We refer to Ref.~\cite{Passarelli_2019} for more information regarding the adopted algorithm for the computation of the quantum-jump trajectory. Suitably arranging the latter, exploiting the Gaussian nature of the states we deal with, we actually apply the algorithm directly on the covariance matrix related to the Gaussian state, which results in an enormous advantage in terms of computational cost of our calculations. Specifically, we look at the time evolution of $\mathbf{G}(t)|_{\mathrm{\scriptstyle traj}}$. In the following we will name $\mathbf{G}(t)|_{\mathrm{\scriptstyle traj}}$ as $\mathbf{G}(t)$, for brevity. Nevertheless, it is important to stress that, in the main text, $\mathbf{G}(t)$ is the average over many realizations of $\mathbf{G}(t)|_{\mathrm{\scriptstyle traj}}$.
\subsubsection{Non-Hermitian evolution}
Let us now consider the non-Hermitian contribution to the evolution ruled by $\hat{H}_{eff}$. An efficient way of updating the state when the non-Hermitian evolution occurs is considering
\begin{widetext}
\begin{equation}
    \textbf{G}(t+\text{d}t) = \braket{\psi(t)|\left(\hat{\mathds{1}}+i\hat{H}_{\mathrm{\scriptstyle eff}}^\dagger\text{d}t+\Lambda(t)\delta t\right){\hat{c}^{\dagger}}_{j}\hat{c}^{\phantom \dagger}_{i}\left(\hat{\mathds{1}}-i\hat{H}_{\mathrm{\scriptstyle eff}}\text{d}t +\Lambda(t)\text{d}t\right)|\psi(t)}+ o(\text{d}t)
\end{equation}
\end{widetext}
is the part of equation \eqref{eq:stocastic_schrod} ruling the non-Hermitian evolution of the correlation function including the normalization of the state, valid in $o(\text{d}t)$ limit, where we have written
\begin{subequations}
\begin{flalign}
    &\hat{\Lambda}=\frac{\gamma}{2}\sum_k \hat{L}^{\dagger}_{k} \hat{L}_{k}\\
    &\Lambda=\braket{\psi(t)|\hat{\Lambda}|\psi(t)}\\
    &\hat{H}_{\mathrm{\scriptstyle eff}}=\hat{H}_{\mathrm{\scriptscriptstyle SSH}}-i \hat{\Lambda}.
\end{flalign}
\end{subequations}
This leads to
\begin{widetext}
\begin{equation}
\begin{split}
    G_{ij}(t+\text{d}t) = G_{ij}(t)+i \text{d}t \braket{\psi(t)|[\hat{H}_{\mathrm{\scriptscriptstyle SSH}}, {\hat{c}^{\dagger}}_{j}\hat{c}^{\phantom \dagger}_{i}]|\psi(t)}-\text{d}t \braket{\psi(t)|\left\{\hat{\Lambda},{\hat{c}^{\dagger}}_{j}\hat{c}^{\phantom \dagger}_{i}\right\}|\psi(t)}+2\text{d}t\Lambda G_{ij}(t)+o(\text{d}t).
\end{split}
\end{equation}
\end{widetext}
What makes symmetry-breaking and symmetry-preserving case different is $\hat{\Lambda}$, so we will write them in the following. For the unitary part, which is common to both dynamics, we have
\begin{equation}
    \braket{\left[\hat{H}_{\mathrm{\scriptscriptstyle SSH}},{\hat{c}^{\dagger}}_{j}\hat{c}^{\phantom \dagger}_{i}\right]}=\sum_{\alpha}\left(G_{i,\alpha} H_{\alpha,j}-H_{i,\alpha}G_{\alpha,j}\right)
\end{equation}

\subsection{Evolution of the norm}
In order to consider the occurrence of the quantum jump when $\braket{\psi(t^*)|\psi(t^*)} > r $, where $r$ is the random number uniformly distributed in $[0,1]$, we simultaneously compute the time-evolution of the norm $n(t)=\braket{\psi(t)|\psi(t)}$ when it evolves under $\hat{H}_{eff}$
\begin{equation} 
n(t+\text{d}t)=n(\text{d}t)-2\text{d}t n(t)\Lambda(t)+o(\text{d}t),
\end{equation}
so that
\begin{equation}
    \frac{n(t+\text{d}t)-n(t)}{dt}=-2\Lambda(t)n(t),
\end{equation}
which in $\text{d}t \rightarrow 0$ limit is
\begin{equation}
    \frac{dn(t)}{dt}=-2\Lambda(t)n(t),
\end{equation}
where, again, $\Lambda$ must be computed in the two dissipative cases.

\subsubsection{SPD}

\paragraph{Non-Hermitian evolution.}
In the SPD case we have 
\begin{equation}
    \hat{\Lambda}=\frac{\gamma}{2}\left[\sum_{\alpha=1}^{L}(-1)^\alpha{\hat{c}^{\dagger}}_{\alpha}\hat{c}^{\phantom \dagger}_{\alpha}\right]+L\frac{\gamma}{4}
\end{equation}
which leads to a non-Hermitian evolution of $\mathbf{G}$ of the form
\begin{equation}
    \frac{d\mathbf{G}}{dt}=i(\mathbf{G}\mathbf{H}-\mathbf{H}\mathbf{G})-\gamma\left[\frac{\mathbf{G}\textbf{S}+\textbf{S}\mathbf{G}}{2}-\mathbf{G}\textbf{S}\mathbf{G}\right],
\end{equation}
where $\textbf{S}=\mathbf{1}_{\scriptscriptstyle o}-\mathbf{1}_{\scriptscriptstyle e}$.

\paragraph{Quantum Jump.}
In the symmetry-preserving case we have jump operators of the form \eqref{eq:spd_jump}. We discretize the time at intervals $\text{d}t$. At each time step, we call $\ket{\psi_t}$ the state before the time step is performed, define $\hat{c}^{\phantom \dagger}_{2j-1} = \hat{c}^{\phantom \dagger}_{j,\mathrm{\scriptstyle A}}$, $\hat{c}^{\phantom \dagger}_{2j} = \hat{c}^{\phantom \dagger}_{j,B}$ and the time-dependent covariance matrix and anomalous covariance matrix, respectively as
\begin{equation}
  G_{i,j}(t) \equiv \braket{\psi_t|{\hat{c}^{\dagger}}_{j} \hat{c}^{\phantom \dagger}_{i}|\psi_t}\,,\quad F_{i\,j}(t) \equiv \braket{\psi_t|\hat{c}^{\phantom \dagger}_{j} \hat{c}^{\phantom \dagger}_{i}|\psi_t}\,.
\end{equation}
In order to perform the time step, we act on $\ket{\psi_t}$. There are some mutually exclusive possibilities
\begin{itemize}
  \item For one $l=1,\,\ldots,\,N$, with probability $$\text{dp}_l = \gamma \text{d}t \braket{\psi_t|{\hat{c}^{\dagger}}_{l\,A} \hat{c}^{\phantom \dagger}_{l\,A}|\psi_t} = \gamma \text{d}t G_{2l-1,\,2l-1}(t)\,,$$ we apply the transformation
  \begin{equation}
    \ket{\psi_t} \longrightarrow \ket{\psi_{t+\text{d}t}} = \frac{\hat{c}^{\phantom \dagger}_{2l-1}\ket{\psi_t}}{||\hat{c}^{\phantom \dagger}_{2l-1}\ket{\psi_t}||}\,.
  \end{equation}
  Assuming that the state remains Gaussian and that the Wick's theorem holds, this equation translates into an evolution equation for the covariance matrix
  \begin{widetext}
  \begin{equation}
    G_{i\,j}(t+\text{d}t) = \braket{\psi_{t+\text{d}t}|{\hat{c}^{\dagger}}_{j} \hat{c}^{\phantom \dagger}_{i}|\psi_{t+\text{d}t}}= \frac{\braket{\psi_t|{\hat{c}^{\dagger}}_{2l-1} {\hat{c}^{\dagger}}_{j} \hat{c}^{\phantom \dagger}_{i} \hat{c}^{\phantom \dagger}_{2l-1}|\psi_t}}{G_{2l-1,\,2l-1}(t)} = G_{i\,j}(t) + \frac{F_{2l-1,\,j}^*(t)F_{2l-1,\,i}(t)-G_{i,\,2l-1}(t)G_{2l-1,j}(t)}{G_{2l-1,\,2l-1}(t)}
\end{equation}
\begin{equation}
    F_{i\,j}(t+\text{d}t) = \braket{\psi_{t+\text{d}t}|\hat{c}^{\phantom \dagger}_{j} \hat{c}^{\phantom \dagger}_{i}|\psi_{t+\text{d}t}}= \frac{\braket{\psi_t|{\hat{c}^{\dagger}}_{2l-1} \hat{c}^{\phantom \dagger}_{j} \hat{c}^{\phantom \dagger}_{i} \hat{c}^{\phantom \dagger}_{2l-1}|\psi_t}}{G_{2l-1,\,2l-1}(t)} = F_{i\,j}(t) + \frac{G_{j,\,2l-1}(t)F_{2l-1,\,i}(t) - G_{i,\,2l-1}(t) F_{2l-1,\,j}(t)}{G_{2l-1,\,2l-1}(t)}\,.
\end{equation}
   \end{widetext}
  \item For one $l=1,\,\ldots,\,N$, with probability $$\text{dq}_l = \gamma \text{d}t \braket{\psi_t|\hat{c}^{\phantom \dagger}_{l\,B} {\hat{c}^{\dagger}}_{l\,B}|\psi_t} = \gamma \text{d}t[1- G_{2l,\,2l}(t)]\,,$$ we apply the transformation
  \begin{equation*}
    \ket{\psi_t} \longrightarrow \ket{\psi_{t+\text{d}t}} = \frac{{\hat{c}^{\dagger}}_{2l}\ket{\psi_t}}{||{\hat{c}^{\dagger}}_{2l}\ket{\psi_t}||}\,.
  \end{equation*}
  This reflects into a transformation for the correlation functions
  \begin{widetext}
  \begin{equation}
    G_{i\,j}(t+\text{d}t) = \braket{\psi_{t+\text{d}t}|{\hat{c}^{\dagger}}_{j} \hat{c}^{\phantom \dagger}_{i}|\psi_{t+\text{d}t}}= \frac{\braket{\psi_t|\hat{c}^{\phantom \dagger}_{2l} {\hat{c}^{\dagger}}_{j} \hat{c}^{\phantom \dagger}_{i} {\hat{c}^{\dagger}}_{2l}|\psi_t}}{1-G_{2l,\,2l}(t)} =G_{i\,j}(t) + \frac{[\delta_{2l\,j} - G_{2l\,j}(t)][\delta_{i\,2l}-G_{i\,2l}(t)]-F_{i\,2l}(t)F_{j\,2l}^*(t)}{1-G_{2l,\,2l}(t)}
  \end{equation}
  \begin{equation}
      F_{i\,j}(t+\Delta t) = \braket{\psi_{t+\Delta t}|\hat{c}^{\phantom \dagger}_{j} \hat{c}^{\phantom \dagger}_{i}|\psi_{t+\Delta t}}= \frac{\braket{\psi_t|\hat{c}^{\phantom \dagger}_{2l} \hat{c}^{\phantom \dagger}_{j} \hat{c}^{\phantom \dagger}_{i} {\hat{c}^{\dagger}}_{2l}|\psi_t}}{1-G_{2l,\,2l}(t)} =F_{i\,j}(t) + \frac{F_{j\,2l}(t)[\delta_{j\,2l}-G_{i\,2l}]-F_{i\,2l}[\delta_{j\,2l}-G_{j\,2l}(t)]}{1-G_{2l,\,2l}(t)}\,.
  \end{equation}
  \end{widetext}
\end{itemize}

\subsubsection{SBD}
\paragraph{Non-Hermitian evolution.}
In the symmetry-breaking case we have
\begin{equation}
    \hat{\Lambda}=\frac{\gamma}{2}\left[\sum_{\alpha=0}^{L-2}{\hat{c}^{\dagger}}_{\alpha}\hat{c}^{\phantom \dagger}_{\alpha}+{\hat{c}^{\dagger}}_{\alpha+1}\hat{c}^{\phantom \dagger}_{\alpha+1}+{\hat{c}^{\dagger}}_{\alpha+1}\hat{c}^{\phantom \dagger}_{\alpha}+{\hat{c}^{\dagger}}_{\alpha}\hat{c}^{\phantom \dagger}_{\alpha+1}\right].
\end{equation}
From which it is straightforward to derive the non-Hermitian evolution of \textbf{G} as done in the SPD dynamics.

\paragraph{Quantum Jump.}
In the symmetry-breaking case we have the jump operators of the form \eqref{eq:sbd_jump}. In this case, we can write $\hat{c}^{\phantom \dagger}_{2j-1}=\hat{c}^{\phantom \dagger}_{jA}$, $\hat{c}^{\phantom \dagger}_{2j}=\hat{c}^{\phantom \dagger}_{jB}$.
The mutually exclusive possibilities are
\begin{itemize}
    \item For one $l=1,...,N$ with probability 
    \begin{widetext}
    \begin{equation}   
        \text{dp}_l=\gamma \text{d}t\braket{\psi_t|\hat{L}^{\dagger}_{2l-1} \hat{L}_{2l-1}|\psi_t}=\gamma \text{d}t\left[G_{2l-1,2l-1}+G_{2l,2l}+G_{2l,2l-1}+G_{2l-1,2l}\right]=\gamma \text{d}t N
    \end{equation}
    \end{widetext}
    we apply the transformation 
    \begin{equation*}
        \ket{\psi_t}\rightarrow\ket{\psi_{t+\text{d}t}}=\frac{\hat{L}_{2l-1}\ket{\psi_t}}{||\hat{L}_{2l-1}\ket{\psi_t}||}=\frac{\left(\hat{c}^{\phantom \dagger}_{2l-1}+\hat{c}^{\phantom \dagger}_{2l}\right)\ket{\psi_t}}{||\hat{c}^{\phantom \dagger}_{2l-1}+\hat{c}^{\phantom \dagger}_{2l}\ket{\psi_t}||}
    \end{equation*}
    So that the correlation matrix, neglecting the anomalous correlations, becomes
    \begin{widetext}
    \begin{equation}
            G_{i,j}(t+\text{d}t)=G_{ij}-\frac{1}{N}\left\{G_{i,2l-1}G_{2l-1,j}+G_{i,2l}G_{2l,j}+G_{i,2l}G_{2l-1,j}+G_{i,2l-1}G_{2l,j}\right\}
    \end{equation}
    \end{widetext}

    \item For one $l=1,...,N$ with probability 
    \begin{widetext}
    \begin{equation}   
        \text{dq}_l=\gamma \text{d}t\left[G_{2l,2l}+G_{2l+1,2l+1}+G_{2l+1,2l}+G_{2l+1,2l}\right]=\gamma \text{d}t N
    \end{equation}
    \end{widetext}
    we apply the transformation 
    \begin{equation*}
        \ket{\psi_t}\rightarrow\ket{\psi_{t+\text{d}t}}=\frac{\hat{L}_{2l}\ket{\psi_t}}{||\hat{L}_{2l}\ket{\psi_t}||}=\frac{\left(\hat{c}^{\phantom \dagger}_{2l}+\hat{c}^{\phantom \dagger}_{2l+1}\right)\ket{\psi_t}}{||\hat{c}^{\phantom \dagger}_{2l}+\hat{c}^{\phantom \dagger}_{2l+1}\ket{\psi_t}||}
    \end{equation*}
    So that the correlation matrix, still neglecting the anomalous correlations, becomes
    \begin{widetext}
    \begin{equation}
            G_{i,j}(t+\text{d}t)=G_{ij}-\frac{1}{N}\left\{G_{i,2l}G_{2l,j}+G_{i,2l+1}G_{2l+1,j}+G_{i,2l+1}G_{2l,j}+G_{i,2l}G_{2l+1,j}\right\}.
    \end{equation}
    \end{widetext}
\end{itemize}

\section{\label{app:effect_of_jumps} Effect of the first jump on a $4$-sites chain}

In the following, we report the effect of the first jump on the ground state of the Hamiltonian in the perfectly-dimerized topological state for a simple example of 4-sites chain. Although extremely simplistic, this example helps understanding the origin of some of the peaks highlighted in the histograms of the index-resolved statistical analysis, in particular the peak at $\Delta S^D = -2$ for the global SPD dynamics and that at $\Delta S^D\simeq-0.38$ for the global SBD dynamics. The simplification of considering the jump as occurring as soon as the initial state is prepared is justified by the observation that the non-Hermitian evolution statistically does not drastically change the amount of entanglement of the initial state.

Let us consider a chain of two unit cells and four sites. We naturally choose a bipartition of the system such that the first two sites belong to $\mathcal{A}$ and the last two belong to $B$. Let us consider the local jump operator that acts on the first site of the chain so to give the non-normalized state 
\begin{equation}
\begin{split}
    \ket{\psi_\textrm{jump}^{\textrm{SPD}}}&=\hat{c}_1 \ket{\mathrm{GS}}=\hat{c}_1 \frac{1}{2}\left(\hat{c}^\dagger_1+\hat{c}^\dagger_4\right)\left(\hat{c}^\dagger_2+\hat{c}^\dagger_3\right)\ket{0}=\\
    &=\frac{1}{2}\left(\hat{c}^\dagger_2+\hat{c}^\dagger_3\right)\ket{0},
    \end{split}
\end{equation}
which we will write in the following as 
\begin{equation}
\ket{\psi_\textrm{jump}^{\textrm{SPD}}}=\frac{1}{2}\left(\ket{1_2}+\ket{1_3}\right).
\end{equation}
In this notation, each state has implicit zero-occupation values on the non-written sites, e.g., $\ket{1_2}$ is implicitly equivalent to $\ket{0_1 1_2 0_3 0_4}$. The corresponding and suitably normalized density matrix will be 
\begin{equation}
\begin{split}
    &\hat{\rho}^{\textrm{SPD}}=\ket{\psi_\textrm{jump}^{\textrm{SPD}}}\bra{\psi_\textrm{jump}^{\textrm{SPD}}}=\\
    &=\frac{1}{2}\left(\ket{1_2}\bra{1_2}+\ket{1_2}\bra{1_3}+\ket{1_3}\bra{1_2}+\ket{1_3}\bra{1_3}\right).
    \end{split}
\end{equation}
 We can then compute the reduced density matrix, so to know the effect of the jump on one of the terms that make up the DEE. In doing the partial trace of this fermionic system we do not encounter ambiguities, since the states we deal with are always superposition of kets of a number of fermions of same (odd) parity~\cite{Friis_2016, Friis_2013}. Hence, we trace out the degrees of freedom of sites $3$ and $4$ and obtain 
\begin{equation}
    \hat{\rho}_\mathcal{A}^{\textrm{SPD}}=\frac{1}{2}\left(\ket{0_1 1_2}\bra{0_1 1_2 }+\ket{0_1 0_2}\bra{0_1 0_2}\right),
\end{equation}
and its eigenvalues are $\{1/2,1/2,0,0\}$ so that
$S_\mathcal{A}=1$ and $\Delta S_\mathcal{A}=-1$. Tracing out the degrees of freedom of sites $1$,$2$ and $3$ we can also compute the variation of $S_{\mathcal{A}\cup\mathcal{B}}$ since
\begin{equation}
    \hat{\rho}_{\mathcal{A}\cup\mathcal{B}}^{\textrm{SPD}}=\frac{1}{2}\left(\ket{0_3}\bra{0_3}+\ket{1_3}\bra{1_3}\right)
\end{equation}
so that $S_{\mathcal{A}\cup\mathcal{B}}=1$ and $\Delta S_{\mathcal{A}\cup\mathcal{B}}=0$. We can straightforwardly show that $\Delta S_\mathcal{B}=-1$ and $\Delta S_{\mathcal{A}\cap\mathcal{B}}=0$ so to understand why the peak at $\Delta S^D=-2$ in Figs.~\ref{fig:time_stat_global}(a) and \ref{fig:index_stat}(a) is due to the jump localized at one of the two edge sites.

On the other hand, if the SBD jump operator acts on the first site, without loss of generality, it simultaneously affects the second site, so that the effect of the first jump on the boundary is completely different. In this case, after the jump, we have the non-normalized state
\begin{equation}
\begin{split}
    &\ket{\psi_{\textrm{jump}}^{\textrm{SBD}}}= \frac{1}{2}\left(\hat{c}_1+\hat{c}_2\right)\left(\hat{c}_1^\dagger+\hat{c}_4^\dagger\right)\left(\hat{c}_2^\dagger+\hat{c}_3^\dagger\right)\ket{0}=\\
    &=\frac{1}{2}\left[\left(\hat{c}_2^\dagger+\hat{c}_3^\dagger\right)-\left(\hat{c}_1^\dagger+\hat{c}_4^\dagger\right)\right]\ket{0}.
\end{split}
\end{equation}
In this case, there is still a trace of the long-range-entangled Bell-like state $\frac{1}{\sqrt{2}}\left(\hat{c}_1^\dagger+\hat{c}_4^\dagger\right)$, which causes a totally different and non-discrete variation in the entanglement contributions of the DEE. The corresponding normalized density matrix is
\begin{equation}
\begin{split}
    &\hat{\rho}_\textrm{jump}^\textrm{SBD}=\frac{1}{4}\left[\ket{1_2}\bra{1_2}+\ket{1_2}\bra{1_3}-\ket{1_2}\bra{1_1}+\right.\\
    &-\ket{1_2}\bra{1_4}+\ket{1_3}\bra{1_2}+\ket{1_3}\bra{1_3}-\ket{1_3}\bra{1_1}+\\
    &-\ket{1_3}\bra{1_4}-\ket{1_1}\bra{1_2}-\ket{1_1}\bra{1_3}+\ket{1_1}\bra{1_1}+\\
    &+\ket{1_1}\bra{1_4}-\ket{1_4}\bra{1_2}-\ket{1_4}\bra{1_3}+\ket{1_4}\bra{1_1}+\\
    &\left.+\ket{1_4}\bra{1_4}\right].
\end{split}
\end{equation}
By tracing out the degrees of freedom of sites $3$ and $4$, we get
\begin{equation}
\begin{split}
    &\hat{\rho}_{\mathcal{A}}^\textrm{SBD}=\frac{1}{4}\left[2\ket{0_1 0_2}\bra{0_1 0_2}+\ket{0_1 1_2}\bra{0_1 1_2}+\right.\\
    &\left.-\ket{0_1 1_2}\bra{1_1 0_2}-\ket{1_1 0_2}\bra{0_1 1_2}+\ket{1_1 0_2}\bra{1_1 0_2}\right],
\end{split}
\end{equation}
whose eigenvlaues are $\{1/2,1/2,0,0\}$ so that $S_\mathcal{A}=1$ and $\Delta S_\mathcal{A}=-1$. This result is interesting since, after the occurrence of the jump on the boundary, one would expect $S_\mathcal{A}$ to go to $0$ owing to the action on both the sites $1$ and $2$ that are part of two distinct Bell-like pairs. On the contrary, the peculiar form of $\ket{\psi_\textrm{jump}^\textrm{SBD}}$ mixes the things in an unexpectable way. Furthermore, we can trace out the degrees of freedom of sites $1$, $2$ and $3$ and see that
\begin{equation}
    \hat{\rho}_{\mathcal{A}\cup\mathcal{B}}^\textrm{SBD}=\frac{1}{4}\left[3\ket{0_3}\bra{0_3}+\ket{1_3}\bra{1_3}\right]
\end{equation}
so that $S_{\mathcal{A}\cup\mathcal{B}}=-\frac{3}{4}\log_2(3)+2\simeq0.81$ and $\Delta S_{\mathcal{A}\cup\mathcal{B}}=-\frac{3}{4}\log_2(3)+1\simeq-0.19$. Again, counter-intuitively, the variation is not quantized, and we have numerically verified that, in the limit of perfectly-dimerized chain, this is not a size-dependent result. Again, it is straightforward to show that $\Delta S_\mathcal{B}\simeq-1$ and $\Delta S_{\mathcal{A}\cap\mathcal{B}}=-\frac{3}{4}\log_2(3)+1\simeq-0.19$ so that $\Delta S^D=2(-\frac{3}{4}\log_2(3)+2)\simeq1.62$. This result is mainly useful to understand that the variation of the DEE in the SBD dynamics is not discrete when the first jump occurs. At the same time, analyzing a single trajectory at larger sizes shows that the variation, although still not discrete, is different and such that $\Delta S^D=2-2(-\frac{3}{4}\log_2(3)+2)\simeq-0.38$ because $\Delta S_\mathcal{A}=-\frac{3}{4}\log_2(3)+1\simeq-0.19$, $\Delta S_\mathcal{B}=-\frac{3}{4}\log_2(3)+1\simeq-0.19$, $\Delta S_{\mathcal{A}\cup\mathcal{B}}\simeq 0$ and $\Delta S_{\mathcal{A}\cap\mathcal{B}}\simeq 0$. This considerations help us in the interpretation of the peak at $\Delta S^D=2-2(-\frac{3}{4}\log_2(3)+2)\simeq-0.38$ in Figs.~\ref{fig:time_stat_global}(c) and \ref{fig:index_stat}(b). At the same time, we have numerically verified that the peak at $\Delta S^D=-2$ for the SBD dynamics is due to jumps occurring at the boundary after several other jumps, and thus are not directly involved in the initial destruction of the long-range entangled Bell-like state.


\begin{thebibliography}{79}%
\makeatletter
\providecommand \@ifxundefined [1]{%
 \@ifx{#1\undefined}
}%
\providecommand \@ifnum [1]{%
 \ifnum #1\expandafter \@firstoftwo
 \else \expandafter \@secondoftwo
 \fi
}%
\providecommand \@ifx [1]{%
 \ifx #1\expandafter \@firstoftwo
 \else \expandafter \@secondoftwo
 \fi
}%
\providecommand \natexlab [1]{#1}%
\providecommand \enquote  [1]{``#1''}%
\providecommand \bibnamefont  [1]{#1}%
\providecommand \bibfnamefont [1]{#1}%
\providecommand \citenamefont [1]{#1}%
\providecommand \href@noop [0]{\@secondoftwo}%
\providecommand \href [0]{\begingroup \@sanitize@url \@href}%
\providecommand \@href[1]{\@@startlink{#1}\@@href}%
\providecommand \@@href[1]{\endgroup#1\@@endlink}%
\providecommand \@sanitize@url [0]{\catcode `\\12\catcode `\$12\catcode
  `\&12\catcode `\#12\catcode `\^12\catcode `\_12\catcode `\%12\relax}%
\providecommand \@@startlink[1]{}%
\providecommand \@@endlink[0]{}%
\providecommand \url  [0]{\begingroup\@sanitize@url \@url }%
\providecommand \@url [1]{\endgroup\@href {#1}{\urlprefix }}%
\providecommand \urlprefix  [0]{URL }%
\providecommand \Eprint [0]{\href }%
\providecommand \doibase [0]{https://doi.org/}%
\providecommand \selectlanguage [0]{\@gobble}%
\providecommand \bibinfo  [0]{\@secondoftwo}%
\providecommand \bibfield  [0]{\@secondoftwo}%
\providecommand \translation [1]{[#1]}%
\providecommand \BibitemOpen [0]{}%
\providecommand \bibitemStop [0]{}%
\providecommand \bibitemNoStop [0]{.\EOS\space}%
\providecommand \EOS [0]{\spacefactor3000\relax}%
\providecommand \BibitemShut  [1]{\csname bibitem#1\endcsname}%
\let\auto@bib@innerbib\@empty
%</preamble>
\bibitem [{\citenamefont {Bernevig}\ and\ \citenamefont
  {Hughes}(2013)}]{bernevig2013topological}%
  \BibitemOpen
  \bibfield  {author} {\bibinfo {author} {\bibfnamefont {B.}~\bibnamefont
  {Bernevig}}\ and\ \bibinfo {author} {\bibfnamefont {T.}~\bibnamefont
  {Hughes}},\ }\href {https://books.google.it/books?id=wOn7JHSSxrsC} {\emph
  {\bibinfo {title} {{Topological Insulators and Topological
  Superconductors}}}}\ (\bibinfo  {publisher} {Princeton University Press},\
  \bibinfo {year} {2013})\BibitemShut {NoStop}%
\bibitem [{\citenamefont {Ashida}\ \emph {et~al.}(2020)\citenamefont {Ashida},
  \citenamefont {Gong},\ and\ \citenamefont {Ueda}}]{Ashida_2020}%
  \BibitemOpen
  \bibfield  {author} {\bibinfo {author} {\bibfnamefont {Y.}~\bibnamefont
  {Ashida}}, \bibinfo {author} {\bibfnamefont {Z.}~\bibnamefont {Gong}},\ and\
  \bibinfo {author} {\bibfnamefont {M.}~\bibnamefont {Ueda}},\ }\bibfield
  {title} {\bibinfo {title} {{Non-Hermitian physics}},\ }\href
  {https://doi.org/10.1080/00018732.2021.1876991} {\bibfield  {journal}
  {\bibinfo  {journal} {Advances in Physics}\ }\textbf {\bibinfo {volume}
  {69}},\ \bibinfo {pages} {249–435} (\bibinfo {year} {2020})}\BibitemShut
  {NoStop}%
\bibitem [{\citenamefont {Okuma}\ and\ \citenamefont
  {Sato}(2023)}]{okuma2023nonhermitian}%
  \BibitemOpen
  \bibfield  {author} {\bibinfo {author} {\bibfnamefont {N.}~\bibnamefont
  {Okuma}}\ and\ \bibinfo {author} {\bibfnamefont {M.}~\bibnamefont {Sato}},\
  }\bibfield  {title} {\bibinfo {title} {{Non-Hermitian Topological Phenomena:
  A Review}},\ }\href
  {https://doi.org/10.1146/annurev-conmatphys-040521-033133} {\bibfield
  {journal} {\bibinfo  {journal} {Annual Review of Condensed Matter Physics}\
  }\textbf {\bibinfo {volume} {14}},\ \bibinfo {pages} {83–107} (\bibinfo
  {year} {2023})}\BibitemShut {NoStop}%
\bibitem [{\citenamefont {Diehl}\ \emph {et~al.}(2011)\citenamefont {Diehl},
  \citenamefont {Rico}, \citenamefont {Baranov},\ and\ \citenamefont
  {Zoller}}]{Diehl_2011}%
  \BibitemOpen
  \bibfield  {author} {\bibinfo {author} {\bibfnamefont {S.}~\bibnamefont
  {Diehl}}, \bibinfo {author} {\bibfnamefont {E.}~\bibnamefont {Rico}},
  \bibinfo {author} {\bibfnamefont {M.~A.}\ \bibnamefont {Baranov}},\ and\
  \bibinfo {author} {\bibfnamefont {P.}~\bibnamefont {Zoller}},\ }\bibfield
  {title} {\bibinfo {title} {{Topology by dissipation in atomic quantum
  wires}},\ }\href {https://doi.org/10.1038/nphys2106} {\bibfield  {journal}
  {\bibinfo  {journal} {Nature Physics}\ }\textbf {\bibinfo {volume} {7}},\
  \bibinfo {pages} {971–977} (\bibinfo {year} {2011})}\BibitemShut {NoStop}%
\bibitem [{\citenamefont {Bardyn}\ \emph {et~al.}(2013)\citenamefont {Bardyn},
  \citenamefont {Baranov}, \citenamefont {Kraus}, \citenamefont {Rico},
  \citenamefont {İmamoğlu}, \citenamefont {Zoller},\ and\ \citenamefont
  {Diehl}}]{bardyn2013topologybydissipation}%
  \BibitemOpen
  \bibfield  {author} {\bibinfo {author} {\bibfnamefont {C.-E.}\ \bibnamefont
  {Bardyn}}, \bibinfo {author} {\bibfnamefont {M.~A.}\ \bibnamefont {Baranov}},
  \bibinfo {author} {\bibfnamefont {C.~V.}\ \bibnamefont {Kraus}}, \bibinfo
  {author} {\bibfnamefont {E.}~\bibnamefont {Rico}}, \bibinfo {author}
  {\bibfnamefont {A.}~\bibnamefont {İmamoğlu}}, \bibinfo {author}
  {\bibfnamefont {P.}~\bibnamefont {Zoller}},\ and\ \bibinfo {author}
  {\bibfnamefont {S.}~\bibnamefont {Diehl}},\ }\bibfield  {title} {\bibinfo
  {title} {{Topology by dissipation}},\ }\href
  {https://doi.org/10.1088/1367-2630/15/8/085001} {\bibfield  {journal}
  {\bibinfo  {journal} {New Journal of Physics}\ }\textbf {\bibinfo {volume}
  {15}},\ \bibinfo {pages} {085001} (\bibinfo {year} {2013})}\BibitemShut
  {NoStop}%
\bibitem [{\citenamefont {Lieu}\ \emph {et~al.}(2020)\citenamefont {Lieu},
  \citenamefont {McGinley},\ and\ \citenamefont {Cooper}}]{cooper}%
  \BibitemOpen
  \bibfield  {author} {\bibinfo {author} {\bibfnamefont {S.}~\bibnamefont
  {Lieu}}, \bibinfo {author} {\bibfnamefont {M.}~\bibnamefont {McGinley}},\
  and\ \bibinfo {author} {\bibfnamefont {N.~R.}\ \bibnamefont {Cooper}},\
  }\bibfield  {title} {\bibinfo {title} {{Tenfold Way for Quadratic
  Lindbladians}},\ }\href {https://doi.org/10.1103/PhysRevLett.124.040401}
  {\bibfield  {journal} {\bibinfo  {journal} {Phys. Rev. Lett.}\ }\textbf
  {\bibinfo {volume} {124}},\ \bibinfo {pages} {040401} (\bibinfo {year}
  {2020})}\BibitemShut {NoStop}%
\bibitem [{\citenamefont {Altland}\ \emph {et~al.}(2021)\citenamefont
  {Altland}, \citenamefont {Fleischhauer},\ and\ \citenamefont
  {Diehl}}]{Altland_2021}%
  \BibitemOpen
  \bibfield  {author} {\bibinfo {author} {\bibfnamefont {A.}~\bibnamefont
  {Altland}}, \bibinfo {author} {\bibfnamefont {M.}~\bibnamefont
  {Fleischhauer}},\ and\ \bibinfo {author} {\bibfnamefont {S.}~\bibnamefont
  {Diehl}},\ }\bibfield  {title} {\bibinfo {title} {{Symmetry Classes of Open
  Fermionic Quantum Matter}},\ }\href
  {https://doi.org/10.1103/PhysRevX.11.021037} {\bibfield  {journal} {\bibinfo
  {journal} {Phys. Rev. X}\ }\textbf {\bibinfo {volume} {11}},\ \bibinfo
  {pages} {021037} (\bibinfo {year} {2021})}\BibitemShut {NoStop}%
\bibitem [{\citenamefont {Altland}\ and\ \citenamefont
  {Zirnbauer}(1997)}]{altland}%
  \BibitemOpen
  \bibfield  {author} {\bibinfo {author} {\bibfnamefont {A.}~\bibnamefont
  {Altland}}\ and\ \bibinfo {author} {\bibfnamefont {M.~R.}\ \bibnamefont
  {Zirnbauer}},\ }\bibfield  {title} {\bibinfo {title} {{Nonstandard symmetry
  classes in mesoscopic normal-superconducting hybrid structures}},\ }\href
  {https://doi.org/10.1103/physrevb.55.1142} {\bibfield  {journal} {\bibinfo
  {journal} {Physical Review B}\ }\textbf {\bibinfo {volume} {55}},\ \bibinfo
  {pages} {1142} (\bibinfo {year} {1997})}\BibitemShut {NoStop}%
\bibitem [{\citenamefont {Kawabata}\ \emph {et~al.}(2019)\citenamefont
  {Kawabata}, \citenamefont {Shiozaki}, \citenamefont {Ueda},\ and\
  \citenamefont {Sato}}]{Kawabata_2019}%
  \BibitemOpen
  \bibfield  {author} {\bibinfo {author} {\bibfnamefont {K.}~\bibnamefont
  {Kawabata}}, \bibinfo {author} {\bibfnamefont {K.}~\bibnamefont {Shiozaki}},
  \bibinfo {author} {\bibfnamefont {M.}~\bibnamefont {Ueda}},\ and\ \bibinfo
  {author} {\bibfnamefont {M.}~\bibnamefont {Sato}},\ }\bibfield  {title}
  {\bibinfo {title} {{Symmetry and Topology in Non-Hermitian Physics}},\ }\href
  {https://doi.org/10.1103/PhysRevX.9.041015} {\bibfield  {journal} {\bibinfo
  {journal} {Phys. Rev. X}\ }\textbf {\bibinfo {volume} {9}},\ \bibinfo {pages}
  {041015} (\bibinfo {year} {2019})}\BibitemShut {NoStop}%
\bibitem [{\citenamefont {Bernard}\ and\ \citenamefont
  {LeClair}(2002)}]{Bernard_2002}%
  \BibitemOpen
  \bibfield  {author} {\bibinfo {author} {\bibfnamefont {D.}~\bibnamefont
  {Bernard}}\ and\ \bibinfo {author} {\bibfnamefont {A.}~\bibnamefont
  {LeClair}},\ }\bibinfo {title} {{A Classification of Non-Hermitian Random
  Matrices}},\ in\ \href {https://doi.org/10.1007/978-94-010-0514-2_19} {\emph
  {\bibinfo {booktitle} {Statistical Field Theories}}}\ (\bibinfo  {publisher}
  {Springer Netherlands},\ \bibinfo {year} {2002})\ p.\ \bibinfo {pages}
  {207–214}\BibitemShut {NoStop}%
\bibitem [{\citenamefont {Prosen}(2012)}]{Prosen_2012}%
  \BibitemOpen
  \bibfield  {author} {\bibinfo {author} {\bibfnamefont {T.}~\bibnamefont
  {Prosen}},\ }\bibfield  {title} {\bibinfo {title}
  {{$\mathbb{P}\mathbb{T}$-Symmetric Quantum Liouvillean Dynamics}},\ }\href
  {https://doi.org/10.1103/PhysRevLett.109.090404} {\bibfield  {journal}
  {\bibinfo  {journal} {Phys. Rev. Lett.}\ }\textbf {\bibinfo {volume} {109}},\
  \bibinfo {pages} {090404} (\bibinfo {year} {2012})}\BibitemShut {NoStop}%
\bibitem [{\citenamefont {Buca}\ and\ \citenamefont
  {Prosen}(2012)}]{Buča_2012}%
  \BibitemOpen
  \bibfield  {author} {\bibinfo {author} {\bibfnamefont {B.}~\bibnamefont
  {Buca}}\ and\ \bibinfo {author} {\bibfnamefont {T.}~\bibnamefont {Prosen}},\
  }\bibfield  {title} {\bibinfo {title} {{A note on symmetry reductions of the
  Lindblad equation: transport in constrained open spin chains}},\ }\href
  {https://doi.org/10.1088/1367-2630/14/7/073007} {\bibfield  {journal}
  {\bibinfo  {journal} {New Journal of Physics}\ }\textbf {\bibinfo {volume}
  {14}},\ \bibinfo {pages} {073007} (\bibinfo {year} {2012})}\BibitemShut
  {NoStop}%
\bibitem [{\citenamefont {Albert}\ and\ \citenamefont
  {Jiang}(2014)}]{Liang_2014}%
  \BibitemOpen
  \bibfield  {author} {\bibinfo {author} {\bibfnamefont {V.~V.}\ \bibnamefont
  {Albert}}\ and\ \bibinfo {author} {\bibfnamefont {L.}~\bibnamefont {Jiang}},\
  }\bibfield  {title} {\bibinfo {title} {{Symmetries and conserved quantities
  in Lindblad master equations}},\ }\href
  {https://doi.org/10.1103/PhysRevA.89.022118} {\bibfield  {journal} {\bibinfo
  {journal} {Phys. Rev. A}\ }\textbf {\bibinfo {volume} {89}},\ \bibinfo
  {pages} {022118} (\bibinfo {year} {2014})}\BibitemShut {NoStop}%
\bibitem [{\citenamefont {Huber}\ \emph {et~al.}(2020)\citenamefont {Huber},
  \citenamefont {Kirton}, \citenamefont {Rotter},\ and\ \citenamefont
  {Rabl}}]{Rabl_2020}%
  \BibitemOpen
  \bibfield  {author} {\bibinfo {author} {\bibfnamefont {J.}~\bibnamefont
  {Huber}}, \bibinfo {author} {\bibfnamefont {P.}~\bibnamefont {Kirton}},
  \bibinfo {author} {\bibfnamefont {S.}~\bibnamefont {Rotter}},\ and\ \bibinfo
  {author} {\bibfnamefont {P.}~\bibnamefont {Rabl}},\ }\bibfield  {title}
  {\bibinfo {title} {{Emergence of PT-symmetry breaking in open quantum
  systems}},\ }\href {https://doi.org/10.21468/SciPostPhys.9.4.052} {\bibfield
  {journal} {\bibinfo  {journal} {SciPost Phys.}\ }\textbf {\bibinfo {volume}
  {9}},\ \bibinfo {pages} {052} (\bibinfo {year} {2020})}\BibitemShut {NoStop}%
\bibitem [{\citenamefont {Kawasaki}\ \emph {et~al.}(2022)\citenamefont
  {Kawasaki}, \citenamefont {Mochizuki},\ and\ \citenamefont
  {Obuse}}]{Kawasaki_2022}%
  \BibitemOpen
  \bibfield  {author} {\bibinfo {author} {\bibfnamefont {M.}~\bibnamefont
  {Kawasaki}}, \bibinfo {author} {\bibfnamefont {K.}~\bibnamefont
  {Mochizuki}},\ and\ \bibinfo {author} {\bibfnamefont {H.}~\bibnamefont
  {Obuse}},\ }\bibfield  {title} {\bibinfo {title} {{Topological phases
  protected by shifted sublattice symmetry in dissipative quantum systems}},\
  }\href {https://doi.org/10.1103/PhysRevB.106.035408} {\bibfield  {journal}
  {\bibinfo  {journal} {Phys. Rev. B}\ }\textbf {\bibinfo {volume} {106}},\
  \bibinfo {pages} {035408} (\bibinfo {year} {2022})}\BibitemShut {NoStop}%
\bibitem [{\citenamefont {Mao}\ \emph {et~al.}(2024)\citenamefont {Mao},
  \citenamefont {Yang},\ and\ \citenamefont {Zhai}}]{Mao_2024}%
  \BibitemOpen
  \bibfield  {author} {\bibinfo {author} {\bibfnamefont {L.}~\bibnamefont
  {Mao}}, \bibinfo {author} {\bibfnamefont {F.}~\bibnamefont {Yang}},\ and\
  \bibinfo {author} {\bibfnamefont {H.}~\bibnamefont {Zhai}},\ }\bibfield
  {title} {\bibinfo {title} {{Symmetry-preserving quadratic Lindbladian and
  dissipation driven topological transitions in Gaussian states}},\ }\href
  {https://doi.org/10.1088/1361-6633/ad44d4} {\bibfield  {journal} {\bibinfo
  {journal} {Reports on Progress in Physics}\ }\textbf {\bibinfo {volume}
  {87}},\ \bibinfo {pages} {070501} (\bibinfo {year} {2024})}\BibitemShut
  {NoStop}%
\bibitem [{\citenamefont {Kunst}\ \emph {et~al.}(2018)\citenamefont {Kunst},
  \citenamefont {Edvardsson}, \citenamefont {Budich},\ and\ \citenamefont
  {Bergholtz}}]{Kunst_2018}%
  \BibitemOpen
  \bibfield  {author} {\bibinfo {author} {\bibfnamefont {F.~K.}\ \bibnamefont
  {Kunst}}, \bibinfo {author} {\bibfnamefont {E.}~\bibnamefont {Edvardsson}},
  \bibinfo {author} {\bibfnamefont {J.~C.}\ \bibnamefont {Budich}},\ and\
  \bibinfo {author} {\bibfnamefont {E.~J.}\ \bibnamefont {Bergholtz}},\
  }\bibfield  {title} {\bibinfo {title} {{Biorthogonal Bulk-Boundary
  Correspondence in Non-Hermitian Systems}},\ }\href
  {https://doi.org/10.1103/PhysRevLett.121.026808} {\bibfield  {journal}
  {\bibinfo  {journal} {Phys. Rev. Lett.}\ }\textbf {\bibinfo {volume} {121}},\
  \bibinfo {pages} {026808} (\bibinfo {year} {2018})}\BibitemShut {NoStop}%
\bibitem [{\citenamefont {Gong}\ \emph {et~al.}(2018)\citenamefont {Gong},
  \citenamefont {Ashida}, \citenamefont {Kawabata}, \citenamefont {Takasan},
  \citenamefont {Higashikawa},\ and\ \citenamefont {Ueda}}]{Gong_2018}%
  \BibitemOpen
  \bibfield  {author} {\bibinfo {author} {\bibfnamefont {Z.}~\bibnamefont
  {Gong}}, \bibinfo {author} {\bibfnamefont {Y.}~\bibnamefont {Ashida}},
  \bibinfo {author} {\bibfnamefont {K.}~\bibnamefont {Kawabata}}, \bibinfo
  {author} {\bibfnamefont {K.}~\bibnamefont {Takasan}}, \bibinfo {author}
  {\bibfnamefont {S.}~\bibnamefont {Higashikawa}},\ and\ \bibinfo {author}
  {\bibfnamefont {M.}~\bibnamefont {Ueda}},\ }\bibfield  {title} {\bibinfo
  {title} {{Topological Phases of Non-Hermitian Systems}},\ }\href
  {https://doi.org/10.1103/PhysRevX.8.031079} {\bibfield  {journal} {\bibinfo
  {journal} {Phys. Rev. X}\ }\textbf {\bibinfo {volume} {8}},\ \bibinfo {pages}
  {031079} (\bibinfo {year} {2018})}\BibitemShut {NoStop}%
\bibitem [{\citenamefont {Shen}\ \emph {et~al.}(2018)\citenamefont {Shen},
  \citenamefont {Zhen},\ and\ \citenamefont {Fu}}]{Shen_2018}%
  \BibitemOpen
  \bibfield  {author} {\bibinfo {author} {\bibfnamefont {H.}~\bibnamefont
  {Shen}}, \bibinfo {author} {\bibfnamefont {B.}~\bibnamefont {Zhen}},\ and\
  \bibinfo {author} {\bibfnamefont {L.}~\bibnamefont {Fu}},\ }\bibfield
  {title} {\bibinfo {title} {{Topological Band Theory for Non-Hermitian
  Hamiltonians}},\ }\href {https://doi.org/10.1103/PhysRevLett.120.146402}
  {\bibfield  {journal} {\bibinfo  {journal} {Phys. Rev. Lett.}\ }\textbf
  {\bibinfo {volume} {120}},\ \bibinfo {pages} {146402} (\bibinfo {year}
  {2018})}\BibitemShut {NoStop}%
\bibitem [{\citenamefont {Viyuela}\ \emph {et~al.}(2014)\citenamefont
  {Viyuela}, \citenamefont {Rivas},\ and\ \citenamefont
  {Martin-Delgado}}]{Delgado_2014}%
  \BibitemOpen
  \bibfield  {author} {\bibinfo {author} {\bibfnamefont {O.}~\bibnamefont
  {Viyuela}}, \bibinfo {author} {\bibfnamefont {A.}~\bibnamefont {Rivas}},\
  and\ \bibinfo {author} {\bibfnamefont {M.~A.}\ \bibnamefont
  {Martin-Delgado}},\ }\bibfield  {title} {\bibinfo {title} {{Uhlmann Phase as
  a Topological Measure for One-Dimensional Fermion Systems}},\ }\href
  {https://doi.org/10.1103/PhysRevLett.112.130401} {\bibfield  {journal}
  {\bibinfo  {journal} {Phys. Rev. Lett.}\ }\textbf {\bibinfo {volume} {112}},\
  \bibinfo {pages} {130401} (\bibinfo {year} {2014})}\BibitemShut {NoStop}%
\bibitem [{\citenamefont {Viyuela}\ \emph {et~al.}(2018)\citenamefont
  {Viyuela}, \citenamefont {Rivas}, \citenamefont {Gasparinetti}, \citenamefont
  {Wallraff}, \citenamefont {Filipp},\ and\ \citenamefont
  {Martin-Delgado}}]{Viyuela_2018}%
  \BibitemOpen
  \bibfield  {author} {\bibinfo {author} {\bibfnamefont {O.}~\bibnamefont
  {Viyuela}}, \bibinfo {author} {\bibfnamefont {A.}~\bibnamefont {Rivas}},
  \bibinfo {author} {\bibfnamefont {S.}~\bibnamefont {Gasparinetti}}, \bibinfo
  {author} {\bibfnamefont {A.}~\bibnamefont {Wallraff}}, \bibinfo {author}
  {\bibfnamefont {S.}~\bibnamefont {Filipp}},\ and\ \bibinfo {author}
  {\bibfnamefont {M.~A.}\ \bibnamefont {Martin-Delgado}},\ }\bibfield  {title}
  {\bibinfo {title} {{Observation of topological Uhlmann phases with
  superconducting qubits}},\ }\href {https://doi.org/10.1038/s41534-017-0056-9}
  {\bibfield  {journal} {\bibinfo  {journal} {npj Quantum Information}\
  }\textbf {\bibinfo {volume} {4}},\ \bibinfo {pages} {10} (\bibinfo {year}
  {2018})}\BibitemShut {NoStop}%
\bibitem [{\citenamefont {Carollo}\ \emph {et~al.}(2017)\citenamefont
  {Carollo}, \citenamefont {Spagnolo},\ and\ \citenamefont
  {Valenti}}]{carollo2017uhlmanncurvaturedissipativephase}%
  \BibitemOpen
  \bibfield  {author} {\bibinfo {author} {\bibfnamefont {A.}~\bibnamefont
  {Carollo}}, \bibinfo {author} {\bibfnamefont {B.}~\bibnamefont {Spagnolo}},\
  and\ \bibinfo {author} {\bibfnamefont {D.}~\bibnamefont {Valenti}},\
  }\bibfield  {title} {\bibinfo {title} {{Uhlmann curvature in dissipative
  phase transitions}},\ }\href {https://arxiv.org/abs/1710.07560} {\  (\bibinfo
  {year} {2017})},\ \Eprint {https://arxiv.org/abs/1710.07560}
  {arXiv:1710.07560} \BibitemShut {NoStop}%
\bibitem [{\citenamefont {Bardyn}\ \emph {et~al.}(2018)\citenamefont {Bardyn},
  \citenamefont {Wawer}, \citenamefont {Altland}, \citenamefont
  {Fleischhauer},\ and\ \citenamefont {Diehl}}]{Bardyn_2018}%
  \BibitemOpen
  \bibfield  {author} {\bibinfo {author} {\bibfnamefont {C.-E.}\ \bibnamefont
  {Bardyn}}, \bibinfo {author} {\bibfnamefont {L.}~\bibnamefont {Wawer}},
  \bibinfo {author} {\bibfnamefont {A.}~\bibnamefont {Altland}}, \bibinfo
  {author} {\bibfnamefont {M.}~\bibnamefont {Fleischhauer}},\ and\ \bibinfo
  {author} {\bibfnamefont {S.}~\bibnamefont {Diehl}},\ }\bibfield  {title}
  {\bibinfo {title} {{Probing the Topology of Density Matrices}},\ }\href
  {https://doi.org/10.1103/PhysRevX.8.011035} {\bibfield  {journal} {\bibinfo
  {journal} {Phys. Rev. X}\ }\textbf {\bibinfo {volume} {8}},\ \bibinfo {pages}
  {011035} (\bibinfo {year} {2018})}\BibitemShut {NoStop}%
\bibitem [{\citenamefont {Unanyan}\ \emph {et~al.}(2020)\citenamefont
  {Unanyan}, \citenamefont {Kiefer-Emmanouilidis},\ and\ \citenamefont
  {Fleischhauer}}]{Unayan_2020}%
  \BibitemOpen
  \bibfield  {author} {\bibinfo {author} {\bibfnamefont {R.}~\bibnamefont
  {Unanyan}}, \bibinfo {author} {\bibfnamefont {M.}~\bibnamefont
  {Kiefer-Emmanouilidis}},\ and\ \bibinfo {author} {\bibfnamefont
  {M.}~\bibnamefont {Fleischhauer}},\ }\bibfield  {title} {\bibinfo {title}
  {{Finite-Temperature Topological Invariant for Interacting Systems}},\ }\href
  {https://doi.org/10.1103/PhysRevLett.125.215701} {\bibfield  {journal}
  {\bibinfo  {journal} {Phys. Rev. Lett.}\ }\textbf {\bibinfo {volume} {125}},\
  \bibinfo {pages} {215701} (\bibinfo {year} {2020})}\BibitemShut {NoStop}%
\bibitem [{\citenamefont {Huang}\ and\ \citenamefont
  {Diehl}(2024)}]{huang2024mixedstatetopologicalorder}%
  \BibitemOpen
  \bibfield  {author} {\bibinfo {author} {\bibfnamefont {Z.-M.}\ \bibnamefont
  {Huang}}\ and\ \bibinfo {author} {\bibfnamefont {S.}~\bibnamefont {Diehl}},\
  }\bibfield  {title} {\bibinfo {title} {{Mixed state topological order
  parameters for symmetry protected fermion matter}},\ }\href
  {https://arxiv.org/abs/2401.10993} {\  (\bibinfo {year} {2024})},\ \Eprint
  {https://arxiv.org/abs/2401.10993} {arXiv:2401.10993} \BibitemShut {NoStop}%
\bibitem [{\citenamefont {Chen}\ \emph {et~al.}(2024)\citenamefont {Chen},
  \citenamefont {Chesi},\ and\ \citenamefont
  {Choi}}]{chen2024universalentanglementrevivaltopological}%
  \BibitemOpen
  \bibfield  {author} {\bibinfo {author} {\bibfnamefont {D.}~\bibnamefont
  {Chen}}, \bibinfo {author} {\bibfnamefont {S.}~\bibnamefont {Chesi}},\ and\
  \bibinfo {author} {\bibfnamefont {M.-S.}\ \bibnamefont {Choi}},\ }\href
  {https://arxiv.org/abs/2410.17562} {\bibinfo {title} {{Universal Entanglement
  Revival of Topological Origin}}} (\bibinfo {year} {2024}),\ \Eprint
  {https://arxiv.org/abs/2410.17562} {arXiv:2410.17562} \BibitemShut {NoStop}%
\bibitem [{\citenamefont {Nava}\ \emph {et~al.}(2023)\citenamefont {Nava},
  \citenamefont {Campagnano}, \citenamefont {Sodano},\ and\ \citenamefont
  {Giuliano}}]{nava2023}%
  \BibitemOpen
  \bibfield  {author} {\bibinfo {author} {\bibfnamefont {A.}~\bibnamefont
  {Nava}}, \bibinfo {author} {\bibfnamefont {G.}~\bibnamefont {Campagnano}},
  \bibinfo {author} {\bibfnamefont {P.}~\bibnamefont {Sodano}},\ and\ \bibinfo
  {author} {\bibfnamefont {D.}~\bibnamefont {Giuliano}},\ }\bibfield  {title}
  {\bibinfo {title} {{Lindblad master equation approach to the topological
  phase transition in the disordered Su-Schrieffer-Heeger model}},\ }\href
  {https://doi.org/10.1103/PhysRevB.107.035113} {\bibfield  {journal} {\bibinfo
   {journal} {Phys. Rev. B}\ }\textbf {\bibinfo {volume} {107}},\ \bibinfo
  {pages} {035113} (\bibinfo {year} {2023})}\BibitemShut {NoStop}%
\bibitem [{\citenamefont {Cinnirella}\ \emph {et~al.}(2024)\citenamefont
  {Cinnirella}, \citenamefont {Nava}, \citenamefont {Campagnano},\ and\
  \citenamefont {Giuliano}}]{campagnano2024}%
  \BibitemOpen
  \bibfield  {author} {\bibinfo {author} {\bibfnamefont {E.~G.}\ \bibnamefont
  {Cinnirella}}, \bibinfo {author} {\bibfnamefont {A.}~\bibnamefont {Nava}},
  \bibinfo {author} {\bibfnamefont {G.}~\bibnamefont {Campagnano}},\ and\
  \bibinfo {author} {\bibfnamefont {D.}~\bibnamefont {Giuliano}},\ }\bibfield
  {title} {\bibinfo {title} {{Fate of high winding number topological phases in
  the disordered extended Su-Schrieffer-Heeger model}},\ }\href
  {https://doi.org/10.1103/PhysRevB.109.035114} {\bibfield  {journal} {\bibinfo
   {journal} {Phys. Rev. B}\ }\textbf {\bibinfo {volume} {109}},\ \bibinfo
  {pages} {035114} (\bibinfo {year} {2024})}\BibitemShut {NoStop}%
\bibitem [{\citenamefont {Carmichael}(2013)}]{carmichael2013statistical}%
  \BibitemOpen
  \bibfield  {author} {\bibinfo {author} {\bibfnamefont {H.}~\bibnamefont
  {Carmichael}},\ }\href {https://books.google.it/books?id=5hv2CAAAQBAJ} {\emph
  {\bibinfo {title} {{Statistical Methods in Quantum Optics 1: Master Equations
  and Fokker-Planck Equations}}}},\ Theoretical and Mathematical Physics\
  (\bibinfo  {publisher} {Springer Berlin Heidelberg},\ \bibinfo {year}
  {2013})\BibitemShut {NoStop}%
\bibitem [{\citenamefont {Daley}(2014)}]{Daley_2014}%
  \BibitemOpen
  \bibfield  {author} {\bibinfo {author} {\bibfnamefont {A.~J.}\ \bibnamefont
  {Daley}},\ }\bibfield  {title} {\bibinfo {title} {{Quantum trajectories and
  open many-body quantum systems}},\ }\href
  {https://doi.org/10.1080/00018732.2014.933502} {\bibfield  {journal}
  {\bibinfo  {journal} {Advances in Physics}\ }\textbf {\bibinfo {volume}
  {63}},\ \bibinfo {pages} {77–149} (\bibinfo {year} {2014})}\BibitemShut
  {NoStop}%
\bibitem [{\citenamefont {Plenio}\ and\ \citenamefont
  {Knight}(1998)}]{Plenio_1998}%
  \BibitemOpen
  \bibfield  {author} {\bibinfo {author} {\bibfnamefont {M.~B.}\ \bibnamefont
  {Plenio}}\ and\ \bibinfo {author} {\bibfnamefont {P.~L.}\ \bibnamefont
  {Knight}},\ }\bibfield  {title} {\bibinfo {title} {{The quantum-jump approach
  to dissipative dynamics in quantum optics}},\ }\href
  {https://doi.org/10.1103/revmodphys.70.101} {\bibfield  {journal} {\bibinfo
  {journal} {Reviews of Modern Physics}\ }\textbf {\bibinfo {volume} {70}},\
  \bibinfo {pages} {101–144} (\bibinfo {year} {1998})}\BibitemShut {NoStop}%
\bibitem [{\citenamefont {M{\o}lmer}\ \emph {et~al.}(1993)\citenamefont
  {M{\o}lmer}, \citenamefont {Castin},\ and\ \citenamefont
  {Dalibard}}]{Molmer:93}%
  \BibitemOpen
  \bibfield  {author} {\bibinfo {author} {\bibfnamefont {K.}~\bibnamefont
  {M{\o}lmer}}, \bibinfo {author} {\bibfnamefont {Y.}~\bibnamefont {Castin}},\
  and\ \bibinfo {author} {\bibfnamefont {J.}~\bibnamefont {Dalibard}},\
  }\bibfield  {title} {\bibinfo {title} {{Monte Carlo wave-function method in
  quantum optics}},\ }\href {https://doi.org/10.1364/JOSAB.10.000524}
  {\bibfield  {journal} {\bibinfo  {journal} {J. Opt. Soc. Am. B}\ }\textbf
  {\bibinfo {volume} {10}},\ \bibinfo {pages} {524} (\bibinfo {year}
  {1993})}\BibitemShut {NoStop}%
\bibitem [{\citenamefont {Breuer}\ and\ \citenamefont
  {Petruccione}(2002)}]{breuer}%
  \BibitemOpen
  \bibfield  {author} {\bibinfo {author} {\bibfnamefont {H.~P.}\ \bibnamefont
  {Breuer}}\ and\ \bibinfo {author} {\bibfnamefont {F.}~\bibnamefont
  {Petruccione}},\ }\href@noop {} {\emph {\bibinfo {title} {{The theory of open
  quantum systems}}}}\ (\bibinfo  {publisher} {Oxford University Press},\
  \bibinfo {address} {Great Clarendon Street},\ \bibinfo {year}
  {2002})\BibitemShut {NoStop}%
\bibitem [{\citenamefont {Li}\ \emph {et~al.}(2018)\citenamefont {Li},
  \citenamefont {Chen},\ and\ \citenamefont
  {Fisher}}]{li2018quantumzenoeffect}%
  \BibitemOpen
  \bibfield  {author} {\bibinfo {author} {\bibfnamefont {Y.}~\bibnamefont
  {Li}}, \bibinfo {author} {\bibfnamefont {X.}~\bibnamefont {Chen}},\ and\
  \bibinfo {author} {\bibfnamefont {M.~P.~A.}\ \bibnamefont {Fisher}},\
  }\bibfield  {title} {\bibinfo {title} {{Quantum Zeno effect and the many-body
  entanglement transition}},\ }\href
  {https://doi.org/10.1103/PhysRevB.98.205136} {\bibfield  {journal} {\bibinfo
  {journal} {Phys. Rev. B}\ }\textbf {\bibinfo {volume} {98}},\ \bibinfo
  {pages} {205136} (\bibinfo {year} {2018})}\BibitemShut {NoStop}%
\bibitem [{\citenamefont {Skinner}\ \emph {et~al.}(2019)\citenamefont
  {Skinner}, \citenamefont {Ruhman},\ and\ \citenamefont
  {Nahum}}]{skinner2019measurementinducedphase}%
  \BibitemOpen
  \bibfield  {author} {\bibinfo {author} {\bibfnamefont {B.}~\bibnamefont
  {Skinner}}, \bibinfo {author} {\bibfnamefont {J.}~\bibnamefont {Ruhman}},\
  and\ \bibinfo {author} {\bibfnamefont {A.}~\bibnamefont {Nahum}},\ }\bibfield
   {title} {\bibinfo {title} {{Measurement-Induced Phase Transitions in the
  Dynamics of Entanglement}},\ }\href
  {https://doi.org/10.1103/PhysRevX.9.031009} {\bibfield  {journal} {\bibinfo
  {journal} {Phys. Rev. X}\ }\textbf {\bibinfo {volume} {9}},\ \bibinfo {pages}
  {031009} (\bibinfo {year} {2019})}\BibitemShut {NoStop}%
\bibitem [{\citenamefont {Potter}\ and\ \citenamefont
  {Vasseur}(2022)}]{vasseur2022entanglementhybrid}%
  \BibitemOpen
  \bibfield  {author} {\bibinfo {author} {\bibfnamefont {A.~C.}\ \bibnamefont
  {Potter}}\ and\ \bibinfo {author} {\bibfnamefont {R.}~\bibnamefont
  {Vasseur}},\ }\bibinfo {title} {{{Entanglement Dynamics in Hybrid Quantum
  Circuits}}},\ in\ \href {https://doi.org/10.1007/978-3-031-03998-0_9} {\emph
  {\bibinfo {booktitle} {Entanglement in Spin Chains: From Theory to Quantum
  Technology Applications}}},\ \bibinfo {editor} {edited by\ \bibinfo {editor}
  {\bibfnamefont {A.}~\bibnamefont {Bayat}}, \bibinfo {editor} {\bibfnamefont
  {S.}~\bibnamefont {Bose}},\ and\ \bibinfo {editor} {\bibfnamefont
  {H.}~\bibnamefont {Johannesson}}}\ (\bibinfo  {publisher} {Springer
  International Publishing},\ \bibinfo {address} {Cham},\ \bibinfo {year}
  {2022})\ pp.\ \bibinfo {pages} {211--249}\BibitemShut {NoStop}%
\bibitem [{\citenamefont {Fazio}\ \emph {et~al.}(2024)\citenamefont {Fazio},
  \citenamefont {Keeling}, \citenamefont {Mazza},\ and\ \citenamefont
  {Schirò}}]{Fazio2024}%
  \BibitemOpen
  \bibfield  {author} {\bibinfo {author} {\bibfnamefont {R.}~\bibnamefont
  {Fazio}}, \bibinfo {author} {\bibfnamefont {J.}~\bibnamefont {Keeling}},
  \bibinfo {author} {\bibfnamefont {L.}~\bibnamefont {Mazza}},\ and\ \bibinfo
  {author} {\bibfnamefont {M.}~\bibnamefont {Schirò}},\ }\href
  {https://arxiv.org/abs/2409.10300} {\bibinfo {title} {{Many-Body Open Quantum
  Systems}}} (\bibinfo {year} {2024}),\ \Eprint
  {https://arxiv.org/abs/2409.10300} {arXiv:2409.10300 [quant-ph]} \BibitemShut
  {NoStop}%
\bibitem [{\citenamefont {Zeng}\ \emph {et~al.}(2019)\citenamefont {Zeng},
  \citenamefont {Chen}, \citenamefont {Zhou},\ and\ \citenamefont
  {Wen}}]{zeng2018quantum}%
  \BibitemOpen
  \bibfield  {author} {\bibinfo {author} {\bibfnamefont {B.}~\bibnamefont
  {Zeng}}, \bibinfo {author} {\bibfnamefont {X.}~\bibnamefont {Chen}}, \bibinfo
  {author} {\bibfnamefont {D.}~\bibnamefont {Zhou}},\ and\ \bibinfo {author}
  {\bibfnamefont {X.}~\bibnamefont {Wen}},\ }\href
  {https://books.google.it/books?id=BPnPxgEACAAJ} {\emph {\bibinfo {title}
  {{Quantum Information Meets Quantum Matter: From Quantum Entanglement to
  Topological Phases of Many-Body Systems}}}},\ Quantum science and technology\
  (\bibinfo  {publisher} {Springer New York},\ \bibinfo {year}
  {2019})\BibitemShut {NoStop}%
\bibitem [{\citenamefont {Chen}\ \emph {et~al.}(2015)\citenamefont {Chen},
  \citenamefont {Ji}, \citenamefont {Li}, \citenamefont {Poon}, \citenamefont
  {Shen}, \citenamefont {Yu}, \citenamefont {Zeng},\ and\ \citenamefont
  {Zhou}}]{Chen_2015}%
  \BibitemOpen
  \bibfield  {author} {\bibinfo {author} {\bibfnamefont {J.}~\bibnamefont
  {Chen}}, \bibinfo {author} {\bibfnamefont {Z.}~\bibnamefont {Ji}}, \bibinfo
  {author} {\bibfnamefont {C.-K.}\ \bibnamefont {Li}}, \bibinfo {author}
  {\bibfnamefont {Y.-T.}\ \bibnamefont {Poon}}, \bibinfo {author}
  {\bibfnamefont {Y.}~\bibnamefont {Shen}}, \bibinfo {author} {\bibfnamefont
  {N.}~\bibnamefont {Yu}}, \bibinfo {author} {\bibfnamefont {B.}~\bibnamefont
  {Zeng}},\ and\ \bibinfo {author} {\bibfnamefont {D.}~\bibnamefont {Zhou}},\
  }\bibfield  {title} {\bibinfo {title} {{Discontinuity of maximum entropy
  inference and quantum phase transitions}},\ }\href
  {https://doi.org/10.1088/1367-2630/17/8/083019} {\bibfield  {journal}
  {\bibinfo  {journal} {New Journal of Physics}\ }\textbf {\bibinfo {volume}
  {17}},\ \bibinfo {pages} {083019} (\bibinfo {year} {2015})}\BibitemShut
  {NoStop}%
\bibitem [{\citenamefont {Isakov}\ \emph {et~al.}(2011)\citenamefont {Isakov},
  \citenamefont {Hastings},\ and\ \citenamefont {Melko}}]{Isakov_2011}%
  \BibitemOpen
  \bibfield  {author} {\bibinfo {author} {\bibfnamefont {S.~V.}\ \bibnamefont
  {Isakov}}, \bibinfo {author} {\bibfnamefont {M.~B.}\ \bibnamefont
  {Hastings}},\ and\ \bibinfo {author} {\bibfnamefont {R.~G.}\ \bibnamefont
  {Melko}},\ }\bibfield  {title} {\bibinfo {title} {{Topological entanglement
  entropy of a Bose–Hubbard spin liquid}},\ }\href
  {https://doi.org/10.1038/nphys2036} {\bibfield  {journal} {\bibinfo
  {journal} {Nature Physics}\ }\textbf {\bibinfo {volume} {7}},\ \bibinfo
  {pages} {772–775} (\bibinfo {year} {2011})}\BibitemShut {NoStop}%
\bibitem [{\citenamefont {Jiang}\ \emph {et~al.}(2012)\citenamefont {Jiang},
  \citenamefont {Wang},\ and\ \citenamefont {Balents}}]{Jiang_2012}%
  \BibitemOpen
  \bibfield  {author} {\bibinfo {author} {\bibfnamefont {H.-C.}\ \bibnamefont
  {Jiang}}, \bibinfo {author} {\bibfnamefont {Z.}~\bibnamefont {Wang}},\ and\
  \bibinfo {author} {\bibfnamefont {L.}~\bibnamefont {Balents}},\ }\bibfield
  {title} {\bibinfo {title} {{Identifying topological order by entanglement
  entropy}},\ }\href {https://doi.org/10.1038/nphys2465} {\bibfield  {journal}
  {\bibinfo  {journal} {Nature Physics}\ }\textbf {\bibinfo {volume} {8}},\
  \bibinfo {pages} {902–905} (\bibinfo {year} {2012})}\BibitemShut {NoStop}%
\bibitem [{\citenamefont {Kitaev}\ and\ \citenamefont
  {Preskill}(2006)}]{Kitaev_2006}%
  \BibitemOpen
  \bibfield  {author} {\bibinfo {author} {\bibfnamefont {A.}~\bibnamefont
  {Kitaev}}\ and\ \bibinfo {author} {\bibfnamefont {J.}~\bibnamefont
  {Preskill}},\ }\bibfield  {title} {\bibinfo {title} {{Topological
  Entanglement Entropy}},\ }\href
  {https://doi.org/10.1103/PhysRevLett.96.110404} {\bibfield  {journal}
  {\bibinfo  {journal} {Phys. Rev. Lett.}\ }\textbf {\bibinfo {volume} {96}},\
  \bibinfo {pages} {110404} (\bibinfo {year} {2006})}\BibitemShut {NoStop}%
\bibitem [{\citenamefont {Levin}\ and\ \citenamefont {Wen}(2006)}]{Levin_2006}%
  \BibitemOpen
  \bibfield  {author} {\bibinfo {author} {\bibfnamefont {M.}~\bibnamefont
  {Levin}}\ and\ \bibinfo {author} {\bibfnamefont {X.-G.}\ \bibnamefont
  {Wen}},\ }\bibfield  {title} {\bibinfo {title} {{Detecting Topological Order
  in a Ground State Wave Function}},\ }\href
  {https://doi.org/10.1103/PhysRevLett.96.110405} {\bibfield  {journal}
  {\bibinfo  {journal} {Phys. Rev. Lett.}\ }\textbf {\bibinfo {volume} {96}},\
  \bibinfo {pages} {110405} (\bibinfo {year} {2006})}\BibitemShut {NoStop}%
\bibitem [{\citenamefont {Torre}\ \emph {et~al.}(2024)\citenamefont {Torre},
  \citenamefont {Odavić}, \citenamefont {Fromholz}, \citenamefont
  {Giampaolo},\ and\ \citenamefont
  {Franchini}}]{torre2023longrangeentanglementtopologicalexcitations}%
  \BibitemOpen
  \bibfield  {author} {\bibinfo {author} {\bibfnamefont {G.}~\bibnamefont
  {Torre}}, \bibinfo {author} {\bibfnamefont {J.}~\bibnamefont {Odavić}},
  \bibinfo {author} {\bibfnamefont {P.}~\bibnamefont {Fromholz}}, \bibinfo
  {author} {\bibfnamefont {S.~M.}\ \bibnamefont {Giampaolo}},\ and\ \bibinfo
  {author} {\bibfnamefont {F.}~\bibnamefont {Franchini}},\ }\bibfield  {title}
  {\bibinfo {title} {{Long-range entanglement and topological excitations}},\
  }\href {https://doi.org/10.21468/SciPostPhysCore.7.3.050} {\bibfield
  {journal} {\bibinfo  {journal} {SciPost Phys. Core}\ }\textbf {\bibinfo
  {volume} {7}},\ \bibinfo {pages} {050} (\bibinfo {year} {2024})}\BibitemShut
  {NoStop}%
\bibitem [{\citenamefont {Arora}\ \emph {et~al.}(2023)\citenamefont {Arora},
  \citenamefont {Kejriwal},\ and\ \citenamefont
  {Muralidharan}}]{arora2023conclusivedetectionmajoranazero}%
  \BibitemOpen
  \bibfield  {author} {\bibinfo {author} {\bibfnamefont {A.}~\bibnamefont
  {Arora}}, \bibinfo {author} {\bibfnamefont {A.}~\bibnamefont {Kejriwal}},\
  and\ \bibinfo {author} {\bibfnamefont {B.}~\bibnamefont {Muralidharan}},\
  }\bibfield  {title} {\bibinfo {title} {{On the conclusive detection of
  Majorana zero modes: conductance spectroscopy, disconnected entanglement
  entropy and the fermion parity noise}},\ }\href
  {https://arxiv.org/abs/2303.03837} {\  (\bibinfo {year} {2023})},\ \Eprint
  {https://arxiv.org/abs/2303.03837} {arXiv:2303.03837} \BibitemShut {NoStop}%
\bibitem [{\citenamefont {Pichler}\ \emph {et~al.}(2016)\citenamefont
  {Pichler}, \citenamefont {Zhu}, \citenamefont {Seif}, \citenamefont
  {Zoller},\ and\ \citenamefont {Hafezi}}]{Zoller_PRX_2016}%
  \BibitemOpen
  \bibfield  {author} {\bibinfo {author} {\bibfnamefont {H.}~\bibnamefont
  {Pichler}}, \bibinfo {author} {\bibfnamefont {G.}~\bibnamefont {Zhu}},
  \bibinfo {author} {\bibfnamefont {A.}~\bibnamefont {Seif}}, \bibinfo {author}
  {\bibfnamefont {P.}~\bibnamefont {Zoller}},\ and\ \bibinfo {author}
  {\bibfnamefont {M.}~\bibnamefont {Hafezi}},\ }\bibfield  {title} {\bibinfo
  {title} {{Measurement Protocol for the Entanglement Spectrum of Cold
  Atoms}},\ }\href {https://doi.org/10.1103/PhysRevX.6.041033} {\bibfield
  {journal} {\bibinfo  {journal} {Phys. Rev. X}\ }\textbf {\bibinfo {volume}
  {6}},\ \bibinfo {pages} {041033} (\bibinfo {year} {2016})}\BibitemShut
  {NoStop}%
\bibitem [{\citenamefont {Elben}\ \emph {et~al.}(2018)\citenamefont {Elben},
  \citenamefont {Vermersch}, \citenamefont {Dalmonte}, \citenamefont {Cirac},\
  and\ \citenamefont {Zoller}}]{Zoller_PRL_2018}%
  \BibitemOpen
  \bibfield  {author} {\bibinfo {author} {\bibfnamefont {A.}~\bibnamefont
  {Elben}}, \bibinfo {author} {\bibfnamefont {B.}~\bibnamefont {Vermersch}},
  \bibinfo {author} {\bibfnamefont {M.}~\bibnamefont {Dalmonte}}, \bibinfo
  {author} {\bibfnamefont {J.~I.}\ \bibnamefont {Cirac}},\ and\ \bibinfo
  {author} {\bibfnamefont {P.}~\bibnamefont {Zoller}},\ }\bibfield  {title}
  {\bibinfo {title} {{R\'enyi Entropies from Random Quenches in Atomic Hubbard
  and Spin Models}},\ }\href {https://doi.org/10.1103/PhysRevLett.120.050406}
  {\bibfield  {journal} {\bibinfo  {journal} {Phys. Rev. Lett.}\ }\textbf
  {\bibinfo {volume} {120}},\ \bibinfo {pages} {050406} (\bibinfo {year}
  {2018})}\BibitemShut {NoStop}%
\bibitem [{\citenamefont {Brydges}\ \emph {et~al.}(2019)\citenamefont
  {Brydges}, \citenamefont {Elben}, \citenamefont {Jurcevic}, \citenamefont
  {Vermersch}, \citenamefont {Maier}, \citenamefont {Lanyon}, \citenamefont
  {Zoller}, \citenamefont {Blatt},\ and\ \citenamefont
  {Roos}}]{Zoller_Science_2019}%
  \BibitemOpen
  \bibfield  {author} {\bibinfo {author} {\bibfnamefont {T.}~\bibnamefont
  {Brydges}}, \bibinfo {author} {\bibfnamefont {A.}~\bibnamefont {Elben}},
  \bibinfo {author} {\bibfnamefont {P.}~\bibnamefont {Jurcevic}}, \bibinfo
  {author} {\bibfnamefont {B.}~\bibnamefont {Vermersch}}, \bibinfo {author}
  {\bibfnamefont {C.}~\bibnamefont {Maier}}, \bibinfo {author} {\bibfnamefont
  {B.~P.}\ \bibnamefont {Lanyon}}, \bibinfo {author} {\bibfnamefont
  {P.}~\bibnamefont {Zoller}}, \bibinfo {author} {\bibfnamefont
  {R.}~\bibnamefont {Blatt}},\ and\ \bibinfo {author} {\bibfnamefont {C.~F.}\
  \bibnamefont {Roos}},\ }\bibfield  {title} {\bibinfo {title} {{Probing Rényi
  entanglement entropy via randomized measurements}},\ }\href
  {https://doi.org/10.1126/science.aau4963} {\bibfield  {journal} {\bibinfo
  {journal} {Science}\ }\textbf {\bibinfo {volume} {364}},\ \bibinfo {pages}
  {260} (\bibinfo {year} {2019})}\BibitemShut {NoStop}%
\bibitem [{\citenamefont {Fromholz}\ \emph {et~al.}(2020)\citenamefont
  {Fromholz}, \citenamefont {Magnifico}, \citenamefont {Vitale}, \citenamefont
  {Mendes-Santos},\ and\ \citenamefont
  {Dalmonte}}]{Dalmonte_PhysRevB.101.085136}%
  \BibitemOpen
  \bibfield  {author} {\bibinfo {author} {\bibfnamefont {P.}~\bibnamefont
  {Fromholz}}, \bibinfo {author} {\bibfnamefont {G.}~\bibnamefont {Magnifico}},
  \bibinfo {author} {\bibfnamefont {V.}~\bibnamefont {Vitale}}, \bibinfo
  {author} {\bibfnamefont {T.}~\bibnamefont {Mendes-Santos}},\ and\ \bibinfo
  {author} {\bibfnamefont {M.}~\bibnamefont {Dalmonte}},\ }\bibfield  {title}
  {\bibinfo {title} {{Entanglement topological invariants for one-dimensional
  topological superconductors}},\ }\href
  {https://doi.org/10.1103/PhysRevB.101.085136} {\bibfield  {journal} {\bibinfo
   {journal} {Phys. Rev. B}\ }\textbf {\bibinfo {volume} {101}},\ \bibinfo
  {pages} {085136} (\bibinfo {year} {2020})}\BibitemShut {NoStop}%
\bibitem [{\citenamefont {Micallo}\ \emph {et~al.}(2020)\citenamefont
  {Micallo}, \citenamefont {Vitale}, \citenamefont {Dalmonte},\ and\
  \citenamefont {Fromholz}}]{Micallo_2020}%
  \BibitemOpen
  \bibfield  {author} {\bibinfo {author} {\bibfnamefont {T.}~\bibnamefont
  {Micallo}}, \bibinfo {author} {\bibfnamefont {V.}~\bibnamefont {Vitale}},
  \bibinfo {author} {\bibfnamefont {M.}~\bibnamefont {Dalmonte}},\ and\
  \bibinfo {author} {\bibfnamefont {P.}~\bibnamefont {Fromholz}},\ }\bibfield
  {title} {\bibinfo {title} {{Topological entanglement properties of
  disconnected partitions in the Su-Schrieffer-Heeger model}},\ }\href
  {https://doi.org/10.21468/SciPostPhysCore.3.2.012} {\bibfield  {journal}
  {\bibinfo  {journal} {SciPost Phys. Core}\ }\textbf {\bibinfo {volume} {3}},\
  \bibinfo {pages} {012} (\bibinfo {year} {2020})}\BibitemShut {NoStop}%
\bibitem [{\citenamefont {Mondal}\ \emph {et~al.}(2022)\citenamefont {Mondal},
  \citenamefont {Sen},\ and\ \citenamefont {Dutta}}]{Mondal_2022}%
  \BibitemOpen
  \bibfield  {author} {\bibinfo {author} {\bibfnamefont {S.}~\bibnamefont
  {Mondal}}, \bibinfo {author} {\bibfnamefont {D.}~\bibnamefont {Sen}},\ and\
  \bibinfo {author} {\bibfnamefont {A.}~\bibnamefont {Dutta}},\ }\bibfield
  {title} {\bibinfo {title} {{Disconnected entanglement entropy as a marker of
  edge modes in a periodically driven Kitaev chain}},\ }\href
  {https://doi.org/10.1088/1361-648x/aca7f7} {\bibfield  {journal} {\bibinfo
  {journal} {Journal of Physics: Condensed Matter}\ }\textbf {\bibinfo {volume}
  {35}},\ \bibinfo {pages} {085601} (\bibinfo {year} {2022})}\BibitemShut
  {NoStop}%
\bibitem [{\citenamefont {Su}\ \emph {et~al.}(1979)\citenamefont {Su},
  \citenamefont {Schrieffer},\ and\ \citenamefont {Heeger}}]{ssh_1979}%
  \BibitemOpen
  \bibfield  {author} {\bibinfo {author} {\bibfnamefont {W.~P.}\ \bibnamefont
  {Su}}, \bibinfo {author} {\bibfnamefont {J.~R.}\ \bibnamefont {Schrieffer}},\
  and\ \bibinfo {author} {\bibfnamefont {A.~J.}\ \bibnamefont {Heeger}},\
  }\bibfield  {title} {\bibinfo {title} {{Solitons in Polyacetylene}},\ }\href
  {https://doi.org/10.1103/PhysRevLett.42.1698} {\bibfield  {journal} {\bibinfo
   {journal} {Phys. Rev. Lett.}\ }\textbf {\bibinfo {volume} {42}},\ \bibinfo
  {pages} {1698} (\bibinfo {year} {1979})}\BibitemShut {NoStop}%
\bibitem [{\citenamefont {Su}\ \emph {et~al.}(1980)\citenamefont {Su},
  \citenamefont {Schrieffer},\ and\ \citenamefont {Heeger}}]{ssh_1980}%
  \BibitemOpen
  \bibfield  {author} {\bibinfo {author} {\bibfnamefont {W.~P.}\ \bibnamefont
  {Su}}, \bibinfo {author} {\bibfnamefont {J.~R.}\ \bibnamefont {Schrieffer}},\
  and\ \bibinfo {author} {\bibfnamefont {A.~J.}\ \bibnamefont {Heeger}},\
  }\bibfield  {title} {\bibinfo {title} {{Soliton excitations in
  polyacetylene}},\ }\href {https://doi.org/10.1103/PhysRevB.22.2099}
  {\bibfield  {journal} {\bibinfo  {journal} {Phys. Rev. B}\ }\textbf {\bibinfo
  {volume} {22}},\ \bibinfo {pages} {2099} (\bibinfo {year}
  {1980})}\BibitemShut {NoStop}%
\bibitem [{\citenamefont {Hasan}\ and\ \citenamefont
  {Kane}(2010)}]{Hasan_2010}%
  \BibitemOpen
  \bibfield  {author} {\bibinfo {author} {\bibfnamefont {M.~Z.}\ \bibnamefont
  {Hasan}}\ and\ \bibinfo {author} {\bibfnamefont {C.~L.}\ \bibnamefont
  {Kane}},\ }\bibfield  {title} {\bibinfo {title} {{Colloquium: Topological
  insulators}},\ }\href {https://doi.org/10.1103/revmodphys.82.3045} {\bibfield
   {journal} {\bibinfo  {journal} {Reviews of Modern Physics}\ }\textbf
  {\bibinfo {volume} {82}},\ \bibinfo {pages} {3045–3067} (\bibinfo {year}
  {2010})}\BibitemShut {NoStop}%
\bibitem [{\citenamefont {Qi}\ and\ \citenamefont {Zhang}(2011)}]{Qi_2011}%
  \BibitemOpen
  \bibfield  {author} {\bibinfo {author} {\bibfnamefont {X.-L.}\ \bibnamefont
  {Qi}}\ and\ \bibinfo {author} {\bibfnamefont {S.-C.}\ \bibnamefont {Zhang}},\
  }\bibfield  {title} {\bibinfo {title} {{Topological insulators and
  superconductors}},\ }\href {https://doi.org/10.1103/revmodphys.83.1057}
  {\bibfield  {journal} {\bibinfo  {journal} {Reviews of Modern Physics}\
  }\textbf {\bibinfo {volume} {83}},\ \bibinfo {pages} {1057–1110} (\bibinfo
  {year} {2011})}\BibitemShut {NoStop}%
\bibitem [{\citenamefont {Asb{\'{o}}th}\ \emph {et~al.}(2016)\citenamefont
  {Asb{\'{o}}th}, \citenamefont {Oroszl{\'{a}}ny},\ and\ \citenamefont
  {P{\'{a}}lyi}}]{Asb_th_2016}%
  \BibitemOpen
  \bibfield  {author} {\bibinfo {author} {\bibfnamefont {J.~K.}\ \bibnamefont
  {Asb{\'{o}}th}}, \bibinfo {author} {\bibfnamefont {L.}~\bibnamefont
  {Oroszl{\'{a}}ny}},\ and\ \bibinfo {author} {\bibfnamefont {A.}~\bibnamefont
  {P{\'{a}}lyi}},\ }\href {https://doi.org/10.1007/978-3-319-25607-8} {\emph
  {\bibinfo {title} {{A Short Course on Topological Insulators}}}}\ (\bibinfo
  {publisher} {Springer International Publishing},\ \bibinfo {year}
  {2016})\BibitemShut {NoStop}%
\bibitem [{\citenamefont {Le~Gal}\ \emph {et~al.}(2024)\citenamefont {Le~Gal},
  \citenamefont {Turkeshi},\ and\ \citenamefont
  {Schir\`o}}]{gal2024entanglement}%
  \BibitemOpen
  \bibfield  {author} {\bibinfo {author} {\bibfnamefont {Y.}~\bibnamefont
  {Le~Gal}}, \bibinfo {author} {\bibfnamefont {X.}~\bibnamefont {Turkeshi}},\
  and\ \bibinfo {author} {\bibfnamefont {M.}~\bibnamefont {Schir\`o}},\
  }\bibfield  {title} {\bibinfo {title} {{Entanglement Dynamics in Monitored
  Systems and the Role of Quantum Jumps}},\ }\href
  {https://doi.org/10.1103/PRXQuantum.5.030329} {\bibfield  {journal} {\bibinfo
   {journal} {PRX Quantum}\ }\textbf {\bibinfo {volume} {5}},\ \bibinfo {pages}
  {030329} (\bibinfo {year} {2024})}\BibitemShut {NoStop}%
\bibitem [{\citenamefont {Lindblad}(1976)}]{lindblad}%
  \BibitemOpen
  \bibfield  {author} {\bibinfo {author} {\bibfnamefont {G.}~\bibnamefont
  {Lindblad}},\ }\bibfield  {title} {\bibinfo {title} {{On the generators of
  quantum dynamical semigroups}},\ }\href@noop {} {\bibfield  {journal}
  {\bibinfo  {journal} {Communications in Mathematical Physics}\ }\textbf
  {\bibinfo {volume} {48}},\ \bibinfo {pages} {119 } (\bibinfo {year}
  {1976})}\BibitemShut {NoStop}%
\bibitem [{\citenamefont {Simon}\ \emph {et~al.}(2018)\citenamefont {Simon},
  \citenamefont {Osawa},\ and\ \citenamefont {Sergienko}}]{Simon_2019}%
  \BibitemOpen
  \bibfield  {author} {\bibinfo {author} {\bibfnamefont {D.~S.}\ \bibnamefont
  {Simon}}, \bibinfo {author} {\bibfnamefont {S.}~\bibnamefont {Osawa}},\ and\
  \bibinfo {author} {\bibfnamefont {A.~V.}\ \bibnamefont {Sergienko}},\
  }\bibfield  {title} {\bibinfo {title} {{Topological boundaries and bulk
  wavefunctions in the Su–Schreiffer–Heeger model}},\ }\href
  {https://doi.org/10.1088/1361-648X/aaf0bf} {\bibfield  {journal} {\bibinfo
  {journal} {Journal of Physics: Condensed Matter}\ }\textbf {\bibinfo {volume}
  {31}},\ \bibinfo {pages} {045001} (\bibinfo {year} {2018})}\BibitemShut
  {NoStop}%
\bibitem [{\citenamefont {Berry}(1984)}]{M.V.Berry03081984}%
  \BibitemOpen
  \bibfield  {author} {\bibinfo {author} {\bibfnamefont {M.~V.}\ \bibnamefont
  {Berry}},\ }\bibfield  {title} {\bibinfo {title} {{Quantal Phase Factors
  Accompanying Adiabatic Changes}},\ }\href
  {https://doi.org/10.1098/rspa.1984.0023} {\bibfield  {journal} {\bibinfo
  {journal} {Proceedings of the Royal Society of London. A. Mathematical and
  Physical Sciences}\ }\textbf {\bibinfo {volume} {392}},\ \bibinfo {pages}
  {45} (\bibinfo {year} {1984})}\BibitemShut {NoStop}%
\bibitem [{\citenamefont {Zak}(1989)}]{Zak1989}%
  \BibitemOpen
  \bibfield  {author} {\bibinfo {author} {\bibfnamefont {J.}~\bibnamefont
  {Zak}},\ }\bibfield  {title} {\bibinfo {title} {{Berry's phase for energy
  bands in solids}},\ }\href {https://doi.org/10.1103/PhysRevLett.62.2747}
  {\bibfield  {journal} {\bibinfo  {journal} {Phys. Rev. Lett.}\ }\textbf
  {\bibinfo {volume} {62}},\ \bibinfo {pages} {2747} (\bibinfo {year}
  {1989})}\BibitemShut {NoStop}%
\bibitem [{\citenamefont {Schomerus}(2013)}]{Schomerus:13}%
  \BibitemOpen
  \bibfield  {author} {\bibinfo {author} {\bibfnamefont {H.}~\bibnamefont
  {Schomerus}},\ }\bibfield  {title} {\bibinfo {title} {{Topologically
  protected midgap states in complex photonic lattices}},\ }\href
  {https://doi.org/10.1364/OL.38.001912} {\bibfield  {journal} {\bibinfo
  {journal} {Opt. Lett.}\ }\textbf {\bibinfo {volume} {38}},\ \bibinfo {pages}
  {1912} (\bibinfo {year} {2013})}\BibitemShut {NoStop}%
\bibitem [{\citenamefont {Weimann}\ \emph {et~al.}(2017)\citenamefont
  {Weimann}, \citenamefont {Kremer}, \citenamefont {Plotnik}, \citenamefont
  {Lumer}, \citenamefont {Nolte}, \citenamefont {Makris}, \citenamefont
  {Segev}, \citenamefont {Rechtsman},\ and\ \citenamefont
  {Szameit}}]{Weimann2017TopologicallyPB}%
  \BibitemOpen
  \bibfield  {author} {\bibinfo {author} {\bibfnamefont {S.}~\bibnamefont
  {Weimann}}, \bibinfo {author} {\bibfnamefont {M.}~\bibnamefont {Kremer}},
  \bibinfo {author} {\bibfnamefont {Y.}~\bibnamefont {Plotnik}}, \bibinfo
  {author} {\bibfnamefont {Y.}~\bibnamefont {Lumer}}, \bibinfo {author}
  {\bibfnamefont {S.}~\bibnamefont {Nolte}}, \bibinfo {author} {\bibfnamefont
  {K.~G.}\ \bibnamefont {Makris}}, \bibinfo {author} {\bibfnamefont
  {M.}~\bibnamefont {Segev}}, \bibinfo {author} {\bibfnamefont {M.~C.}\
  \bibnamefont {Rechtsman}},\ and\ \bibinfo {author} {\bibfnamefont
  {A.}~\bibnamefont {Szameit}},\ }\bibfield  {title} {\bibinfo {title}
  {{Topologically protected bound states in photonic parity-time-symmetric
  crystals}},\ }\href {https://api.semanticscholar.org/CorpusID:26931680}
  {\bibfield  {journal} {\bibinfo  {journal} {Nature materials}\ }\textbf
  {\bibinfo {volume} {16 4}},\ \bibinfo {pages} {433} (\bibinfo {year}
  {2017})}\BibitemShut {NoStop}%
\bibitem [{Note1()}]{Note1}%
  \BibitemOpen
  \bibinfo {note} {Other choices could also be considered, but with less
  experimental relevance and a more complicated interpretation in terms of
  mutual information~\cite {Dalmonte_PhysRevB.101.085136}.}\BibitemShut {Stop}%
\bibitem [{\citenamefont {Peschel}(2003)}]{IngoPeschel_2003}%
  \BibitemOpen
  \bibfield  {author} {\bibinfo {author} {\bibfnamefont {I.}~\bibnamefont
  {Peschel}},\ }\bibfield  {title} {\bibinfo {title} {{Calculation of reduced
  density matrices from correlation functions}},\ }\href
  {https://doi.org/10.1088/0305-4470/36/14/101} {\bibfield  {journal} {\bibinfo
   {journal} {Journal of Physics A: Mathematical and General}\ }\textbf
  {\bibinfo {volume} {36}},\ \bibinfo {pages} {L205} (\bibinfo {year}
  {2003})}\BibitemShut {NoStop}%
\bibitem [{\citenamefont {Polkovnikov}\ \emph {et~al.}(2011)\citenamefont
  {Polkovnikov}, \citenamefont {Sengupta}, \citenamefont {Silva},\ and\
  \citenamefont {Vengalattore}}]{Polkovnikov_RMP11}%
  \BibitemOpen
  \bibfield  {author} {\bibinfo {author} {\bibfnamefont {A.}~\bibnamefont
  {Polkovnikov}}, \bibinfo {author} {\bibfnamefont {K.}~\bibnamefont
  {Sengupta}}, \bibinfo {author} {\bibfnamefont {A.}~\bibnamefont {Silva}},\
  and\ \bibinfo {author} {\bibfnamefont {M.}~\bibnamefont {Vengalattore}},\
  }\bibfield  {title} {\bibinfo {title} {{Nonequilibrium dynamics of closed
  interacting quantum systems}},\ }\href@noop {} {\bibfield  {journal}
  {\bibinfo  {journal} {Rev. Mod. Phys.}\ }\textbf {\bibinfo {volume} {83}},\
  \bibinfo {pages} {863} (\bibinfo {year} {2011})}\BibitemShut {NoStop}%
\bibitem [{\citenamefont {Calabrese}(2020)}]{Caleb_SciPostPhysLectNotes20}%
  \BibitemOpen
  \bibfield  {author} {\bibinfo {author} {\bibfnamefont {P.}~\bibnamefont
  {Calabrese}},\ }\bibfield  {title} {\bibinfo {title} {{Entanglement spreading
  in non-equilibrium integrable systems}},\ }\href
  {https://doi.org/10.21468/SciPostPhysLectNotes.20} {\bibfield  {journal}
  {\bibinfo  {journal} {SciPost Phys. Lect. Notes}\ ,\ \bibinfo {pages} {20}}
  (\bibinfo {year} {2020})}\BibitemShut {NoStop}%
\bibitem [{\citenamefont {Wang}\ \emph {et~al.}(2015)\citenamefont {Wang},
  \citenamefont {Xu}, \citenamefont {Wang},\ and\ \citenamefont
  {Wu}}]{PhysRevB.91.115118}%
  \BibitemOpen
  \bibfield  {author} {\bibinfo {author} {\bibfnamefont {D.}~\bibnamefont
  {Wang}}, \bibinfo {author} {\bibfnamefont {S.}~\bibnamefont {Xu}}, \bibinfo
  {author} {\bibfnamefont {Y.}~\bibnamefont {Wang}},\ and\ \bibinfo {author}
  {\bibfnamefont {C.}~\bibnamefont {Wu}},\ }\bibfield  {title} {\bibinfo
  {title} {{Detecting edge degeneracy in interacting topological insulators
  through entanglement entropy}},\ }\href
  {https://doi.org/10.1103/PhysRevB.91.115118} {\bibfield  {journal} {\bibinfo
  {journal} {Phys. Rev. B}\ }\textbf {\bibinfo {volume} {91}},\ \bibinfo
  {pages} {115118} (\bibinfo {year} {2015})}\BibitemShut {NoStop}%
\bibitem [{\citenamefont {Mbeng}\ \emph {et~al.}(2024)\citenamefont {Mbeng},
  \citenamefont {Russomanno},\ and\ \citenamefont {Santoro}}]{Mbeng_2024}%
  \BibitemOpen
  \bibfield  {author} {\bibinfo {author} {\bibfnamefont {G.~B.}\ \bibnamefont
  {Mbeng}}, \bibinfo {author} {\bibfnamefont {A.}~\bibnamefont {Russomanno}},\
  and\ \bibinfo {author} {\bibfnamefont {G.~E.}\ \bibnamefont {Santoro}},\
  }\bibfield  {title} {\bibinfo {title} {{The quantum Ising chain for
  beginners}},\ }\href {https://doi.org/10.21468/SciPostPhysLectNotes.82}
  {\bibfield  {journal} {\bibinfo  {journal} {SciPost Phys. Lect. Notes}\ ,\
  \bibinfo {pages} {82}} (\bibinfo {year} {2024})}\BibitemShut {NoStop}%
\bibitem [{Note2()}]{Note2}%
  \BibitemOpen
  \bibinfo {note} {The standard definition by Kitaev, which in row-vector form
  would read: \begin {equation} \begin {split} &\protect \check {\protect
  \mathbf {c}}^{\protect \phantom \dagger }_{} = (\protect \check {c}_{1},
  \protect \check {c}_{2}, \protect \check {c}_{3}, \protect \check {c}_{4},
  \protect \cdots , \protect \check {c}_{2N-1}, \protect \check {c}_{2L}
  )^{\scriptscriptstyle {\protect \mathrm {T}}} \equiv \\ &\equiv ( \protect
  \check {c}_{1,1}, \protect \check {c}_{2,1}, \protect \check {c}_{1,2},
  \protect \check {c}_{2,2}, \protect \cdots , \protect \check {c}_{1,N},
  \protect \check {c}_{2,L})^{\scriptscriptstyle {\protect \mathrm {T}}}
  \protect \tmspace +\thickmuskip {.2777em}, \end {split} \end {equation} mixes
  the different blocks of the Nambu fermions in a way that makes the algebra
  more complicated.}\BibitemShut {Stop}%
\bibitem [{\citenamefont {Prosen}(2010)}]{Prosen_2010}%
  \BibitemOpen
  \bibfield  {author} {\bibinfo {author} {\bibfnamefont {T.}~\bibnamefont
  {Prosen}},\ }\bibfield  {title} {\bibinfo {title} {{Spectral theorem for the
  Lindblad equation for quadratic open fermionic systems}},\ }\href
  {https://doi.org/10.1088/1742-5468/2010/07/P07020} {\bibfield  {journal}
  {\bibinfo  {journal} {Journal of Statistical Mechanics: Theory and
  Experiment}\ }\textbf {\bibinfo {volume} {2010}},\ \bibinfo {pages} {P07020}
  (\bibinfo {year} {2010})}\BibitemShut {NoStop}%
\bibitem [{\citenamefont {Prosen}(2008)}]{prosen}%
  \BibitemOpen
  \bibfield  {author} {\bibinfo {author} {\bibfnamefont {T.}~\bibnamefont
  {Prosen}},\ }\bibfield  {title} {\bibinfo {title} {{Third quantization: a
  general method to solve master equations for quadratic open Fermi systems}},\
  }\href {https://doi.org/10.1088/1367-2630/10/4/043026} {\bibfield  {journal}
  {\bibinfo  {journal} {New Journal of Physics}\ }\textbf {\bibinfo {volume}
  {10}},\ \bibinfo {pages} {043026} (\bibinfo {year} {2008})}\BibitemShut
  {NoStop}%
\bibitem [{\citenamefont {Kitaev}(2009)}]{Kitaev_2009}%
  \BibitemOpen
  \bibfield  {author} {\bibinfo {author} {\bibfnamefont {A.}~\bibnamefont
  {Kitaev}},\ }\bibfield  {title} {\bibinfo {title} {{Periodic table for
  topological insulators and superconductors}},\ }\href
  {https://doi.org/10.1063/1.3149495} {\bibfield  {journal} {\bibinfo
  {journal} {AIP Conference Proceedings}\ }\textbf {\bibinfo {volume} {1134}},\
  \bibinfo {pages} {22} (\bibinfo {year} {2009})}\BibitemShut {NoStop}%
\bibitem [{\citenamefont {Chiu}\ \emph {et~al.}(2016)\citenamefont {Chiu},
  \citenamefont {Teo}, \citenamefont {Schnyder},\ and\ \citenamefont
  {Ryu}}]{chiu}%
  \BibitemOpen
  \bibfield  {author} {\bibinfo {author} {\bibfnamefont {C.-K.}\ \bibnamefont
  {Chiu}}, \bibinfo {author} {\bibfnamefont {J.~C.~Y.}\ \bibnamefont {Teo}},
  \bibinfo {author} {\bibfnamefont {A.~P.}\ \bibnamefont {Schnyder}},\ and\
  \bibinfo {author} {\bibfnamefont {S.}~\bibnamefont {Ryu}},\ }\bibfield
  {title} {\bibinfo {title} {{Classification of topological quantum matter with
  symmetries}},\ }\href {https://doi.org/10.1103/RevModPhys.88.035005}
  {\bibfield  {journal} {\bibinfo  {journal} {Rev. Mod. Phys.}\ }\textbf
  {\bibinfo {volume} {88}},\ \bibinfo {pages} {035005} (\bibinfo {year}
  {2016})}\BibitemShut {NoStop}%
\bibitem [{\citenamefont {Surace}\ and\ \citenamefont
  {Tagliacozzo}(2022)}]{Surace_2022}%
  \BibitemOpen
  \bibfield  {author} {\bibinfo {author} {\bibfnamefont {J.}~\bibnamefont
  {Surace}}\ and\ \bibinfo {author} {\bibfnamefont {L.}~\bibnamefont
  {Tagliacozzo}},\ }\bibfield  {title} {\bibinfo {title} {{Fermionic Gaussian
  states: an introduction to numerical approaches}},\ }\href
  {https://doi.org/10.21468/SciPostPhysLectNotes.54} {\bibfield  {journal}
  {\bibinfo  {journal} {SciPost Phys. Lect. Notes}\ ,\ \bibinfo {pages} {54}}
  (\bibinfo {year} {2022})}\BibitemShut {NoStop}%
\bibitem [{\citenamefont {Sieberer}\ \emph {et~al.}(2023)\citenamefont
  {Sieberer}, \citenamefont {Buchhold}, \citenamefont {Marino},\ and\
  \citenamefont {Diehl}}]{sieberer2023universalitydrivenopenquantum}%
  \BibitemOpen
  \bibfield  {author} {\bibinfo {author} {\bibfnamefont {L.~M.}\ \bibnamefont
  {Sieberer}}, \bibinfo {author} {\bibfnamefont {M.}~\bibnamefont {Buchhold}},
  \bibinfo {author} {\bibfnamefont {J.}~\bibnamefont {Marino}},\ and\ \bibinfo
  {author} {\bibfnamefont {S.}~\bibnamefont {Diehl}},\ }\bibfield  {title}
  {\bibinfo {title} {{Universality in driven open quantum matter}},\ }\href
  {https://arxiv.org/abs/2312.03073} {\  (\bibinfo {year} {2023})},\ \Eprint
  {https://arxiv.org/abs/2312.03073} {arXiv:2312.03073} \BibitemShut {NoStop}%
\bibitem [{\citenamefont {Passarelli}\ \emph {et~al.}(2019)\citenamefont
  {Passarelli}, \citenamefont {Cataudella},\ and\ \citenamefont
  {Lucignano}}]{Passarelli_2019}%
  \BibitemOpen
  \bibfield  {author} {\bibinfo {author} {\bibfnamefont {G.}~\bibnamefont
  {Passarelli}}, \bibinfo {author} {\bibfnamefont {V.}~\bibnamefont
  {Cataudella}},\ and\ \bibinfo {author} {\bibfnamefont {P.}~\bibnamefont
  {Lucignano}},\ }\bibfield  {title} {\bibinfo {title} {{Improving quantum
  annealing of the ferromagnetic $p$-spin model through pausing}},\ }\href
  {https://doi.org/10.1103/PhysRevB.100.024302} {\bibfield  {journal} {\bibinfo
   {journal} {Phys. Rev. B}\ }\textbf {\bibinfo {volume} {100}},\ \bibinfo
  {pages} {024302} (\bibinfo {year} {2019})}\BibitemShut {NoStop}%
\bibitem [{\citenamefont {Friis}(2016)}]{Friis_2016}%
  \BibitemOpen
  \bibfield  {author} {\bibinfo {author} {\bibfnamefont {N.}~\bibnamefont
  {Friis}},\ }\bibfield  {title} {\bibinfo {title} {{Reasonable fermionic
  quantum information theories require relativity}},\ }\href
  {https://doi.org/10.1088/1367-2630/18/3/033014} {\bibfield  {journal}
  {\bibinfo  {journal} {New Journal of Physics}\ }\textbf {\bibinfo {volume}
  {18}},\ \bibinfo {pages} {033014} (\bibinfo {year} {2016})}\BibitemShut
  {NoStop}%
\bibitem [{\citenamefont {Friis}\ \emph {et~al.}(2013)\citenamefont {Friis},
  \citenamefont {Lee},\ and\ \citenamefont {Bruschi}}]{Friis_2013}%
  \BibitemOpen
  \bibfield  {author} {\bibinfo {author} {\bibfnamefont {N.}~\bibnamefont
  {Friis}}, \bibinfo {author} {\bibfnamefont {A.~R.}\ \bibnamefont {Lee}},\
  and\ \bibinfo {author} {\bibfnamefont {D.~E.}\ \bibnamefont {Bruschi}},\
  }\bibfield  {title} {\bibinfo {title} {{Fermionic-mode entanglement in
  quantum information}},\ }\href {https://doi.org/10.1103/PhysRevA.87.022338}
  {\bibfield  {journal} {\bibinfo  {journal} {Phys. Rev. A}\ }\textbf {\bibinfo
  {volume} {87}},\ \bibinfo {pages} {022338} (\bibinfo {year}
  {2013})}\BibitemShut {NoStop}%
\end{thebibliography}
\end{document}